\documentclass[12pt]{article}
\pdfoutput=1

\usepackage{putex}
\usepackage{graphicx}
\usepackage{caption}
\usepackage{amsmath, amsbsy, mathtools}
\usepackage{array}
\usepackage{subcaption}
\usepackage{epstopdf}
\usepackage{enumerate}
\usepackage{cite}
\usepackage{youngtab}
\usepackage{tensor}
\usepackage{slashed}
\usepackage[aligntableaux=center]{ytableau}
\usepackage[utf8]{inputenc}
\usepackage[
      colorlinks=true,
      linkcolor=blue,
      urlcolor=blue,
      filecolor=blacsk,
      citecolor=red,
      ]{hyperref}
\usepackage{dsfont}
\usepackage{slashed}
\usepackage{bm}
\usepackage{bbm}
\usepackage{booktabs}
\usepackage{comment}
\usepackage{placeins}

\newcommand{\grp}[1]{\mathrm{#1}}

\newcommand{\grU}{\grp{U}}
\newcommand{\grSU}{\grp{SU}}
\newcommand{\grSO}{\grp{SO}}

\newcommand {\be} {\begin {equation}}
\newcommand {\ee} {\end {equation}}

\newcommand {\bes} {\begin {equation*}}
\newcommand {\ees} {\end {equation*}}

\newcommand{\es}[2] {\begin{equation} \label{#1} \begin{split} #2 \end{split} \end{equation}}

\newcommand{\Z}{\mathbb{Z}}

\newcommand{\beq}{\begin{equation}}
\newcommand{\eeq}{\end{equation}}

\def\ie{\begin{equation}\begin{aligned}}
\def\fe{\end{aligned}\end{equation}}

\newcommand{\m}{\mu}

\numberwithin{equation}{section}

\def\<{\langle}
\def\>{\rangle}

\newcommand{\ket}[1]{|#1\rangle}

\let\tr\relax
\DeclareMathOperator{\tr}{tr}
\DeclareMathOperator{\SU}{SU}

\DeclareMathOperator{\spn}{span}
\DeclareMathOperator{\Lie}{Lie}

\def\V{\mathcal{V}}
\def\H{\mathcal{H}}
\def\Q{\mathcal{Q}}
\def\W{\mathcal{W}}
\let\O\relax
\def\O{\mathcal{O}}

\usepackage{tikz}
\usetikzlibrary{calc}
\tikzstyle{tensor}=[circle,draw=black!50,thick, minimum size = 40pt]

\tikzstyle{square}=[fill=white, draw=black, shape=rectangle, rounded corners, minimum width=1.cm, minimum height=1.cm, inner sep=0pt, outer sep=0pt,thick]
\tikzstyle{circle}=[fill=white, draw=black, shape=circle, minimum width=1.cm, minimum height=1.cm, inner sep=0pt, outer sep=0pt, thick]
\tikzstyle{shade}=[dotted,fill=gray!30,rounded corners=5mm]
\tikzset{
	diagram/.style={
		baseline={([yshift=-.5ex]current bounding box.center)},
		every node/.style={scale=0.7},
		thick
	}
}

\graphicspath{{}}

\begin{document}

\preprint{PUPT-2657}

\institution{PU}{Joseph Henry Laboratories, Princeton University, Princeton, NJ 08544, USA}
\institution{PCTS}{Princeton Center for Theoretical Science, Princeton University, Princeton, NJ 08544, USA}

\title{Infinite matrix product states for $(1+1)$-dimensional gauge theories}

\authors{Ross Dempsey,\worksat{\PU} Anna-Maria E. Gl\"uck,\worksat{\PU} Silviu S.~Pufu,\worksat{\PU,\PCTS}  
 \\[10pt] and Benjamin T. S\o{}gaard\worksat{\PU}}

\abstract{We present a matrix product operator construction that allows us to represent the lattice Hamiltonians of (abelian or non-abelian) gauge theories in a local and manifestly translation-invariant form. In particular, we use symmetric matrix product states and introduce link-enhanced matrix product operators (LEMPOs) that can act on both the physical and virtual spaces of the matrix product states. This construction allows us to study Hamiltonian lattice gauge theories on infinite lattices. As examples, we show how to implement this method to study the massless and massive one-flavor Schwinger model and adjoint QCD$_2$.
}
\date{August 2025}

\maketitle

\tableofcontents

\section{Introduction}

An important problem in modern physics is to understand non-perturbative phenomena of gauge theories, such as color confinement and mass gap generation. $(1+1)$-dimensional gauge theories can serve as tractable toy models where the strongly coupled dynamics can be studied. A simple example of such a model is $\grU(1)$ gauge theory coupled to a massless Dirac fermion, also known as the Schwinger model \cite{Schwinger:1962tp}. This model is solvable: the massless electron and positron bind to form a massive non-interacting boson. Other models, such as adjoint QCD$_2$ \cite{Dalley:1992yy}, have richer dynamics and are not known to have analytic solutions. For these models we can turn to numerical schemes, such as Euclidean or Hamiltonian lattice methods. In this paper, we pursue the latter approach and introduce a novel tensor network approach for solving $(1+1)$-dimensional Hamiltonian lattice gauge theories.

Hamiltonian lattice gauge theory was introduced by Kogut and Susskind in \cite{Kogut:1974ag}. In contrast to Wilson's Euclidean approach to lattice gauge theory \cite{Wilson:1974sk}, in which spacetime is discretized into a lattice and the path integral is reduced to a classical sampling problem, the Hamiltonian approach is based on a discretization of space that leads to a quantum many-body problem. This problem can be formulated on both classical and quantum computers (see, for instance, \cite{Bauer:2022hpo, DiMeglio:2023nsa}). The Hamiltonian lattice offers several advantages; for instance, it does not suffer from a sign problem \cite{Banuls:2016jws, Banuls:2016gid, Banuls:2016hhv}, it reduces the amount of fermion doubling \cite{Kogut:1974ag}, it offers direct access to the spectrum of the theory \cite{Banuls:2013jaa, Buyens:2013yza}, and it allows for the simulation of real-time dynamics \cite{Buyens:2013yza, Pichler:2015yqa, Buyens:2015tea, Buyens:2016hhu}.

The price we pay for these advantages is the exponential scaling of the Hilbert space with the number of lattice sites, where the matter fields live, and with the number of links, where the gauge fields live. This scaling makes it practically impossible to calculate the spectrum on even modestly-sized lattices in three or more spacetime dimensions using exact diagonalization. In $(1+1)$ dimensions, it is usually possible to carry out exact diagonalization for relatively small lattices and low-rank gauge groups (see e.g. \cite{Nagele:2018egu,Dempsey:2022nys,Dempsey:2023fvm,Dempsey:2024alw}), but the most accurate results to date have been obtained using tensor network methods (for reviews, see \cite{Banuls:2018jag, Banuls:2019rao, Banuls:2019bmf, Meurice:2020pxc, Meurice:2022xbk}).

Tensor network methods have been developed extensively for one-dimensional condensed matter lattice models \cite{Orus:2018dya,Banuls:2022vxp}. These methods assume that the ground state wave function of a gapped many-body quantum system is represented by a matrix product state (MPS), which by construction exhibits the area-law entanglement entropy that is expected in such a state. In particular, the ground state is constructed from local rank-three MPS tensors that each have one physical index and two virtual indices.  Adjacent tensors have their virtual indices contracted, and the logarithm of the dimensions of these virtual spaces bound the entanglement entropy \cite{Okunishi:2021but}.  MPSs give a very efficient representation of low-lying energy eigenstates. Together with powerful algorithms such as the density matrix renormalization group (DMRG) \cite{White:1992zz, White:1993zza}, this representation allows for precision studies of lattice systems with hundreds or even thousands of sites \cite{Schollwoeck:2010uqf}. For local Hamiltonians with manifest translation invariance, MPS methods that work directly on an infinite lattice have also been developed (see e.g. \cite{Ostlund:1995zz, Vidal:2006ofj, McCulloch:2008aun}). A modern example of such an algorithm is the variational uniform matrix product states (VUMPS) algorithm \cite{Zauner-Stauber:2017eqw}, which produces similarly precise results without (potentially unwanted) boundary effects.

In this work, we introduce a novel MPS setup for gauge theories that uses {\em symmetric} MPSs \cite{Perez-Garcia:2008hhq, Sanz:2009lax}. Symmetric MPSs have traditionally been used to simulate condensed matter systems with global symmetries, for which each such symmetry is generated by a sum of local (on-site) charges.  The MPS tensors of a symmetric MPS obey local symmetry constraints. In this paper, we show that these constraints can be identified with the Gauss law constraints of a Hamiltonian lattice gauge theory. It follows that the virtual spaces of a symmetric MPS contain information about the state of the gauge field. To extract this information, we will introduce a new type of matrix product operator (MPO) construction, which we call a link-enhanced matrix product operator (LEMPO)\@.  A LEMPO can operate on both the physical and virtual spaces of the symmetric MPS, as opposed to an MPO, which operates only on physical spaces.  The use of LEMPOs provides a natural way of writing the Hamiltonian of a lattice gauge theory, and as we will describe, it avoids some pitfalls of earlier MPS implementations for gauge theories.   In particular, by using LEMPOs, we can apply infinite MPS techniques to generic Hamiltonian lattice gauge theories.

We can contrast our approach with two previous strategies for applying MPSs to gauge theories. The first strategy (see, for instance, \cite{Byrnes:2002gj, Banuls:2013jaa, Banuls:2015sta, Banuls:2016lkq, 
Banuls:2016gid,
Banuls:2016jws,
Funcke:2019zna, 
Papaefstathiou:2021cho,
Honda:2022edn, 
Funcke:2023lli, Dempsey:2023gib, Angelides:2023bme, ArguelloCruz:2024xzi,
Itou:2024psm, 
Papaefstathiou:2024zsu}), is to eliminate the gauge field degrees of freedom in the Hamiltonian using the Gauss law. This process converts the gauge theory Hamiltonian into a Hamiltonian for the matter degrees of freedom with non-local interactions~\cite{Banuls:2013jaa}, at which point finite tensor network methods can be used to obtain results on large lattices. However, this rewriting spoils the manifest translation invariance and locality of the Hamiltonian, so that infinite tensor network methods are not applicable.  (See, however, \cite{Godfrey:2025heu} for a different recent approach that uses infinite MPSs for abelian gauge theories.) In contrast, because the LEMPO representation of the Hamiltonian directly accesses the gauge fields on the links, it is local and translation-invariant, and so infinite tensor network methods can be readily applied. 

Another common strategy is to add additional sites to the MPS that represent the gauge field degrees of freedom. The extra physical Hilbert spaces are infinite-dimensional and must be made finite. This truncation can be achieved by imposing a cut-off by hand (see also \cite{Horn:1981kk, Orland:1989st, Chandrasekharan:1996ih, Rico:2013qya, Silvi:2014pta, Tagliacozzo:2014bta} for quantum link models, which truncate using auxiliary fermions). The Hamiltonian remains local and manifestly translation-invariant, so that infinite tensor network methods can be applied \cite{Buyens:2013yza, Buyens:2015tea, Buyens:2017crb, Fujii:2024reh, Zohar:2015jnb}. However, as we will explain in Section~\ref{subsec:schwinger}, this method involves constraints on the MPS tensors that cannot be framed as local gauge invariance. This makes it difficult to extend the setup beyond abelian theories (see, however, \cite{Banuls:2017ena, Silvi:2019wnf, Cataldi:2023xki}). In contrast, to use LEMPOs we only need to impose the local symmetry constraints that appear in symmetric MPSs, and so it is straightforward to use LEMPOs in both abelian and non-abelian theories. We will give examples of both in this paper.

As we will show, a key advantage of LEMPOs is that they can be implemented within standard tensor network algorithms such as DMRG and VUMPS with relative ease. In addition, by leveraging existing methods for the adjustment of virtual space dimensions in an MPS, we avoid the need for an explicit cutoff on the Hilbert space of the gauge field: this cutoff is chosen dynamically by the optimization algorithms. For the numerical calculations in this paper, we implemented LEMPOs as an extension to the Julia package \texttt{MPSKit.jl} \cite{MPSKit2025}, which already provides algorithms for symmetric MPSs with abelian or non-abelian symmetries. See Appendix~\ref{app:algo} for some implementation details.

To illustrate that our method provides a powerful new tool for probing the physics of gauge theories, we apply infinite matrix product states together with VUMPS and the quasiparticle ansatz \cite{Haegeman:2013} to study some $(1+1)$-dimensional gauge theories of particular recent interest. In general, we find that our method can give precise estimates of many quantities of interest over a wide range of parameter values.

First, we study the massive Schwinger model as a benchmark of our method. This model has been studied previously using both strategies mentioned above (see, for instance, \cite{Banuls:2013jaa, Buyens:2013yza, Rico:2013qya, Funcke:2019zna, 
 Honda:2022edn, Dempsey:2023gib, ArguelloCruz:2024xzi, Fujii:2024reh}), as well as with older methods such as the lattice strong-coupling expansion \cite{Banks:1975gq, Hamer:1997dx, Berruto:1997jv}. Our method can be compared directly with the strong-coupling expansion, since both are formulated on infinite lattices. We show that the two methods agree very well, but that tensor networks allow us to make a much more precise extrapolation to the continuum limit. Indeed, for a wide range of values for the electron mass and the theta-angle (described in Section~\ref{subsec:schwinger}; see \eqref{eq:schwinger_action}), we can reliably extrapolate the ground state energy density, expectation values, and bound state spectra as the lattice spacing goes to zero. We present our results for these quantities and discuss how they realize the physics expected in the massive Schwinger model \cite{Coleman:1976uz}.

Second, as an example of a non-abelian theory, we study $\SU(N_c)$ adjoint QCD$_2$ (with $N_c = 2,3$). The implementation is very similar to that of an abelian theory, because our method is based on symmetric MPS algorithms that apply to either case. Using our method, we calculate ground state properties, low-lying bound state masses, and the tension of fundamental strings to high precision, improving on the results of \cite{Dempsey:2023fvm, Dempsey:2024alw}.  In particular, we study both the trivial and non-trivial flux tube sectors of this theory, and we find numerical evidence for the perimeter-law behavior of the fundamental Wilson loop in the massless case \cite{Gross:1997mx, Komargodski:2020mxz} and for the supersymmetry of the model at the value of the mass found in \cite{Kutasov:1993gq, Boorstein:1993nd, Popov:2022vud,Klebanov:2025mbu}.

The rest of this paper is organized as follows. In Section \ref{sec:lmpo}, we introduce LEMPOs for both abelian and non-abelian $(1+1)$-dimensional gauge theories. We show how the LEMPOs allow for the lattice gauge theory Hamiltonian to be expressed in a local and manifestly translation-invariant form. In Section \ref{sec:results}, we present our numerical results obtained using this method for the Schwinger model and adjoint QCD$_2$. We end with a discussion of our results in Section~\ref{sec:discussion}, where we also make some comments on how this method extends to higher-dimensional gauge theories.  Technical details are relegated to the appendices.

\section{Link-enhanced matrix product operators}\label{sec:lmpo}

In this section, we describe the symmetric MPS construction for abelian and non-abelian gauge theories and explain how a LEMPO  generalizes the notion of an MPO\@.  As a warm-up, in Section~\ref{subsec:schwinger} we introduce symmetric MPSs as well as the idea of operators acting on virtual spaces in the lattice Schwinger model \cite{Banks:1975gq}, and in Section~\ref{subsec:lgt} we generalize these ideas to non-abelian gauge theories. In Section~\ref{subsec:LEMPO}, we show how a general gauge theory Hamiltonian can be expressed as a LEMPO\@.

\subsection{Warm-up: The lattice Schwinger model}\label{subsec:schwinger}

Here, we will motivate the notion of operators acting on virtual bonds in a well-known Hamiltonian lattice gauge theory, the lattice Schwinger model \cite{Banks:1975gq}. We will review how the Gauss law for this model is naturally obeyed by a symmetric MPS\@. By using link operators, we will be able to read off information about the gauge field from a symmetric MPS in a local manner, paving the way for a simpler MPS implementation of this model.

\subsubsection{The Hamiltonian and Gauss law}

The Schwinger model \cite{Schwinger:1962tp} is the relativistic theory of quantum electrodynamics (QED) in $1+1$ dimensions with an electron and positron of mass $m$. In the Lagrangian formalism, its field content consists of the $\grU(1)$ gauge field $A_\mu$ with field strength $F_{\mu\nu} = \partial_\mu A_\nu - \partial_\nu A_\mu$ and the two-component complex spinor field $\psi$.  The action is given by
\begin{equation}\label{eq:schwinger_action}
    S = \int d^2x\,\left(-\frac{1}{4g^2}F_{\mu\nu}F^{\mu\nu} - \frac{\theta}{4\pi}\epsilon^{\mu\nu}F_{\mu\nu} + i\bar\psi \gamma^\mu (\partial_\mu - i A_\mu)\psi - m \bar\psi \psi\right)\,,
\end{equation}
where $g>0$ is the gauge coupling (the elementary charge) and $\theta$ is the theta-angle, which, in $(1+1)$ dimensions, can be interpreted as a background electric field of $\theta/(2\pi)$.  The massless Schwinger model was first introduced in \cite{Schwinger:1962tp} and further studied in \cite{Lowenstein:1971fc, Casher:1974vf, Coleman:1975pw}, where it was also generalized to $m \neq 0$. We defer a discussion of the properties of this model until Section~\ref{subsec:schwinger_results};  for now, our interest is in the Hamiltonian lattice regularization of the Schwinger model and its corresponding MPS implementation.

\begin{figure}[t]
    \centering
    \begin{tikzpicture}[xscale=3,yscale=1.2]
		\draw[ultra thick] (0,0) -- (2,0);
        \draw[ultra thick] (3,0) -- (4,0);
        \draw[ultra thick,dashed] (2,0) -- (3,0);
		
		\foreach \x in {1,2} {
			\draw[ultra thick,-latex] (\x-1, 0) -- ({\x - 1 + .6},0);
			\node at ({\x - 1 + .5}, 0.3) {$U_\x$};
		}

        \foreach \x in {4} {
			\draw[ultra thick,-latex] (\x-1, 0) -- ({\x - 1 + .6},0);
			\node at ({\x - 1 + .5}, 0.3) {$U_{N-1}$};
		} 

        \node[blue] at ({4 - 1}, -0.3) {$Q_{N-1}$};
        \node[green!50!black] at ({5 - 1}, -0.3) {$Q_{N}$};

        \node[green!50!black,fill,circle,minimum size=5,inner sep=0,label=90:{\color{green!50!black} $c_N$}] at (5 - 1, 0) {};

        \node[blue,fill,circle,minimum size=5,inner sep=0,label=90:{\color{blue} $c_{N-1}$}] at (4 - 1, 0) {};
        
		\foreach \x in {1,3} {
			\node[blue,fill,circle,minimum size=5,inner sep=0,label=90:{\color{blue} $c_\x$}] at (\x - 1, 0) {};
			\node[blue] at ({\x - 1}, -0.3) {$Q_\x$};
		}
		\foreach \x in {2} {
			\node[green!50!black,fill,circle,minimum size=5,inner sep=0,label=90:{\color{green!50!black} $c_\x$}] at (\x - 1, 0) {};
			\node[green!50!black] at ({\x - 1}, -0.3) {$Q_\x$};
		}
		\foreach \x in {1,2} {
			\node at (\x-1+0.5, -0.3) {$L_\x$};
		}

        \node at (4-1+0.5, -0.3) {$L_{N-1}$};
        
	\end{tikzpicture}
    \caption{The spatial arrangement of the degrees of freedom in the Schwinger model Hamiltonian on the lattice.}
    \label{fig:lattice_schematic_schwinger}
\end{figure}
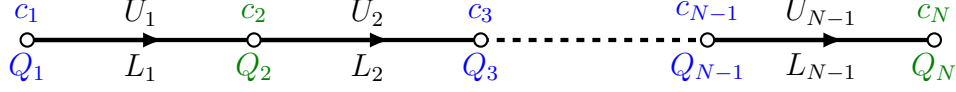

The Hamiltonian lattice regularization of the Schwinger model was first introduced in \cite{Banks:1975gq} using staggered lattice fermions \cite{Kogut:1974ag} on a lattice with an even number $N$ of sites. The lattice model is written in terms of complex lattice fermionic operators $c_n$ (with conjugates $c_n^\dagger$), living on sites and obeying the canonical anti-commutation relations $\{c_n, c_m^\dagger\}=\delta_{nm}$, along with gauge parallel transport operators $U_n$ and their conjugate electric field operators $L_n$, living on links and obeying $[L_n, U_m] = \delta_{nm} U_n$. The degrees of freedom on the lattice are illustrated in Figure~\ref{fig:lattice_schematic_schwinger}.  The operators mentioned above act on a product Hilbert space  $\mathcal H = \mathcal H_\text{matter}\otimes \mathcal H_\text{gauge}$. The fermionic Hilbert space $\mathcal H_\text{matter}$ has dimension $2^N$ and is generated by acting with the creation operators $c^\dagger_n$ on the standard vacuum state $\ket{0}$ that is annihilated by all $c_n$, $c_n\ket{0}=0$. The bosonic Hilbert space $\mathcal H_\text{gauge}$ is the product of the infinite-dimensional spaces $L^2(\grU(1))=L^2(S^1)$ on every link, each one representing the Hilbert space of a particle moving on a circle.  On any given link, we can consider a basis of momentum states $\ket{\ell_n}$, with $\ell_n\in\mathbb Z$.   The operators $U_n$ and $L_n$ act on this basis as
\begin{equation}
    U_n \ket{\ell_n} = \ket{\ell_n+1}\,,\qquad L_n\ket{\ell_n}=\ell_n\ket{\ell_n}\,.
\end{equation}

The dynamics of the lattice Schwinger model is given by the Hamiltonian \cite{Banks:1975gq}
\begin{equation}
    H_{\grU(1)} = \sum_{n=1}^{N-1} \left[\frac{g^2a}{2} \left(L_n+\frac{\theta}{2\pi}\right)^2 - \frac{i}{2a} \left(c_n^\dagger U_n c_{n+1} - c_{n+1}^\dagger U_n^\dagger c_n\right) \right] + m_{\mathrm{lat}} \sum_{n=1}^N (-1)^n c_n^\dagger c_n\,,
    \label{eq:SchwingerH}
\end{equation}
where $m_\text{lat}$ is the lattice mass\footnote{We will use the identification between the lattice and continuum mass parameters $m_\text{lat} = m - \frac{g^2 a}{8}$ as given in \cite{Dempsey:2022nys}.} and $a>0$ is the lattice spacing, supplemented by the Gauss law constraint \cite{Potvin:1985gw, Hamer:1997dx}
\begin{align}
    L_n-L_{n-1} = Q_n \,, \qquad\text{with}\qquad Q_n \equiv c_n^\dagger c_n - \delta_{n, \text{odd}}\,. \label{eq:GaussSchwinger}
\end{align}
Here, $Q_n$ is the local $\grU(1)$ charge on the lattice sites. The staggered shift in the definition of $Q_n$ is such that $Q_n\in\{1, 0\}$ on even sites and $Q_n\in\{0, -1\}$ on odd sites.\footnote{This prescription allows for charge conjugation symmetry when $\theta = 0$ or $\pi$. From the charges on sites, we see that the charge conjugation operator necessarily involves translation by one site.} We are interested in diagonalizing the Hamiltonian in the physical subspace $\H_\text{phys}$ spanned by the states of $\H$ that satisfy the Gauss law \eqref{eq:GaussSchwinger}. The Hamiltonian \eqref{eq:SchwingerH} and Gauss law \eqref{eq:GaussSchwinger} are invariant under gauge transformations parameterized by independent unitary operators $V_n$ living on sites.  For a given choice of $V_n$, the corresponding gauge transformation is implemented by a unitary operator $\mathbb{V}$  acting on $\H$, with the property that on operators $X$ it acts by $\mathbb{V}^\dagger X \mathbb{V} = X'$, with
 \es{GaugeTransf}{
  c_n' = V_n c_n \,, \qquad U_n' = V_n U_n V_{n+1}^\dagger \,, \qquad 
   L_n' = L_n \,.
 }
One can easily see that the Hamiltonian \eqref{eq:SchwingerH} and Gauss law \eqref{eq:GaussSchwinger} are invariant under conjugation by $\mathbb{V}$.

The most common strategy for dealing with the gauge degrees of freedom (see e.g. \cite{Byrnes:2002gj, Banuls:2013jaa}) is to fully gauge-fix the Hamiltonian, removing the $\H_\text{gauge}$ factor from the Hilbert space and formulating the problem on $\H_\text{matter}$ alone. In the Hamiltonian, this amounts to replacing $U_n \mapsto \mathds 1$ and solving the Gauss law \eqref{eq:GaussSchwinger} to rewrite the electric field operators in terms of the matter charges.  If $L_0$ and $L_N$ are the electric fields at the left and right boundary of our chain,\footnote{In general, $L_0$ and $L_N$ are operators acting on an auxiliary Hilbert space representing the boundary conditions. We will take this Hilbert space to be one-dimensional and take $L_0 = L_N = 0$.} which we will set to zero in the discussion below, the solution of the Gauss law \eqref{eq:GaussSchwinger} is
\begin{equation}
    L_n = L_0 + \sum_{k=1}^n Q_k\,, \qquad 
    \label{eq:Lnsum}
\end{equation}
with
 \begin{equation}
     L_N -L_0= \sum_{k=1}^N Q_k \label{eq:gauge_constraint}
 \end{equation}
interpreted as a global constraint. For details of the gauge-fixing procedure in the quantum theory, see Appendix~\ref{app:quantum_gauge}.

One can then use a Jordan-Wigner transformation to write the fermion operators in terms of Pauli matrices $\{\sigma^x_n, \sigma^y_n, \sigma^z_n\}$ (or the combination $\sigma^\pm_n = \frac{\sigma^x_n \pm i\sigma^y_n}{\sqrt{2}}$) acting on two-dimensional Hilbert spaces at each site $n$:
\begin{equation}\label{eq:jordan_wigner}
    c_n = (-i\sigma^z_1)\ldots (-i\sigma^z_{n-1})\sigma^-_n\,,\qquad c_n^\dagger = (i\sigma^z_1)\ldots (i\sigma^z_{n-1})\sigma^+_n\,.
\end{equation}
After fixing the gauge and performing the Jordan-Wigner transformation, the Hamiltonian takes the form
\begin{equation}\label{eq:schwinger_bosonized}
    H_{\grU(1),\text{bos}} = \sum_{n=1}^{N-1} \left[\frac{g^2a}{2} \left(\frac{\theta}{2\pi} + \sum_{k=1}^nQ_k\right)^2 + \frac{1}{2a} \left(\sigma_n^+ \sigma^-_{n+1} + \sigma_{n+1}^+ \sigma_n^-\right) \right] + \frac{m_{\mathrm{lat}}}{2} \sum_{n=1}^N (-1)^n \sigma^z_n\,,
\end{equation}
with the charge given by $Q_n = \frac12 (\sigma_n^z+(-1)^n)$. The only remnant of the Gauss law is the constraint \eqref{eq:gauge_constraint}, which for $L_0 = L_N = 0$ is:
 \begin{equation}
    Q = 0 \,,\qquad\text{where} \qquad Q \equiv \sum_{n=1}^N Q_n = \sum_{n=1}^N\sigma^z_n \,. \label{eq:zero_charge}
 \end{equation} 
 We can think of \eqref{eq:schwinger_bosonized} as a Hamiltonian on the $2^N$-dimensional space of spin states.  This Hamiltonian has a global $\grU(1)$ symmetry generated by $Q$, and to study our gauge theory, we need to work in the zero-charge sector of this system.

\subsubsection{MPS implementation}
 
As discussed at the beginning of Section~\ref{subsec:schwinger_results} below, the lattice Schwinger model is gapped for generic values of $a$ and $m_\text{lat}$, which implies that MPSs can be used to efficiently approximate low-lying energy eigenstates.  MPSs for the lattice Schwinger model were introduced in \cite{Banuls:2013jaa}, and they have been used in many other works (see, for instance, \cite{Buyens:2013yza, Rico:2013qya, Funcke:2019zna, 
 Honda:2022edn, Dempsey:2023gib, ArguelloCruz:2024xzi, Fujii:2024reh}). We will now briefly review how the ground state of a Hamiltonian such as \eqref{eq:schwinger_bosonized} can be represented using MPSs, and how to impose the zero-charge constraint \eqref{eq:zero_charge} in this framework.

In general, for a system with $N$ sites that host some quantum degrees of freedom (for us, on each site we have a qubit), the most general state is
 \begin{equation}
     \ket{\psi} = \sum_{\{i_n\}} \psi_{i_1 \ldots i_N} \ket{i_1}\otimes\cdots\otimes\ket{i_N} \,, \label{eq:generic_state}
 \end{equation}
where $\{\ket{i_n}\}$ is a basis for the Hilbert space $\H_n$ of the degrees of freedom on site $n$.  Any such state can be written in MPS form, as follows. An MPS is specified by tensors $A_n$ at each site, with physical indices $i_n$ that are contracted with the kets of each local Hilbert space and virtual indices $\alpha_n$.  For given $n$, the virtual indices $\alpha_n$ label a basis for a virtual Hilbert space ${\cal V}_n$ of dimension $D_n$, and they are contracted between neighboring sites \cite{Perez-Garcia:2006nqo}:
\begin{equation}\label{eq:mps_def}
	\begin{split}
		\ket{A_1,\ldots,A_N}
		&=\sum_{\{i_n,\alpha_n\}} (v_0)^{\alpha_0} (A_1)_{i_1}^{\alpha_0,\alpha_1}  \cdots (A_N)^{\alpha_{N-1},\alpha_N}_{i_{N}} (v_{N+1})^{\alpha_N} 		\times \ket{i_1}\otimes\cdots\otimes\ket{i_N}\\
		&=\begin{tikzpicture}[diagram]
			\node[style=square] at (0,0) (A1) {$A_{n-1}$};
			\node[style=square] at (1.5,0) (A2) {$A_{n}$};
			\node[style=square] at (3,0) (A3) {$A_{n+1}$};
			\node at (0,-1) {$\ket{i_{n-1}}$};
			\node at (1.5,-1) {$\ket{i_n}$};
			\node at (3,-1) {$\ket{i_{n+1}}$};
			\node at (-1.,0) {$\cdots$};
			\node at (4.,0) {$\cdots$};
			\node at (-1.,-1.) {$\cdots$};
			\node at (4.,-1.) {$\cdots$};
			\draw (A1) -- (A2) -- (A3);
			\draw (A1) -- ++(0,-.8);
			\draw (A2) -- ++(0,-.8);
			\draw (A3) -- ++(0,-.8);
			\draw (A1) -- ++(-0.75,0);
			\draw (A3) -- ++(0.75,0);
		\end{tikzpicture}\,.
	\end{split}
\end{equation}
We fix $D_0 = D_N = 1$ and set the sole components of the vectors $v_0$ and $v_{N+1}$ to 1. 

To write a state \eqref{eq:generic_state} in the many-body Hilbert space $\H_\mathrm{MB} = \otimes_{n=1}^N \H_n$ as an MPS, we can iteratively applying matrix factorizations such as the singular value decomposition (see Appendix~\ref{app:factoring} for details). The number of components needed to describe an MPS is $\sum_n \dim(\H_n) D_{n-1}D_n$. For a generic state in $\H_\text{MB}$, the MPS representation requires values of $D_n$ that scale exponentially with the system size.  However, gapped ground states of one-dimensional chains have an entanglement structure that allows them to be well-approximated by matrix product states for which $\max_{n}(D_n)$ does not scale with system size $N$ \cite{Hastings:2006ple, Verstraete:2006mdr}. Thus, in searching for a gapped ground states, MPSs provide efficient variational ansätze.

To minimize the energy of an MPS ansatz, we can use variational algorithms such as DMRG \cite{White:1992zz, White:1993zza, McCulloch:2007azj, Schollwock:2005zz, Schollwoeck:2010uqf}\@. However, these methods alone do not incorporate the constraint \eqref{eq:zero_charge}. One could impose this constraint energetically, but it is more elegant and more efficient to use {\em symmetric matrix product states} \cite{Perez-Garcia:2008hhq, Sanz:2009lax,Singh_u1} instead. A symmetric MPS is an MPS for which both the physical and virtual Hilbert spaces form representations of a group $G$, and for which the tensors $A_n$ are invariant under $G$.  More explicitly, if $u_n^{(g)}$ and $\mathcal{U}_n^{(g)}$ are unitary operators that represent the action of a group element $g \in G$ on $\H_n$ and $\mathcal{V}_n$, respectively, the condition that $A_n$ is invariant under $G$ is
 \begin{equation}
     \left(u_n^{(g)}\right)^{i_n i_n'} A_n^{i_n'} = \mathcal{U}^{(g)\dagger}_{n-1} A_n^{i_n} \mathcal{U}_n^{(g)} \,. \label{eq:group_invariance}
 \end{equation}
If an MPS of the form \eqref{eq:mps_def} is made up of tensors obeying the condition \eqref{eq:group_invariance}, it is easy to check that the MPS itself also transforms in a definite representation of $G$, with the unitary operator that represents the group element $g$ simply given by the tensor product $u^{(g)} \equiv \otimes_{n=1}^N u_n^{(g)}$.  In fact, in can be shown that the converse is also true:  provided that a state $\ket{\psi}$ transforms (trivially or non-trivially) in a representation of a group $G$ that has an action on each site, then such a state $\ket{\psi}$ can be represented by a symmetric MPS\@.  For a simple proof, see Appendix~\ref{app:factoring}.

In our Schwinger model application, we are interested in the case where $G = \grU(1)$ is generated by $Q = \sum_{n=1}^N Q_n$.  As explained around \eqref{eq:zero_charge}, gauge invariance requires us to work in the $Q=0$ sector of $\H_\mathrm{matter}$.  Since states in this sector transform trivially under $G$, we can use symmetric MPSs to study them. We can parameterize the group element $g = e^{i \varphi}$ by an angle $\varphi$ so that $u^{(g)} = e^{i \varphi Q}$. As this is a continuous symmetry, we consider the infinitesimal limit $\varphi \ll 1$, which defines the generators of the symmetry
\begin{equation}\label{eq:inf_gen}
    u^{(g)}_n = \mathds{1} + i\varphi Q_n+\mathcal O(\varphi^2)\,,\qquad  \mathcal {U}^{(g)}_n = I + i\varphi \mathcal Q_n+\mathcal O(\varphi^2)\,,
\end{equation}
for the physical and virtual spaces, respectively. In terms of the infinitesimal generators, the constraint \eqref{eq:group_invariance} becomes
\begin{equation}
    \left[\left(\Q_{n-1}\right)_{\alpha_{n-1}{\alpha_{n-1}'}}+\left(Q_n\right)^{i_n i_n'}-\left(\Q_{n}\right)_{\alpha_n \alpha_n'}\right] \left(A_n\right)^{i_n'}_{\alpha_{n-1}' \alpha_n'}=0\,,
    \label{eq:charge_conservation_U1}
\end{equation}
which is very reminiscent of the Gauss law \eqref{eq:GaussSchwinger}. Thus, symmetric MPSs do more than just enforce the charge-zero condition: they naturally exhibit the locally constrained structure of a lattice gauge theory, provided we identify the action of $L_n$ on $\H_\mathrm{matter}$ with the action of $\Q_n$ on~$\V_n$. This identification also clarifies the role of the virtual space $\V_n$ in relation to the gauge theory: in rough terms, it carries the gauge field degrees of freedom that we previously gauge-fixed away.

The condition \eqref{eq:charge_conservation_U1} looks simpler if we choose bases for $\H_n$ and $\V_n$ in which $Q_n$ and $\Q_n$ are diagonal. Let us denote the charge of the $(\alpha_n)${th} basis vector of $\V_n$ by $q(\alpha_n)$. Likewise, we choose the standard $\sigma^z$-basis for $\H_n$ so that the charges defined by $Q_n = \frac{1}{2}(\sigma^z_n+(-1)^n)$ are $q(1) = 1$ and $q(2) = 0$ on even sites, and $q(1) = 0$ and $q(2) = -1$ on odd sites. The local charge conservation condition \eqref{eq:charge_conservation_U1} then takes the form of a selection rule on the MPS tensors $A_n$:
\begin{equation}\label{eq:sym_cond}
	\begin{tikzpicture}[diagram]
		\node[style=square] at (0,0) (A1) {$A_{n}$};
		\draw (A1) -- node[below] {$\alpha_n$} ++(1.25,0);
		\draw (A1) -- node[below] {$\alpha_{n-1}$} ++(-1.25,0);
		\draw (A1) -- node[left] {$i_n$} ++(0,-1.25);
		\draw[->,gray] (0.5, .2) -- (1.15, .2) node[midway, above] {$q(\alpha_n)$};
		\draw[<-,gray] (-0.5, .2) -- (-1.15, .2) node[midway, above] {$q(\alpha_{n-1})$};
		\draw[->,gray] (0.2, -1.15) -- (.2, -.5) node[midway, right] {$q(i_n)$};
	\end{tikzpicture}\quad =0 \qquad \text{if}\qquad q(\alpha_{n-1})+q(i_{n})\neq q(\alpha_{n})\,.
\end{equation}

The advantage of the symmetric MPS construction is that we can implement the operator $L_n$ by acting directly on the virtual space of one of the symmetric MPS tensors with $\Q_n$. That is,
\begin{equation}\label{eq:Ln_schwinger}
	\begin{split}
		&L_n \ket{A_1, \ldots, A_{n}, \ldots,  A_N} \\
		&= \ket{A_1, \ldots, A_{n} \Q_n, \ldots,  A_N} = \begin{tikzpicture}[diagram]
			\node[style=square] at (0,0) (A1) {$A_{n-1}$};
			\node[style=square] at (1.5,0) (A2) {$A_{n}$};
			\node[style=circle] at (2.5,0) (C2) {$\mathcal Q_{n}$};
			\node[style=square] at (3.5,0) (A3) {$A_{n+1}$};
			\node at (-1.,0) {$\cdots$};
			\node at (4.5,0) {$\cdots$};		
			\draw (A1) -- (A2) -- (C2) -- (A3);
			\draw (A1) -- ++(0,-.75);
			\draw (A2) -- ++(0,-.75);
			\draw (A3) -- ++(0,-.75);
			\draw (A1) -- ++(-0.75,0);
			\draw (A3) -- ++(0.75,0);
		\end{tikzpicture}\,.
	\end{split}
\end{equation}
Note that \eqref{eq:Ln_schwinger} goes beyond the usual picture in which physical operators can only act on the physical spaces of an MPS\@. In particular, in traditional setups one would have to act with the non-local expression \eqref{eq:Lnsum} for the electric field:
\begin{equation}\label{eq:old_action}
		L_n \ket{A_1, \ldots,  A_N}  = \begin{tikzpicture}[diagram]
			\node[style=square] at (0,0) (A1) {$A_{n-1}$};
			\node[style=square] at (1.5,0) (A2) {$A_{n}$};
			\node[style=square] at (3,0) (A3) {$A_{n+1}$};
			\node at (-1.,0) {$\cdots$};
			\node at (4.,0) {$\cdots$};
			\node at (-1,-1) {$\cdots$};	
			\draw (A1) -- (A2) -- (A3);
			\draw (A1) -- ++(-.75,0);
			\draw (A3) -- ++(.75,0);
			\draw (A1) -- ++(0,-.65);
			\draw (A2) -- ++(0,-.65);
			\draw (0,-1.35) -- ++(0,-.4);
			\draw (1.5,-1.35) -- ++(0,-.4);
			\draw (A3) -- ++(0,-1.75);
			\draw[rounded corners] (-.75, -.65) -- (1.85,-.65) -- (1.85, -1.35) -- (-.75, -1.35);
			\node at (0.55,-1.)  {$\sum_{k=1}^nQ_k$};
		\end{tikzpicture}\,.
\end{equation}
And indeed, for an ordinary MPS there is a good reason to act only on physical spaces: the MPS tensors can be adjusted without changing the physical state by right-multiplying $A_n$ by a matrix $X$ and left-multiplying $A_{n+1}$ by $X^{-1}$, so any physical operator $\mathcal{O}^{L}_n$ we insert on a link must obey $X\mathcal{O}^{L}_n X^{-1} = \mathcal O_n^L$ for it to be well-defined. For an ordinary MPS, $X$ is an arbitrary invertible matrix, so $\mathcal{O}^{L}_n$ can only be a multiple of the identity. However, for a symmetric MPS, $X$ must be a block-diagonal matrix (i.e., $X_{\alpha_n,\beta_n} = 0$ if $q(\alpha_n) \neq q(\beta_n)$). This means we can have a link operator $\mathcal{O}^{L}_n$ that acts as different multiples of the identity on spaces corresponding to different representations, and this is exactly what $\mathcal{Q}_n$ in \eqref{eq:Ln_schwinger} does. By making use of this possibility, we have found a significant simplification: the expressions \eqref{eq:Ln_schwinger} and \eqref{eq:old_action} are exactly equal and differ only by how the action of the electric field is implemented.

It is straightforward to compute the action of a link operator on an MPS\@. Although the state represented by an MPS has all its virtual indices contracted, in practice we always keep track of the $A_n$ tensors separately (as emphasized by the notation $\ket{A_1,\ldots,A_N}$). Thus, nothing prevents us from acting on the virtual spaces; moreover, we see from this example that for a lattice gauge theory state represented by a symmetric MPS, it is natural to do so in order to read off the electric field on a link. 

This approach can be contrasted with an alternative tensor network setup, employed for example in \cite{Buyens:2013yza,Fujii:2024reh}, in which additional MPS tensors are introduced in order to keep track of the state of the gauge field. These tensors do not change the link charge, and they serve only to transmit information about the link charge down to the physical spaces, so in the notation of \eqref{eq:sym_cond} the additional tensors would have to obey the constraint $q(\alpha_{n-1}) = q(\alpha_n) = q(i_n)$. For a $\grU(1)$ theory, this condition is no harder to impose than \eqref{eq:sym_cond}; but since it does not take the form of local charge conservation, this constraint is more difficult to generalize to non-abelian gauge groups. By contrast, the scheme in \eqref{eq:Ln_schwinger} generalizes very naturally, as we will see in the following subsection.

\subsection{Lattice Gauge Theories with Symmetric Matrix Product States}\label{subsec:lgt}

We will now move away from the $\grU(1)$ gauge theory considered in Section~\ref{subsec:schwinger} to the setup of a gauge theory with gauge group $G$ that is generically non-abelian. We will see that symmetric matrix product states can once again be employed and that we can generalize \eqref{eq:Ln_schwinger} to the non-abelian setup.  To be concrete, the class of theories that we aim to study are discretizations of continuum theories with an action of the form
 \begin{equation} \label{eq:Sgauge_general}
    S = \int d^2 x\, \left[ 
     - \frac{1}{2g^2} \tr  \left( F_{\mu\nu} F^{\mu\nu} \right) + {\cal L}_\text{matter} \right] \,,
 \end{equation}
where ${\cal L}_\text{matter}$ is the matter Lagrangian density, $F_{\mu\nu} = \partial_\mu A_\nu - \partial_\nu A_\mu - i[A_\mu, A_\nu]$ is the gauge field strength of the gauge field $A_\mu$, $g$ is the gauge coupling, and the trace is taken in a defining representation of the group (for example, the fundamental representation for $G = \grSU(N_c)$), normalized such that $\tr (T^a T^b) = \frac 12 \delta^{ab}$, where $T^a$ are the Lie algebra generators (with $a=1,\ldots,\dim G$ an adjoint index). More details about the explicit models we study numerically in this paper are given in Section~\ref{sec:results}.

The discretization of a continuum model of this type has a Hamiltonian of the form (cf.~\eqref{eq:SchwingerH})
\begin{equation}\label{eq:lgt_hamiltonian}
    H_G = \frac{g^2 a}{2}\sum_{n=1}^{N-1} L^a_n L^a_n + H_\text{matter}(U_n, \mathrm{matter \  d.o.f.s}) \,,
\end{equation}
and it acts upon a Hilbert space with the following structure. On each site $n=1,\ldots,N$ of our lattice, we will have degrees of freedom that transform in some (possibly reducible) representation $\bm{R}_n$ of $G$. The Hilbert space of these degrees of freedom is thus given by
\begin{equation}
    \H_\text{matter} = \spn\left\lbrace \bigotimes_{n=1}^N \ket{\bm{R}_n; m_n}\,\Bigm\vert\,m_n = 1,\ldots,\dim \bm{R}_n\right\rbrace\,.
\end{equation}
For example, in the previous section we had $G = \grU(1)$, $\bm{R}_n = 1 \oplus 0$ on even sites, and $\bm{R}_n = 0\oplus -1$ on odd sites. The Hilbert space of the gauge fields on each link $n$ (where $n=1,\ldots,N-1$) is, as in the $\grU(1)$ case, the space $L^2(G)$ of square-integrable functions on the group manifold. By the Peter-Weyl theorem, the matrix elements of the irreducible representations of $G$ form an orthonormal basis for the Hilbert space $L^2(G)$. We label the state corresponding to the $(m_{n,L},m_{n,R})$ matrix element in the irrep $\bm{r}_n$ by $\ket{\bm{r}_n;m_{n,L},m_{n,R}}$, so that the Hilbert space of all the links takes the form
\begin{equation}\label{eq:peterweyl}
    \H_\text{gauge} = \spn\left\lbrace \bigotimes_{n=1}^{N-1} \ket{\bm{r}_n;m_{n,L},m_{n,R}}\,\Bigm\vert\, \text{for irreps $\bm{r_n}$ with $m_{n,L},m_{n,R} = 1,\ldots,\dim \bm{r}_n$}\right\rbrace\,.
\end{equation}
The gauge field variables $U_n$, valued in $G$, are conjugate to left- and right-acting electric fields denoted by $L^a_n$ and $R^a_n$, respectively. The two electric fields on a link are related via $L_n^a= U_n^{ab}R_n^b$, where $U_n^{ab}$ is a matrix element of the adjoint representation of $U_n$.\footnote{If $U_n$ is given in the fundamental representation, then in terms of the generators $T^a$ of the Lie algebra we have $U_n^{ab} = 2\mathrm{tr} (T^aU_nT^bU_n^{-1})$.} The operators $U_n$, $L_n^a$, and $R_n^a$ satisfy the commutation relations 
\begin{equation}
\begin{gathered}
[L_m^a, U_n] = T^a U_n \delta_{mn} \,, \qquad [R^a_m, U_n] =  U_n T^a \delta_{mn}\,, \qquad [L^a_m,R^b_n] =0\,,\\
    [L^a_m, L^b_n] = - i f^{abc} L^c_n \delta_{mn} \,, \qquad
 [R^a_m, R^b_n] = i f^{abc} R^c_n \delta_{mn} \,,
\end{gathered} 
\end{equation}
where $f^{abc}$ are the structure constants of the Lie algebra of $G$. The generalization of $L_n^2$ in the electric energy of the Schwinger model \eqref{eq:SchwingerH} is $L^a_n L^a_n=R^a_n R^a_n$, which acts on the basis by computing the quadratic Casimir eigenvalue of the irrep:
\begin{equation}
L^a_n L^a_n\ket{\bm{r}_n;m_{n,L},m_{n,R}}=C_2(\bm r_n)\ket{\bm{r}_n;m_{n,L},m_{n,R}}\,.
\end{equation}

The Gauss law for non-abelian lattice gauge theories is
\begin{equation}\label{eq:gauss_general}
    L^a_n - R^a_{n-1} = Q^a_n\,,
\end{equation}
where $Q^a_n$ are the on-site charge operators on the $n$th site, obeying $[Q_n^a, Q_m^b] =  if^{abc} Q_n^c \delta_{mn}$.  The constraint \eqref{eq:gauss_general} implies that if link $n-1$ carries representation $\bm{r}_{n-1}$, then the representation $\bm{r}_n$ must appear in the tensor product $\bm{r}_{n-1}\otimes \bm{R}_n$.  

The Gauss law \eqref{eq:gauss_general} requires us to impose boundary conditions at the left and right ends of the chain by specifying the Hilbert spaces that $R_0^a$ and $L_N^a$ act on. We take these Hilbert spaces to transform in irreducible representations of $G$, 
 \begin{equation}\label{eq:bdryH}
 \begin{split}
     \H_{\text{gauge}, 0} &= \spn\left\lbrace  \ket{\bm{r}_0;m_{0,R}}\,\Bigm\vert\, \text{with $m_{0,R} = 1,\ldots,\dim \bm{r}_0$}\right\rbrace \,, \\
     \H_{\text{gauge}, N} &= \spn\left\lbrace  \ket{\bm{r}_N;m_{N,L}}\,\Bigm\vert\, \text{with $m_{N,L} = 1,\ldots,\dim \bm{r}_N$}\right\rbrace \,,
 \end{split}
 \end{equation}
and we will take the irreps on links 0 and $N$ to be identical, $\bm{r}_0 = \bm{r}_N$. Choosing $\bm{r}_0$ to be a non-singlet representation corresponds physically to placing a probe particle in representation $\bm{r}_0$ at the left end of the chain and a probe particle in representation $\overline{\bm{r}}_0$ at the right end of the chain. If the matter representation is such that these probe particles cannot be screened, then this allows us to study a non-trivial flux tube sector of the theory \cite{Witten:1978ka}.

As in Section~\ref{subsec:schwinger}, we can gauge-fix, which effectively sets $U_n\mapsto \mathds 1$ in the matter Hamiltonian, and through the Gauss law \eqref{eq:gauss_general} re-express the electric field operators as
\begin{equation}\label{eq:gauss_sol}
    L^a_n=R^a_n = R^a_0 + \sum_{m=1}^n Q_m^a\,.
\end{equation}
These operators act on the Hilbert space that remains after gauge-fixing, $\H_\text{matter} \otimes \H_{\text{gauge},0}\otimes \H_{\text{gauge},N}$, subject to a global constraint (in analogy with \eqref{eq:gauge_constraint}):
\begin{equation}
      Q^a = L^a_N-R^a_0 \,, \qquad\text{where}\qquad Q^a \equiv \sum_{n=1}^N Q^a_n \,.
\end{equation}

For a generic lattice gauge theory defined by \eqref{eq:lgt_hamiltonian} and \eqref{eq:gauss_general}, the gauge-fixed Hamiltonian is of the form 
\begin{equation}
    H_G =  \frac{g^2 a}{2}\sum_{n=1}^{N-1} \left(R_0^a + \sum_{m=1}^n Q_m^a\right)^2 + H_\text{matter}(U_n\mapsto\mathds 1,\mathrm{matter \  d.o.f.s})\,.
    \label{eq:G_hamiltonian}
\end{equation}
Much like in Section~\ref{subsec:schwinger}, we now have a Hamiltonian on the gauge-fixed Hilbert space that we wish to solve in the sector $Q^a - (L^a_N - R^a_0) = 0$ of the global symmetry generated by $Q^a - (L^a_N - R^a_0)$. For this task, we can again use symmetric matrix product states \cite{Sanz:2009lax}.

In analogy to the $\grU(1)$ case, the action of the global symmetry operation $g\in G$ factorizes as $u^{(g)} = \prod_{n=1}^N u_n^{(g)}$, where $u_n^{(g)}$ acts on $\H_n$. We again define a $G$-action on $\V_n$ by the operators $\mathcal{U}_n(g)$, and then require each MPS tensor $A^{i_n}_n$ to obey
\begin{equation}\label{eq:covariance_G}
    \left(u_n^{(g)}\right)^{i_n i'_n} A^{i'_n}_n = \mathcal{U}_{n-1}^{(g)\dagger} A^{i_n}_n \mathcal{U}^{(g)}_n\,.
\end{equation}
In the infinitesimal limit where $g = \mathds{1} + i\varphi^a T^a+\mathcal O(\varphi^2)$, where $T^a$ are generators of $\mathfrak{g} = \Lie(G)$, we have
\begin{equation}
    u_n^{(g)} = \mathds{1} + i\varphi^a Q_n^a +\mathcal O(\varphi^2) \,, \qquad \mathcal{U}_n^{(g)} = I + i\varphi^a \mathcal{Q}_n^a+\mathcal O(\varphi^2) \,,
\end{equation}
and so the condition \eqref{eq:covariance_G} becomes
\begin{equation}
    \left[\left(\Q_{n-1}^a\right)_{\alpha_{n-1}{\alpha_{n-1}'}}+\left(Q^a_n\right)^{i_n i_n'}-\left(\Q_{n}^a\right)_{\alpha_n \alpha_n'}\right]\left(A_n\right)^{i_n'}_{\alpha_{n-1}' \alpha_n'}=0
    \label{eq:charge_conservation_G}
\end{equation}
for all $n=1,\ldots,N$ and $a=1,\ldots,\dim G$\@. Similarly to \eqref{eq:charge_conservation_U1}, the constraint \eqref{eq:charge_conservation_G} mirrors the Gauss law, and consequently the action of $L_n^a$ on $\mathcal H_\text{matter}\otimes \H_{\text{gauge},0}\otimes \H_{\text{gauge},N}$ should be identified with the action of $\mathcal Q^a_n$ on $\V_n$. Thus, we again see that the symmetric MPS naturally reintroduces the notion of $\mathcal H_\text{gauge}$, which was dropped in the gauge-fixing procedure. To see the connection between $\H_\text{gauge}$ and the virtual spaces, one should interpret the indices $m_{n,L}$ and $m_{n,R}$ as the right and left virtual indices of the $n$\textsuperscript{th} and $(n+1)$\textsuperscript{st} uncontracted MPS tensors, respectively.

The condition \eqref{eq:charge_conservation_G} implies that each $A_n$ is an invariant tensor of the group $G$, which means we can write it in terms of Clebsch-Gordan coefficients $C^{\bm{r}_{n-1},\bm{R}_n,\bm{r}_n}$ that describe how $\bm{r}_n$ sits in the tensor product $\bm{r}_{n-1}\otimes\bm{R}_n$.\footnote{When there is multiplicity in the tensor product, the Clebsch-Gordan coefficients also carry multiplicity labels to distinguish different copies of $\bm{r}_n$ in $\bm{r}_{n-1}\otimes\bm{R}_n$.} Indeed, let us choose bases for $\V_n$ and $\H_n$ corresponding to their decompositions into irreps of $G$. We can then replace the virtual index $\alpha_n$ with pairs $(\beta_n, m_n)$, where $\beta_n = 1,\ldots,d_n$ indexes the several (not necessarily distinct) irreps appearing in $\V_n$, the irreps themselves are denoted by $\bm{r}(\beta_n)$, and $m_n=1,\ldots,\dim \bm{r}(\beta_n)$ are indices of these irreps. Likewise, we can decompose $\H_n$ if need be, but to reduce clutter let us assume that $\bm{R}_n$ is irreducible, since this will be the case anyway in Section~\ref{sec:results}. We can then solve the constraint \eqref{eq:charge_conservation_G} by
\begin{equation}\label{eq:sym_cond_G}
	\begin{tikzpicture}[diagram]
		\node[style=square] at (0,0) (A1) {$A_{n}$};
		\draw (A1) -- node[above,transform canvas={xshift=7.}] {$(\beta_n,m_n)$} ++(1.25,0);
		\draw (A1) -- node[above,transform canvas={xshift=-14.}] {$(\beta_{n-1},m_{n-1})$} ++(-1.25,0);
		\draw (A1) -- node[right] {$i_n$} ++(0,-1.25);
	\end{tikzpicture} = (\tilde{A}_n)_{\beta_{n-1},\beta_n} C^{\bm{r}(\beta_{n-1}),\bm{R}_n,\bm{r}(\beta_n)}_{m_{n-1},i_n,m_n}\, .
\end{equation}
(In the previous subsection, we did not need the indices $m_n$ because all irreps of $\grU(1)$ are one-dimensional, and the Clebsch-Gordan symbols are equal to 1 if the charge is conserved and 0 otherwise.) In a symmetric MPS, each tensor is constrained to obey \eqref{eq:sym_cond_G}. We furthermore take the boundary virtual spaces to be $\V_0 = \bm{r}_0$ and $\V_N = \bm{r}_N$. The state described by the symmetric MPS is then guaranteed to satisfy the Gauss law \eqref{eq:gauss_general}. 

To act with the gauge-kinetic term of the Hamiltonian on a state like this, we can follow the example of \eqref{eq:Ln_schwinger}:
\begin{equation}\label{eq:Ln_G}
	\begin{split}
		&L^a_n L^a_n \ket{A_1, \ldots, A_{n}, \ldots,  A_N} \\
		&= \ket{A_1, \ldots, A_{n} \Q^a_n \Q^a_n, \ldots,  A_N} = \begin{tikzpicture}[diagram]
			\node[style=square] at (0,0) (A1) {$A_{n-1}$};
			\node[style=square] at (1.5,0) (A2) {$A_{n}$};
			\node[style=circle] at (2.5,0) (C2) {$\mathcal Q^2_{n}$};
			\node[style=square] at (3.5,0) (A3) {$A_{n+1}$};
			\node at (-1.,0) {$\cdots$};
			\node at (4.5,0) {$\cdots$};		
			\draw (A1) -- (A2) -- (C2) -- (A3);
			\draw (A1) -- ++(0,-.75);
			\draw (A2) -- ++(0,-.75);
			\draw (A3) -- ++(0,-.75);
			\draw (A1) -- ++(-0.75,0);
			\draw (A3) -- ++(0.75,0);
		\end{tikzpicture}\,.
	\end{split}
\end{equation}
Explicitly, $\Q_n^2 \equiv \Q_n^a \Q_n^a$ is a diagonal matrix whose nonzero entry for the index $\alpha_n = (\beta_n, m_n)$ is the quadratic Casimir eigenvalue of $\bm{r}(\beta_n)$. Thus, like in the $\grU(1)$ case, we can directly act with $L_n^a L_n^a$ on the links instead of having to explicitly implement the sum \eqref{eq:gauss_sol}.

\subsection{Link-enhanced Matrix Product Operators for Lattice Gauge Theories} \label{subsec:LEMPO}

In this section, we will show how to combine the construction \eqref{eq:Ln_G} with ordinary MPOs, forming LEMPOs. We will see that the Hamiltonian of a lattice gauge theory can be expressed straightforwardly as a LEMPO\@.

For a lattice gauge theory Hamiltonian of the form \eqref{eq:G_hamiltonian}, where the gauge-fixed $H_\textrm{matter}$ does not depend on link variables, we can represent $H_\text{matter}$ as an ordinary MPO\@. An MPO can be thought of as an operator analog of the factorization in the MPS construction~\eqref{eq:mps_def}. More precisely, we can write an MPO for some operator $\O$ using matrices $W_n$, whose components $(W_n)^{\gamma_{n-1},\gamma_n}$ are operators acting on the physical spaces ${\cal H}_n$. (Each $W_n$ can also be thought of as a rank-four tensor with two physical indices and two virtual indices.) The $W_n$ matrices of operators are multiplied to form the MPO:
\begin{equation}\label{eq:mpo}
\begin{split}
    \O = \sum_{\gamma_0, \gamma_1, \ldots, \gamma_N} &v_i^{\gamma_0}(W_1)^{\gamma_0,\gamma_1}(W_2)^{\gamma_1,\gamma_2}\cdots (W_{N-1})^{\gamma_{N-2},\gamma_{N-1}}(W_N)^{\gamma_{N-1},\gamma_{N}}v_f^{\gamma_{N}}\,.
\end{split}
\end{equation}
Here $\gamma_i$ indexes a virtual space $\W_i$. The virtual indices at the ends of the chain, $\gamma_0$ and $\gamma_{N}$, are contracted with those of the fixed vectors $v_i$ and $v_f$; these vectors are customarily taken to be $(1,0,\ldots,0)$ and $(0,\ldots,0,1)$, respectively. We can think of \eqref{eq:mpo} as a right-to-left process in which we start with $v_f$, act with the sequence of $W_n$ operators, and then arrive at a vector whose first component is our desired operator $\O$, which we then extract using the dot product with $v_i$; see \cite{Crosswhite:2008ucd} for more on this method of constructing MPO representations. 

In order to make the MPO compatible with the $G$-symmetric MPS, we will take $\W_n$ to be in some reducible representation $\bm{r}_{n,1}\oplus\cdots\oplus\bm{r}_{n,p}$ of $G$, in such a way that the $(i,j)$ component of $W_n$ is a multiplet of operators transforming in a representation contained in $\bm{r}_{n-1,i}\otimes\overline{\bm{r}}_{n,j}$. For example, consider the Heisenberg model, with $G = \grSU(2)$ and $\bm{R}_n=\bm{2}$ (i.e., a chain of $N$ spin-$\frac12$s). The hopping term $\sum_n\vec{\sigma}_n\cdot\vec{\sigma}_{n+1}$ in its Hamiltonian can be implemented with the tensors 
\begin{equation}
    W_n = \left(\begin{array}{c|c|c}
        \mathds{1}_{n} & \vec{\sigma}_n &   \\  
        \hline
         &  & \vec{\sigma}_{n} \\
         \hline
         & & \mathds{1}_{n}
    \end{array}\right)
\label{eq:W_hopping}
\end{equation}
at each site (where $\mathds{1}$ is the $2\times 2$ identity matrix). In this example, the virtual space decomposes as $\W_n = \mathbf{1} \mathrel{\oplus} \mathbf{3} \mathrel{\oplus} \mathbf{1}$, and it is straightforward to check that the formula \eqref{eq:mpo} gives $\O = \sum_n \vec{\sigma}_n \cdot \vec{\sigma}_{n+1}$.

The action of an MPO \eqref{eq:mpo} on an MPS can be drawn as
\begin{equation}\label{eq:MPO_action}
		\O \ket{A_1, \ldots,  A_N} = 
		\begin{tikzpicture}[diagram]
			\draw[dotted,fill=gray!30,rounded corners=5mm] (-.6,-1.5) rectangle (.6,1);
			\draw[dotted,fill=gray!30,rounded corners=5mm] (.9,-1.5) rectangle (2.1,1);
			\draw[dotted,fill=gray!30,rounded corners=5mm] (2.4,-1.5) rectangle (3.6,1);
			\node at (0,.65) {$A'_{n-1}$};
			\node at (1.5,.65) {$A'_n$};
			\node at (3,.65) {$A'_{n+1}$};
			\node[style=square] at (0,0) (A1) {$A_{n-1}$};
			\node[style=square] at (1.5,0) (A2) {$A_{n}$};
			\node[style=square] at (3,0) (A3) {$A_{n+1}$};
			\node[style=circle] at (0,-1) (W1) {$W_{n-1}$};
			\node[style=circle] at (1.5,-1) (W2) {$W_{n}$};
			\node[style=circle] at (3,-1) (W3) {$W_{n+1}$};
			\node at (-1.,0) {$\cdots$};
			\node at (4.,0) {$\cdots$};
    			\node at (-1.,-1) {$\cdots$};
			\node at (4.,-1) {$\cdots$};
			\draw (A1) -- (A2) -- (A3);
			\draw (W1) -- (W2) -- (W3);
			\draw (A1) -- (W1);
			\draw (A2) -- (W2);
			\draw (A3) -- (W3);
			\draw (A1) -- ++(-0.75,0);
			\draw (A3) -- ++(0.75,0);
			\draw (W1) -- ++(-0.75,0);
			\draw (W3) -- ++(0.75,0);
			\draw (W1) -- ++(0,-0.75);
			\draw (W2) -- ++(0,-0.75);
			\draw (W3) -- ++(0,-0.75);
		\end{tikzpicture} = \ket{A'_1,\ldots,A'_N}\,.
\end{equation}
Here, we are grouping the tensors as shown in the diagram to form new MPS tensors $A'_n$, and we are interpreting the horizontal lines as new virtual spaces of $A'_n$ given by $\V'_n = \V_n \otimes \W_n$. Thus, the MPO construction allows one to act with operators at the level of MPS tensors rather than the whole state, which enables efficient computation even for large system sizes.

Following this construction, we can also generalize the insertion of a single link operator (i.e., the kind of operator appearing in \eqref{eq:Ln_G}) to an MPO representing an arbitrary operator $\mathcal{O}_n^L$ acting on the links (and, therefore, the virtual indices of the $A_n$ tensors). We define a link MPO by analogy with \eqref{eq:mpo}:
\begin{equation}\label{eq:lmpo}
\begin{split}
    \O^L = \sum_{\delta_0, \delta_1, \ldots, \delta_N} &v_i^{\delta_0}(W_1^L)^{\delta_0,\delta_1}(W_2^L)^{\delta_1,\delta_2}\cdots (W_{N-1}^L)^{\delta_{N-2},\delta_{N-1}}(W_N^L)^{\delta_{N-1},\delta_{N}}v_f^{\delta_{N}}\,,
\end{split}
\end{equation}
but with the difference that the components of $\left(W_n^L\right)^{\delta_{n-1}, \delta_n}$ act on the virtual spaces $\V_n$ rather than the physical spaces $\H_n$. This action can be illustrated as
\begin{equation}
	\O^L\ket{A_1, \ldots,  A_N} = 
	\begin{tikzpicture}[diagram]
		\draw[shade] (-.6,-.5) rectangle (1.6,1.);
		\draw[shade] (1.9,-.5) rectangle (4.1,1.);
		\draw[shade] (4.4,-.5) rectangle (6.6,1.);
		\node at (1,.65) {$A'_{n-1}$};
		\node at (3.5,.65) {$A'_{n}$};
		\node at (6,.65) {$A'_{n+1}$};
		\node[style=square] at (0,0) (A1) {$A_{n-1}$};
		\node[style=circle] at (1,0) (W1) {$W^L_{n-1}$};
		\node[style=square] at (2.5,0) (A2) {$A_{n}$};
		\node[style=circle] at (3.5,0) (W2) {$W^L_{n}$};
		\node[style=square] at (5,0) (A3) {$A_{n+1}$};
		\node[style=circle] at (6,0) (W3) {$W^L_{n+1}$};
		\node at (-1.,0) {$\cdots$};
		\node at (7.,0) {$\cdots$};
		\draw (A1) -- (W1) -- (A2) -- (W2) -- (A3) -- (W3);
		\draw (A1) -- ++(-0.75,0);
		\draw (W3) -- ++(0.75,0);
		\draw (A1) -- ++(0,-0.75);
		\draw (A2) -- ++(0,-0.75);
		\draw (A3) -- ++(0,-0.75);
		\draw (W1) to[bend left=35] (W2);
		\draw (W2) to[bend left=35] (W3);
		\draw (-0.75,.45) to[out=0, in=145] (W1);
		\draw (W3) to[out=35, in=180] (6.75,0.45);
	\end{tikzpicture} = \ket{A'_1,\ldots,A'_N}\,.
\end{equation}
For example, if we wanted to write a nearest-neighbor interaction term like $\sum_n L_n^a L_{n+1}^a$ on the links of an $\grSU(N_c)$ gauge theory, we would use link MPO matrices
\begin{equation}
    W_n^L = \left(\begin{array}{c|c|c}
        I_{n} & \Q_n^a &   \\  
        \hline
         &  & \Q_n^a \\
         \hline
         & & I_{n}
    \end{array}\right),
\label{eq:W_gaugehopping}
\end{equation}
where $I_n$ denotes the identity operator on $\V_n$. The virtual space of this matrix decomposes as $\W_n = \mathbf{1} \mathrel{\oplus} \textbf{adj} \mathrel{\oplus} \mathbf{1}$, because $\Q_n^a$ carries the adjoint index $a$. A local term $\sum_n L_n^a L_{n}^a$ (such as the gauge kinetic term in \eqref{eq:G_hamiltonian}) would be implemented with repeating tensors 
\begin{equation}
    W_n^L = \left(\begin{array}{c|c}
        I_{n} & \Q_n^a \Q_n^a   \\  
        \hline
         &  I_{n} \\
    \end{array}\right)\,,
\label{eq:W_gaugelocal}
\end{equation}
with virtual space $\W_n = \mathbf{1} \oplus \mathbf{1}$.

We can now combine a link MPO with an ordinary MPO to form a LEMPO\@. A LEMPO is built as 
\begin{equation}\label{eq:LEMPO}
\begin{split}
    \O^\mathrm{LEMPO}  = \sum_{\substack{\gamma_0, \gamma_1, \ldots, \gamma_N\\\delta_1, \ldots \delta_{N-1}}} &v_i^{\gamma_0}(W_1)^{\gamma_0,\gamma_1}(W_1^L)^{\gamma_1,\delta_1} (W_2)^{\delta_1,\gamma_2}\times \cdots\\
    &\cdots \times (W_{N-1})^{\delta_{N-2},\gamma_{N-1}} (W_{N-1}^L)^{\gamma_{N-1},\delta_{N-1}} (W_N)^{\delta_{N-1},\gamma_{N}}v_f^{\gamma_{N}}\,,
\end{split}
\end{equation}
where the $W_n$ are operators acting on the physical spaces while the $W_n^L$ act on the virtual spaces of $A_n$. Thus, the number of matrices is twice the number that we would have in an ordinary MPO. This is illustrated in the following:
\begin{equation}
	\O^\mathrm{LEMPO} \ket{A_1, \cdots,  A_N} = 
	\begin{tikzpicture}[diagram]
		\draw[shade] (-.6,-1.5) rectangle (1.6,1.);
		\draw[shade] (1.9,-1.5) rectangle (4.1,1.);
		\draw[shade] (4.4,-1.5) rectangle (6.6,1.);
		\node at (1,.65) {$A'_{n-1}$};
		\node at (3.5,.65) {$A'_{n}$};
		\node at (6,.65) {$A'_{n+1}$};
		\node[style=square] at (0,0) (A1) {$A_{n-1}$};
		\node[style=circle] at (1,0) (W1) {$W^L_{n-1}$};
		\node[style=square] at (2.5,0) (A2) {$A_{n}$};
		\node[style=circle] at (3.5,0) (W2) {$W^L_{n}$};
		\node[style=square] at (5,0) (A3) {$A_{n+1}$};
		\node[style=circle] at (6,0) (W3) {$W^L_{n+1}$};
		\node[style=circle] at (0,-1) (V1) {$W_{n-1}$};
		\node[style=circle] at (2.5,-1) (V2) {$W_{n}$};
		\node[style=circle] at (5,-1) (V3) {$W_{n+1}$};
		\node at (-1.,0) {$\cdots$};
		\node at (7.,0) {$\cdots$};
		\node at (-1.,-1) {$\cdots$};
		\node at (7.,-1) {$\cdots$};
		\draw (A1) -- (W1) -- (A2) -- (W2) -- (A3) -- (W3);
		\draw (A1) -- ++(-0.75,0);
		\draw (W3) -- ++(0.75,0);
		\draw (V1) -- ++(-0.75,0);
		\draw (A1) -- (V1) -- ++(0,-0.75);
		\draw (A2) -- (V2) -- ++(0,-0.75);
		\draw (A3) -- (V3) -- ++(0,-0.75);
		\draw (V1) to[out=0, in=-100] (W1);
		\draw (W1) to[out=-80, in=-180] (V2);
		\draw (V2) to[out=0, in=-100] (W2);
		\draw (W2) to[out=-80, in=-180] (V3);
		\draw (V3) to[out=0, in=-100] (W3);
		\draw (W3) to[out=-80, in=-180] (6.75,-1);
	\end{tikzpicture}
	= \ket{A'_1,\ldots,A'_N}\,.
	\label{eq:HtotLEMPO}
\end{equation}
Here, $W_n$ and $W^L_n$ are similar to the tensors appearing in the MPOs and link MPOs, but they need to be adjusted in order to produce the correct operator using the contraction in~\eqref{eq:LEMPO}. For example, consider an operator that is a sum of matter terms and link terms, such that the corresponding MPO and link MPO can be written with matrices
\begin{equation}
    W_n =  \left(\begin{array}{c|c|c}
        \mathds{1}_{n} & a_n & b_n   \\  
        \hline
         & c_n & d_n   \\  
        \hline
         &  & \mathds{1}_{n} \\
    \end{array}\right)\qquad\text{and}\qquad 
    W_n^L =  \left(\begin{array}{c|c|c}
        I_{n} & e_n & f_n   \\  
        \hline
         & g_n & h_n   \\  
        \hline
         &  & I_{n} \\
    \end{array}\right)\,,
\label{eq:W_matter}
\end{equation}
respectively (where $\mathds{1}_n$ is the identity on $\H_n$ and $I_n$ is the identity on $\V_n$). To build a LEMPO for the full operator, we would combine these matrices into LEMPO tensors by suitably rearranging the components:

\begin{equation}
     W_n^\mathrm{LEMPO} =  \left(\begin{array}{c|c|c|c}
        \mathds{1}_{n} & a_n & & b_n   \\  
        \hline
         & c_n & & d_n   \\  
        \hline
         & &  \mathds{1}_{n}&  \\
         \hline
         & &  & \mathds{1}_{n} \\
    \end{array}\right), \qquad W_n^{L,\text{LEMPO}} = \left(\begin{array}{c|c|c|c}
        I_{n} &  & e_n & f_n   \\  
        \hline
         &  I_{n} & &   \\  
        \hline
         & &  g_{n} &  h_n\\
         \hline
         & &  & I_{n} \\
    \end{array}\right)\,.
\label{eq:W_LEMPO}
\end{equation}
If we denote the original virtual space of $W_n$ as $\W_n = \mathbf{1} \oplus \bm{R} \oplus \mathbf{1}$ and that of $W_n^L$ as $\W_n^L = \mathbf{1} \oplus \bm{R}^L \oplus \mathbf{1}$, the new virtual space is given by $\W_n^\mathrm{LEMPO} = \mathbf{1} \oplus \bm{R} \oplus \bm{R}^L \oplus \mathbf{1}$. In particular, $\dim \W_n^\text{LEMPO} = \dim \W_n + \dim \W_n^L - 2$. 

Finally, note that for a gauge-fixed lattice gauge theory with Hamiltonian \eqref{eq:G_hamiltonian}, the gauge term is local as in \eqref{eq:W_gaugelocal}, such that the tensor $W^L$ in a LEMPO corresponding to \eqref{eq:G_hamiltonian}  will be of the form 
\begin{equation}\label{eq:link_L2}
    W_n^\text{L,LEMPO} = \left( \begin{array}{c|c|c|c}
    I_n & & & \frac{g^2 a}{2}\mathcal Q_n^2 \\
    \hline & I_n & & \\ \hline & & \ddots & \\ \hline & & & I_n 
    \end{array}\right)\,.
\end{equation}
For more technical details on the implementation of LEMPOs, especially in the context of DMRG, see Appendix~\ref{app:algo}.

\subsection{LEMPOs with uniform MPSs}\label{sec:uMPS}
A major advantage of using LEMPOs to simulate gauge theories is that they can be easily generalized for use with uniform MPSs (uMPSs) \cite{Haegeman:2011zz, Vanderstraeten:2019voi, Zauner-Stauber:2017eqw} in the thermodynamic limit. uMPSs are defined on infinite translation-invariant lattices consisting of repeating unit cells, indexed by $n$, of $k$ sites each. The physical Hilbert spaces on these sites, $\H_n^{(j)}$ with $j=1,\ldots,k$, satisfy $\H_m^{(j)} = \H_n^{(j)}$ for all $m$ and $n$. A uMPS is parametrized by $k$ MPS tensors, labeled $A^{(1)}, \ldots, A^{(k)}$. We denote their right virtual spaces by $\V^{(1)}, \ldots, \V^{(k)}$; the left virtual space of $A^{(1)}$ is constrained to be $\V^{(k)}$. The uMPS is then given by a formal infinite contraction of these tensors with the on-site kets $\ket{i_n^{(j)}}$:
\begin{equation}
    \ket{\mathrm{uMPS}(A^{(1)}, \ldots, A^{(k)})} = \sum_{i^{(k)}_n,\gamma^{(k)}_n}\prod_{n=-\infty}^{\infty} \left[ A^{(1)\,\gamma^{(k)}_{n-1} \gamma^{(1)}_n}_{i^{(1)}_n} \cdots A^{(k)\,\gamma^{(k-1)}_n \gamma^{(k)}_n}_{i^{(k)}_n} \times \left(\ket{i^{(1)}_n}\otimes\cdots\otimes\ket{i^{(k)}_n}\right) \right]\,.
    \label{eq:uMPS}
\end{equation}
A uMPS is a good ansatz for the ground state of a translation-invariant gapped Hamiltonian defined in the thermodynamic limit \cite{Perez-Garcia:2006nqo,Verstraete:2006mdr}. 

Traditionally, when using tensor networks for lattice gauge theories, one starts by solving \eqref{eq:gauss_general} to eliminate the link variables from the Hamiltonian. Such an approach necessarily breaks translation symmetry, which precludes the use of uMPSs. For $\grU(1)$ gauge theories, this problem can be circumvented by introducing additional sites to encode the gauge field \cite{Buyens:2013yza,Fujii:2024reh}, but as we discussed at the end of Section~\ref{subsec:schwinger}, this workaround involves a construction that does not easily generalize to non-abelian theories. Our LEMPOs instead solve the translation-invariance problem by directly accessing the link variables of the symmetric MPSs, in a way that works equally well for abelian and non-abelian lattice gauge theories. 

The generalization of LEMPOs to infinite translation-invariant Hamiltonians is straightforward. With symmetric uMPSs \cite{Sanz:2009lax}, we can read off the link operators from the virtual bonds much like in the finite case. Thus, for an infinite translation-invariant Hamiltonian acting on a uMPS of the form \eqref{eq:uMPS}, we can define an infinite LEMPO as a repeating unit cell of physical MPO matrices $W_{(1)}, \ldots, W_{(k)}$ and link MPO matrices $W_{(1)}^L, \ldots, W_{(k)}^L$, written as a formal product
\begin{equation}\label{eq:inf_LEMPO}
\begin{split}
    \O^\mathrm{LEMPO}  = \prod_{n=-\infty}^{\infty} \left((W_{(1)}(n))^{\delta^{(k)}_{n-1},\gamma_n^{(1)}}(W_{(1)}^L(n))^{\gamma_n^{(1)},\delta_n^{(1)}} \cdots (W_{(k)}(n))^{\delta_n^{(k-1)},\gamma_n^{(k)}} (W_{(k)}^L(n))^{\gamma_n^{(k)},\delta_n^{(k)}}\right)\,.
\end{split}
\end{equation}
The dependence of the matrices on $n$ is only to specify which site the operators act upon; the structure of the matrices does not depend on $n$ (see Appendix~\ref{app:adjoint_lempos} for examples). Note also that this contraction requires the left virtual space of $W_{(1)}$ to be identical to the right virtual space of $W_{(k)}$.

To find the ground state of a given infinite Hamiltonian written as a LEMPO, we can use the variational uniform matrix product state (VUMPS) algorithm introduced in \cite{Zauner-Stauber:2017eqw} and also detailed in \cite{Vanderstraeten:2019voi}, which is an efficient extension of DMRG to uMPSs. The VUMPS algorithm involves calculating the energy density of a uMPS (rather than the extensive energy, which of course is infinite), and minimizing the energy density to find the vacuum state. From this vacuum state, we can compute many other quantities (e.g. expectation values or correlation functions of local operators). Similarly, we can employ the quasiparticle ansatz developed in \cite{Haegeman:2013} to find the spectrum of particle excitations above the vacuum state. In Section~\ref{sec:results}, we use these methods to study both abelian and non-abelian lattice gauge theories.

\section{Numerical results}\label{sec:results}

As an application of the formalism we developed, in this section we use LEMPOs and infinite tensor network methods to study two $(1+1)$-dimensional gauge theories of interest. In Section~\ref{subsec:schwinger_results} we study the Schwinger model, obtaining very precise agreement with exact results in the massless case and also detailed spectra in the massive case at various values of the $\theta$ parameter.  In Section~\ref{subsec:adjoint_results}, we study adjoint QCD$_2$ with gauge groups $\SU(2)$ and $\SU(3)$, significantly improving on recent numerical results.

\subsection{Schwinger model}\label{subsec:schwinger_results}

\subsubsection{Brief review of the continuum model}

We start with the Schwinger model, which was introduced briefly in Section~\ref{subsec:schwinger}.  The continuum action was given in \eqref{eq:schwinger_action}, with the discretized lattice Hamiltonian given in \eqref{eq:SchwingerH}.  

In a nutshell, the main properties of the continuum model are as follows.  When $m=0$, this model is exactly solvable \cite{Schwinger:1962tp}, being equivalent to a free boson of mass $M_S = \frac{g}{\sqrt{\pi}}$ for any value of $\theta$. Axial transformations $\psi \to e^{-i \alpha \gamma^5} \psi$, with $\gamma^5 \equiv \gamma^0 \gamma^1$ and with $\alpha \in [0, \pi)$ a continuous parameter,\footnote{Here, we restrict the range of $\alpha$ to $[0, \pi)$ because the transformation $\psi \to - \psi$ corresponding to $\alpha = \pi$ is a gauge transformation.} are symmetries of the classical action.  Quantum mechanically, however, the famous Schwinger anomaly  \cite{Schwinger:1962tp} implies that the axial transformation $\psi \to e^{-i \alpha \gamma^5} \psi$ also changes the $\theta$ parameter to $\theta + 2 \alpha$.  This property implies that all physical observables at non-zero $\theta$ can be obtained from those at $\theta = 0$ by performing an appropriate axial transformation.  In particular, the spectrum of the theory is independent of $\theta$.  Vacuum expectation values, however, may change with $\theta$.  For instance, on the infinite line, it can be shown using the bosonized form of the theory that $\langle \bar \psi \psi \rangle = - g \frac{e^{\gamma}}{2 \pi^{3/2}}$ and $\langle \bar \psi \gamma^5 \psi \rangle = 0$ at $\theta = 0$ \cite{Hetrick:1988yg}.  Since under axial transformations $\bar \psi \psi \to \cos (2 \alpha) \bar \psi \psi - i \sin (2 \alpha) \bar \psi \gamma^5 \psi$ and $\bar \psi \gamma^5 \psi \to -i \sin (2 \alpha) \bar \psi \psi + \cos (2 \alpha) \bar \psi \gamma^5 \psi$, we can infer that for general $\theta$ we have \cite{Hetrick:1988yg}
\begin{equation}\label{eq:schwinger_condensates}
    \langle \bar\psi \psi\rangle/g = -\frac{e^\gamma}{2\pi^{3/2}} \cos\theta\,, \qquad 
    \langle \bar\psi \gamma^5 \psi\rangle/g = i \frac{e^\gamma}{2\pi^{3/2}} \sin\theta \,.
\end{equation}

When $m \neq 0$, the Schwinger model is no longer exactly solvable.  Not only are axial transformations not symmetries anymore, but also, due to the transformation of $\bar \psi \psi$ quoted above, they do not preserve the form of the action \eqref{eq:schwinger_action} unless $\alpha = \pi/2$.  Indeed, the $\Z_2$ axial transformation corresponding to $\alpha = \pi/2$ sends $\bar \psi \psi \to - \bar \psi \psi$, and consequently it maps the theory with parameters $(m, \theta)$ to that with parameters $(-m, \theta + \pi)$. All properties of the theory are invariant under this transformation, so without loss of generality we can restrict our attention to $m \geq 0$.  The spectrum in this region is gapped with a unique ground state everywhere, except when $\theta = \pi$ and $m$ is larger than the critical value $m_\text{crit} \approx 0.33g$.  

At $\theta = \pi$ and $m = m_\text{crit}$, the Schwinger model exhibits a critical point in the Ising universality class \cite{Coleman:1976uz,Byrnes:2002gj}---see \cite{ArguelloCruz:2024xzi,Fujii:2024reh} for recent detailed numerical studies of the critical point.  This critical point corresponds to a transition between one non-degenerate vacuum for $m<m_\text{crit}$ and two degenerate vacua for $m > m_\text{crit}$.  For $m>m_\text{crit}$, the two degenerate vacua are characterized by equal and opposite expectation values $\langle E \rangle$ for the electric field, and therefore they spontaneously break charge conjugation symmetry.  As we take $m/g \to \infty$, we have $\langle E \rangle = \pm 1/2$ in the two vacua, while as $m \to m_\text{crit}$, we have $\langle E \rangle  = \pm C (m - m_\text{crit})^{1/8}$, for some constant $C > 0$.  This power-law scaling follows from the fact that the electric field operator $E$ has non-vanishing overlap with the order parameter of the Ising CFT, which has scaling dimension $1/8$, and $m - m_\text{crit}$ is the coupling constant for the energy operator of the Ising CFT, which has scaling dimension $1$\@.   

At $\theta = \pi$ and $m > m_\text{crit}$, we can construct domain walls (solitons) between the two vacua.  At large $m/g$, such solitons can be thought of as the electrons and positrons, which behave as half-asymptotic particles and have masses approximately equal to $m$ \cite{Coleman:1976uz}. As we lower $m/g$, the masses of the solitons decrease due to quantum corrections, and they must vanish linearly in $m - m_\text{crit}$ as we approach the Ising CFT point.  The existence of the solitons implies that the spectrum of the theory restricted to any of the two vacua should contain a continuum of two-soliton states.  When we are away from $\theta = \pi$ and we have a single non-degenerate vacuum, no such continuum should be present.  One may thus wonder how the two-soliton continuum forms as we take $\theta \to \pi$.  The answer, which will be supported by the numerical results we show below, is that the number of bound states in the theory increases as we take $\theta \to \pi$, and these bound states become dense and form a continuum at $\theta = \pi$.  In fact, Coleman estimated the number of bound states at weak coupling (i.e.~large $m/g$) to be 
\begin{equation}\label{eq:coleman_count}
    \text{\# of particles} \approx \frac{4m^2}{\pi g^2}\frac{1}{1-\theta^2/\pi^2}\left(2\sqrt{3}-\log(2+\sqrt{3})\right) \,.
\end{equation}
One can see that this number diverges as $\theta \to \pi$, as expected.

\subsubsection{Numerical results}

As mentioned above, the Hamiltonian lattice formulation of \eqref{eq:schwinger_action} is given in \eqref{eq:SchwingerH}.  As was argued in \cite{Dempsey:2022nys}, the continuum mass $m$ and the lattice mass $m_\text{lat}$ differ by a mass shift of $\frac{g^2 a}{8}$, with $m_\text{lat} = m - \frac{g^2 a}{8}$.  While the difference between $m_\text{lat}$ and $m$ goes to zero in the continuum limit, in \cite{Dempsey:2022nys} it was found that the use of the mass-shifted $m_\text{lat}$ significantly improves the convergence toward the continuum.  We will therefore use this relation in our work.  Following the discussion in Section \ref{subsec:LEMPO}, we implement the Jordan-Wigner transformed Hamiltonian \eqref{eq:schwinger_bosonized} as a LEMPO with matter and link matrices as
\begin{equation}
    W_n = \left(\begin{array}{c|c|c|c}
        \mathds{1}_n & \sigma_n^+ & \sigma_n^- & \frac{(-1)^n m_\mathrm{lat}}{2}  \sigma_n^z \\
        \hline
         &  &  & \frac{1}{2a}\sigma_n^- \\
        \hline
         &  &  & \frac{1}{2a}\sigma_n^+ \\
        \hline
         &  &  & \mathds{1}_n
    \end{array} \right) \qquad \mathrm{and} \qquad W_n^L = \left(\begin{array}{c|c|c|c}
        I_n &  &  & \frac{g^2a}{2} \left(\Q_n+\frac{\theta}{2\pi}\right)^2 \\
        \hline
         & I_n &  &  \\
        \hline
         &  & I_n &  \\
        \hline
         &  &  & I_n
    \end{array}\right)\,,
\end{equation}
respectively. For the following infinite tensor network results, we employ these LEMPOs combined with the VUMPS algorithm and the quasiparticle ansatz mentioned in Section~\ref{sec:uMPS} to find infinite MPS vacuum states and excited states, respectively. 

Some of the first numerical results for the Schwinger model were obtained using the lattice strong-coupling expansion \cite{Banks:1975gq}, for which the Hamiltonian \eqref{eq:SchwingerH} can be treated in perturbation theory starting from an unperturbed leading piece, taken to be the gauge kinetic term and the fermion mass term, plus a small perturbation, taken to be the fermion kinetic term.  The method as developed in \cite{Banks:1975gq} ignores boundary effects, so it is intrinsically designed to work on an infinite lattice.  Using perturbation theory, it is possible to obtain expansions for various quantities on an infinite lattice in powers of $1/a$ \cite{Banks:1975gq,Hamer:1982mx,Hamer:1997dx}. It has been observed that judiciously-chosen Pad\'e approximants to these expansions can yield reasonably accurate extrapolations to the $a\to 0$ continuum limit \cite{Banks:1975gq,Hamer:1982mx,Hamer:1997dx,Dempsey:2022nys}.

\begin{figure}
    \centering
    \begin{subfigure}[t]{0.9\textwidth}
        \centering
        \includegraphics[width=.9\linewidth]{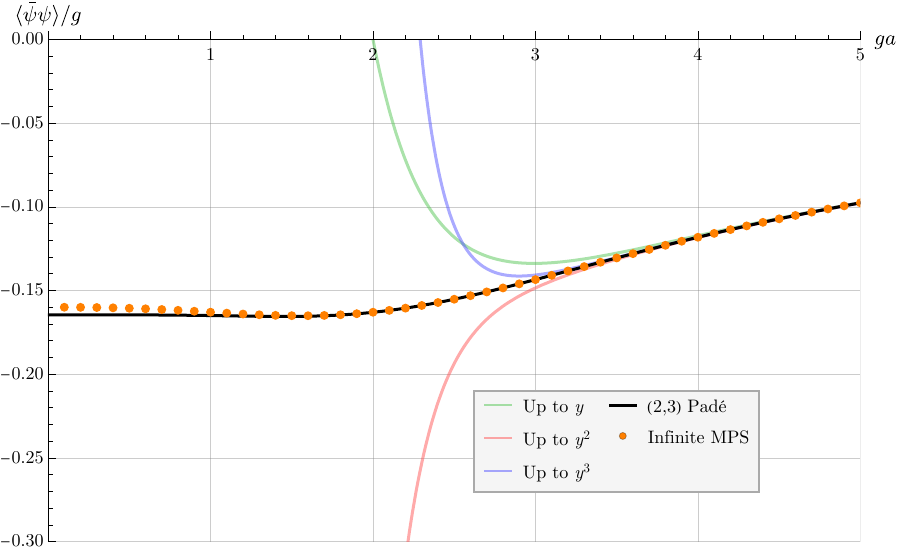}
        \caption{}
        \label{fig:schwinger_massless_sc_condensate}
    \end{subfigure}\\[1em]
    \begin{subfigure}[t]{0.9\textwidth}
        \centering
        \includegraphics[width=.9\linewidth]{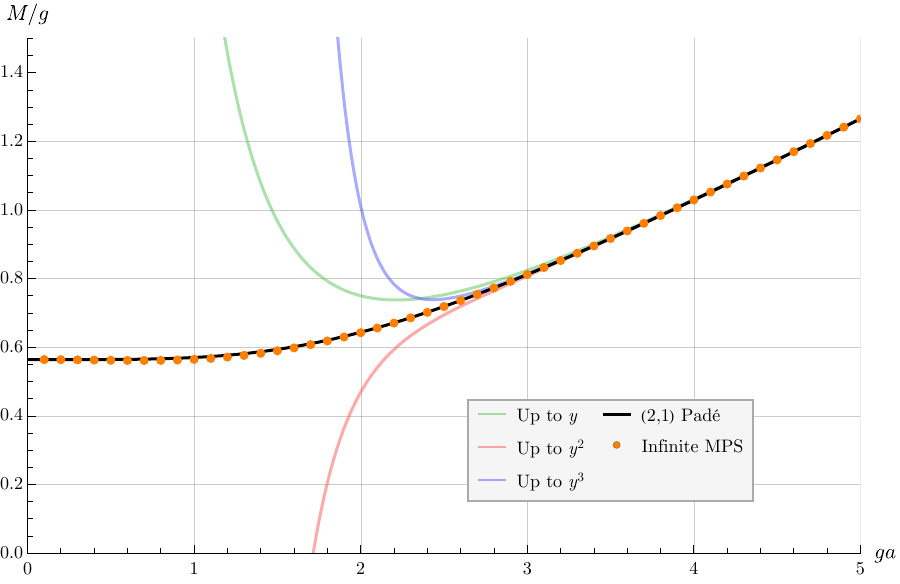}
        \caption{}
        \label{fig:schwinger_massless_sc_gap}
    \end{subfigure}%
    \caption{The lattice strong-coupling expansions \eqref{eq:schwinger_sc}, truncated at various orders in $y \equiv \frac{1}{(ga)^4}$ (green, red, and blue solid lines), compared with our lattice data (orange dots). Pad\'e approximants to the strong-coupling expansions (black solid line) agree rather well with the lattice data even as $a\to 0$.}
    \label{fig:schwinger_sc}
\end{figure}

Since the method we use in this paper also employs an infinite lattice, we can directly compare our numerical results with the lattice strong-coupling expansion. As examples, let us consider the $m=0$, $\theta = 0$ theory and study the mass $M_S$ of the Schwinger boson and the chiral condensate $\langle \bar \psi \psi \rangle$. After using the mass shift, the expansions of \cite{Hamer:1982mx} are
\begin{equation}\label{eq:schwinger_sc}
\begin{split}
    \langle \bar\psi \psi\rangle/g &= -\frac{1}{ga}\left(\frac{1}{2}-8y+288 y^2-\frac{306688 y^3}{25} +\frac{70547968 y^4}{125}-\frac{1374386118656 y^5}{50625}+\ldots\right) \\
    &\approx -\frac{1}{2ga}\left(\frac{9090225+374275520 y+2666321776 y^2}{9090225+956049920 y+28947052656 y^2+227736222720 y^3}\right)^{1/4}\,,\\
    M_S/g &= \frac{ga}{4}\left(1+8y-144y^2+4448y^3+\ldots\right) \approx \frac{ga}{4}\left(\frac{3+190y+2432y^2}{3+94y}\right)^{1/4}\,,
\end{split}
\end{equation}
where $y = \frac{1}{(ga)^4}$, and the approximate equalities are $(2,3)$ and $(2,1)$ Pad\'e approximants, respectively, raised to appropriate fractional powers in order to give finite continuum limits. In Figure~\ref{fig:schwinger_sc}, we plot these expansions and Pad\'e approximants in order to compare them with the results obtained from our tensor network approach. We find excellent agreement with the strong-coupling expansions in the region where they are justified. We also see that the Pad\'e approximants follow the lattice results rather closely even as $a\to 0$.

\begin{figure}
    \centering
    \begin{subfigure}[t]{0.48\textwidth}
        \centering
        \includegraphics[width=\linewidth]{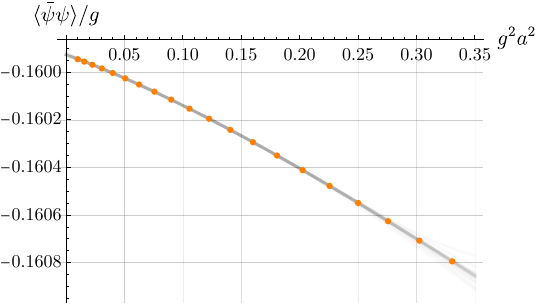}
        \caption{}
        \label{fig:schwinger_massless_condensate}
    \end{subfigure}%
    \hspace{.02\textwidth}%
    \begin{subfigure}[t]{0.48\textwidth}
        \centering
        \includegraphics[width=\linewidth]{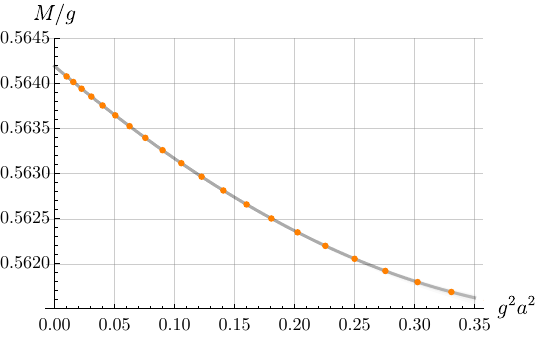}
        \caption{}
        \label{fig:schwinger_massless_gap}
    \end{subfigure}%
    \caption{Precision lattice estimates of the lightest particle mass and the chiral condensate in the massless Schwinger model. We extrapolate to $a\to 0$ using an ensemble of polynomial fits in powers of $g^2 a^2$ (shown in light gray), with degrees between 2 and 5 and using varying subsets of the data, and estimate errors from the variation in these fits. The results agree very well with the exact values; see \eqref{eq:schwinger_numbers} and Table~\ref{tab:schwinger_masses}.}
    \label{fig:schwinger_massless}
\end{figure}

By extrapolating our lattice results to $a\to 0$, we can obtain very precise estimates for continuum quantities. For example, for the chiral condensate we find
\begin{equation}\label{eq:schwinger_numbers}
\begin{split}
    \langle \bar\psi\psi\rangle/g &= -0.159926(3)\quad\text{(lattice)}\qquad = -0.1599288\ldots\quad\text{(exact)}\,.
\end{split}
\end{equation}
We estimate the error by using an ensemble of polynomial fits to the lattice data in powers of $g^2 a^2$, fitting to various subsets of the data and using models of various degrees, and then taking the standard deviation of their $a\to 0$ values. These fits, along with similar fits for the Schwinger boson mass, are shown in Figure~\ref{fig:schwinger_massless}. Note that all the lattice data in Figure~\ref{fig:schwinger_massless} can be obtained in 90 seconds on a single CPU\@.

We can apply the same kind of extrapolation to estimate the masses of particles at various values of the fermion mass $m$. In Table~\ref{tab:schwinger_masses}, we show how our results using infinite tensor networks and the mass shift improve substantially upon previous results obtained with other methods (finite MPS \cite{Banuls:2013jaa}, DMRG \cite{Byrnes:2002gj}, and the strong-coupling expansion \cite{Sriganesh:1999ws}).\footnote{Note that similar mass estimates using infinite MPS (using the method described at the end of Section~\ref{subsec:schwinger}, rather than our new construction) were also given in \cite{Buyens:2013yza}; if their method were used in tandem with the mass shift, it would likely give precisions comparable to what we report in Table~\ref{tab:schwinger_masses}. The advantage of our method is its straightforward applicability to non-abelian theories, as we demonstrate in Section~\ref{subsec:adjoint_results}.}

\begin{table}
    \centering
    \begin{tabular}{cp{3cm}p{3cm}p{3.8cm}p{3cm}}
        \multicolumn{5}{c}{Lightest Particle: $(M^{(1)}-2m)/g$} \\
        \toprule
        $m/g$ & This work & Finite MPS \cite{Banuls:2013jaa} & DMRG \cite{Byrnes:2002gj} & Exact \\
        \midrule
        0     & 0.564191(3)  & 0.56421(9)                     & 0.5642(2)                 & 0.5641896\ldots \\
        1/8   & 0.539503(2) & 0.53953(5)                     & 0.53950(7)                &                 \\
        1/4   & 0.519182(2) & 0.51922(5)                     & 0.51918(5)                &                 \\
        1/2   & 0.487463(6)  & 0.48749(3)                     & 0.48747(2)                &                 \\
        \bottomrule
    \end{tabular}\\[1em]
    \begin{tabular}{cp{3cm}p{3cm}p{7.2cm}}
        \multicolumn{4}{c}{Second-lightest Particle: $(M^{(2)}-2m)/g$} \\
        \toprule
        $m/g$ & This work & Finite MPS \cite{Banuls:2013jaa} & Strong Coupling \cite{Sriganesh:1999ws} \\
        \midrule
        1/8   & 1.21666(3)   & 1.216(3)                      & 1.22(2)                \\
        1/4   & 1.22782(4)   & 1.224(2)                      & 1.24(3)                \\
        1/2   & 1.2002(1)  & 1.200(2)                      & 1.20(3)                \\
        \bottomrule
    \end{tabular}
    \caption{Estimates for the masses of the two lightest particles in the Schwinger model, as calculated in this work and compared with results in \cite{Banuls:2013jaa,Byrnes:2002gj,Sriganesh:1999ws}. Estimates for the same quantities from infinite matrix product states, formulated as described at the end of Section~\ref{subsec:schwinger} rather than with LEMPOs, can be found in \cite{Buyens:2013yza}; we improve on these results as well, but this is likely due to the use of the mass shift \cite{Dempsey:2022nys} along with algorithmic improvements over the past decade.}
    \label{tab:schwinger_masses}
\end{table}

\begin{figure}
    \centering
    \begin{subfigure}[t]{0.48\textwidth}
        \centering
        \includegraphics[width=\linewidth]{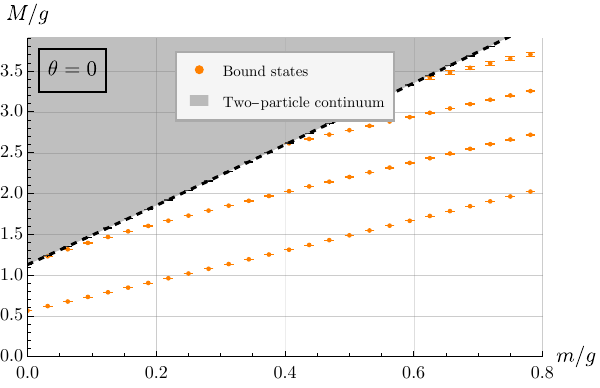}
        \caption{}
    \end{subfigure}%
    \hspace{.02\textwidth}%
    \begin{subfigure}[t]{0.48\textwidth}
        \centering
        \includegraphics[width=\linewidth]{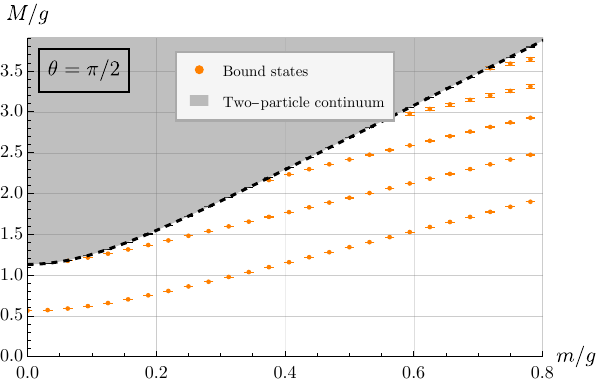}
        \caption{}
    \end{subfigure}\\
    \begin{subfigure}[t]{0.48\textwidth}
        \centering
        \includegraphics[width=\linewidth]{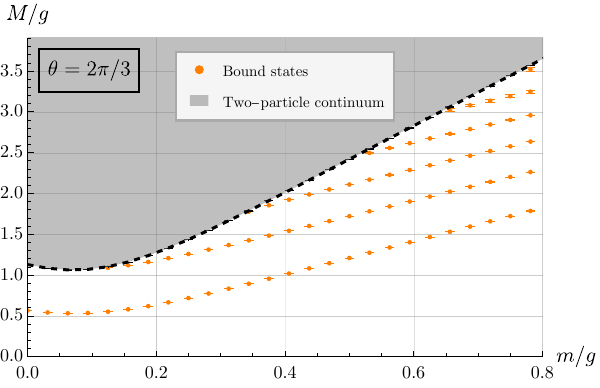}
        \caption{}
    \end{subfigure}%
    \hspace{.02\textwidth}%
    \begin{subfigure}[t]{0.48\textwidth}
        \centering
        \includegraphics[width=\linewidth]{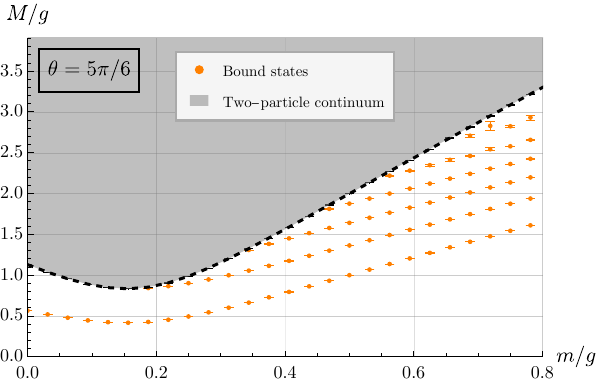}
        \caption{}
    \end{subfigure}\\
    \begin{subfigure}[t]{0.48\textwidth}
        \centering
        \includegraphics[width=\linewidth]{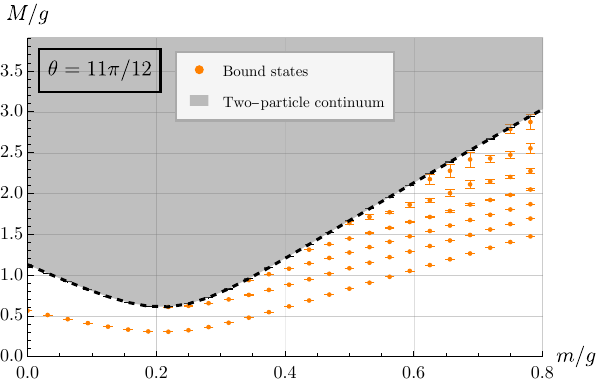}
        \caption{}
    \end{subfigure}%
    \hspace{.02\textwidth}%
    \begin{subfigure}[t]{0.48\textwidth}
        \centering
        \includegraphics[width=\linewidth]{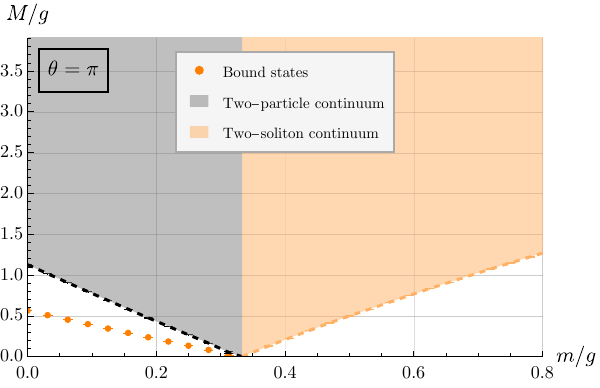}
        \caption{}
        \label{fig:schwinger_spec_pi}
    \end{subfigure}%
    \caption{The spectrum of the Schwinger model as a function of the electron mass $m$, estimated using the quasiparticle ansatz with uniform matrix product states. As $\theta$ approaches $\pi$, the density of states in the $m>m_\text{crit} \approx 0.33g$ region grows, and at $\theta = \pi$ the spectrum becomes a continuum; this can be interpreted as the continuum of two-body states formed from solitons interpolating between the two degenerate vacua in this region.}
    \label{fig:schwinger_spectrum}
\end{figure}

In Figure~\ref{fig:schwinger_spectrum}, we show how the low-lying spectrum as a function of $m$ changes as we increase $\theta$ from 0 to $\pi$. At $m = 0$, the theory is insensitive to $\theta$ and dual to a free boson, so we always have a single particle of mass $M_S = g/\sqrt{\pi}$ and then a two-particle continuum beginning at $2M_S$. As we increase $m$, the mass of the lowest bound state increases monotonically for $\theta < \pi/2$; for $\theta > \pi/2$, it first decreases, reaches a minimum value, and then increases. For $\theta = \pi$, this minimum value is 0, corresponding to the critical point of the model.

Additionally, for $\theta\in [0,\pi)$, as we increase $m$ we see that there are more stable particles below the continuum. A weak-coupling (i.e.~large $m/g$) estimate for the number of stable particles is \eqref{eq:coleman_count} \cite{Coleman:1976uz}, 
but in Figure~\ref{fig:schwinger_spectrum} the values of $m/g$ are too small for this formula to be quantitatively accurate. Nevertheless, we certainly see the qualitative behavior suggested by the factor of $\frac{1}{1-\theta^2/\pi^2}$: as we increase $\theta$ towards $\pi$, the spectrum becomes much more dense. As described above, when $\theta = \pi$ and $m>m_\text{crit}$ the spectrum starts with a continuum of two-soliton states.

We can estimate where this two-soliton continuum begins using the quasiparticle ansatz method. At precisely $\theta = \pi$ this ansatz is not strictly applicable, but for $\theta$ arbitrarily close to $\pi$ the lightest state will indeed be a particle, and if we take the limit of the mass of this particle as $\theta\to\pi$ we should obtain twice the mass of the soliton. Thus, we can argue by continuity that the mass of the lightest state found by the quasiparticle ansatz at $\theta = \pi$ actually indicates where the two-soliton continuum begins, and this is how we determine the curve for $m>m_\text{crit}$ in Figure~\ref{fig:schwinger_spec_pi}.

It is also interesting to compare the energy density $\varepsilon(m, \theta)$ of the vacuum at different values of $\theta$. We define an energy density difference by
\begin{equation}\label{eq:schwinger_sigma}
    \Delta\varepsilon(m, \theta) = \varepsilon(m, \theta) - \varepsilon(m, 0)\,.
\end{equation}
This can be thought of as the string tension between probe particles of fractional charge $\pm\frac{\theta}{2\pi}$.\footnote{Alternatively, if $\theta = 2\pi\frac{p}{q}$ with $p,q\in\mathbb{Z}$, then $\Delta\varepsilon(m,\theta)$ is proportional to the tension of strings between probe particles of charge $\pm p$ in a version of the Schwinger model where the electron has charge $q$ \cite{Dempsey:2022nys,Honda:2022edn}.}
\begin{figure}
    \centering
    \includegraphics[width=0.8\linewidth]{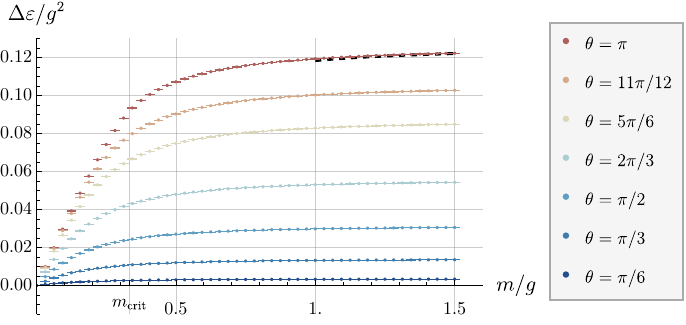}
    \caption{The energy density difference $\Delta \varepsilon(m, \theta)$ in the Schwinger model, defined in \eqref{eq:schwinger_sigma}. The black dashed line is the weak-coupling expression \eqref{eq:schwinger_sc_wc} for $\theta = \pi$. The lattice data also agrees well with the weak-coupling expansion at other values of $\theta$, as well as the small-mass expansion \eqref{eq:schwinger_sigma_lowmass}, but we do not show these approximations in this figure in order to avoid clutter.}
    \label{fig:schwinger_string_tension}
\end{figure}
The infinite MPS methods we are using are particularly well-suited to this quantity, since we obtain the variational ground state by minimizing the energy density $\varepsilon$ of a uniform MPS\@. For each value of $m$ we can extrapolate $\Delta\varepsilon$ as computed on the lattice to the $a\to 0$ continuum limit, and obtain the data in Figure~\ref{fig:schwinger_string_tension}. From \eqref{eq:schwinger_condensates} and $\frac{\partial\varepsilon}{\partial m} = \langle\bar\psi \psi\rangle$ we have the small-mass behavior
\begin{equation}\label{eq:schwinger_sigma_lowmass}
    \Delta\varepsilon(m, \theta)/g^2 = \frac{e^\gamma}{2\pi^{3/2}}(1-\cos\theta) \frac{m}{g} + \ldots\,.
\end{equation}
At large masses, a one-loop calculation \cite{Dempsey:2023gib} gives
\begin{equation}\label{eq:schwinger_sc_wc}
    \Delta\varepsilon(m,\theta)/g^2 = \frac{1}{2}\left(\frac{\theta}{2\pi}\right)^2 \left(1 - \frac{1}{6\pi}\frac{g^2}{m^2} + \ldots\right)
\end{equation}
for $-\pi \leq \theta \leq \pi$; this formula can be extended by periodicity for other values of $\theta$. We plot this curve for $\theta = \pi$ in Figure~\ref{fig:schwinger_string_tension}, showing good agreement with the lattice data for $m/g>1$.

\subsection{Adjoint QCD$_2$}\label{subsec:adjoint_results}

As emphasized in Section~\ref{sec:lmpo}, we can also apply LEMPOs to study non-abelian gauge theories on infinite lattices. As an example, we will now turn to adjoint QCD$_2$ \cite{Dalley:1992yy}, a model of particular recent interest \cite{Cherman:2019hbq,Komargodski:2020mxz,Dempsey:2021xpf,Dempsey:2022uie,Trittmann:2023dar,Cherman:2024onj,Bergner:2024ttq, Narayanan:2025cwl}.

\subsubsection{Brief review of the continuum model}

Adjoint QCD$_2$ is a gauge theory with gauge group $G$ of the form \eqref{eq:Sgauge_general} whose matter content consists of a single Majorana spinor field $\psi$ in the adjoint representation of $G$.  Here, we will take $G = \SU(N_c)$. The continuum action is
\begin{equation}
    S = \int d^2x\,\tr\left(-\frac{1}{2g^2}F_{\mu\nu}F^{\mu\nu} + i\bar\psi \gamma^\mu D_\mu\psi - m\bar\psi \psi\right)\, ,
\end{equation}
where $g$ is the gauge coupling, $m$ is the mass of the adjoint fermion, and $F_{\mu\nu}$ is the field strength of the gauge field $A_\mu$.  The only dimensionless parameter is $m/g$.  Due to the fact that the $\Z_2$ axial transformation $\psi \to \gamma^5 \psi$ sends $m \to -m$, we can restrict our attention to the region $m \geq 0$.   This theory was first studied numerically in the 1990s using discretized light-cone quantization (DLCQ) in the large-$N_c$ limit \cite{Demeterfi:1993rs}, with many other works following up on these studies \cite{Gross:1997mx,Katz:2013qua,Trittmann:2015oka,Trittmann:2023dar,Dempsey:2021xpf}; this method has also been extended to finite $N_c$ \cite{Antonuccio:1998uz,Dempsey:2022uie}.  A lattice Hamiltonian formulation of the theory, which we will also use here, was given in \cite{Dempsey:2023fvm,Dempsey:2024alw}.

For generic values of $m$, this theory is invariant under fermion parity and (for $N_c > 2$) charge conjugation.  Since the matter fermions are not charged under the center of $\grSU(N_c)$, there is also a $\mathbb{Z}_{N_c}$ one-form center symmetry.  This symmetry implies that there are $N_c$ distinct flux tube sectors \cite{Witten:1978ka} (also referred to as universes \cite{Komargodski:2020mxz}), distinguished by the $N_c$-ality $p= 0, \ldots, N_c - 1$ of the electric field.  Selecting one of these flux tube sectors is analogous to choosing a specific value of $\theta$ in the Schwinger model.  Indeed, instead of the $\grSU(N_c)$ gauge theory, one can consider the model with the same action but gauge group $\grSU(N_c) / \Z_{N_c}$;  the Hilbert space of the $\grSU(N_c) / \Z_{N_c}$ gauge theory consists of a single flux tube sector of the $\grSU(N_c)$ gauge theory, selected based on the value of a discrete $\theta$ parameter.   An important quantity is the fundamental string tension $\sigma$, defined as the difference between the energy density of the vacuum of the fundamental flux tube and that of the vacuum of the trivial flux tube,
 \begin{equation} \label{eq:string_tension}
     \sigma \equiv \varepsilon_{p=1} - \varepsilon_{p=0} \,.
 \end{equation}
When $\sigma \neq 0$, the Wilson loop in the fundamental representation exhibits area-law scaling and the theory confines probe fundamental charges, while when $\sigma = 0$ the fundamental Wilson loop exhibits perimeter-law scaling.

When $m = 0$, adjoint QCD$_2$ exhibits extra symmetries. Indeed, for $m = 0$ the discrete $\Z_2$ transformation mentioned above is a symmetry, and it was realized recently that the theory also possesses non-invertible symmetry \cite{Komargodski:2020mxz}.  The infrared physics is described by the topological coset $\grSO(N_c^2 -1)_1 / \grSU(N_c)_{N_c}$, which has vanishing central charge.  There are $2^{N_c - 1}$ degenerate vacua split between the $N_c$ flux tube sectors.  Above each of these vacua, the spectrum of excitations is gapped, as shown in light-cone quantization \cite{Dalley:1992yy, Bhanot:1993xp, Demeterfi:1993rs, Katz:2013qua, Dempsey:2021xpf, Dempsey:2022uie} and confirmed using the Hamiltonian lattice approach \cite{Dempsey:2023fvm, Dempsey:2024alw}.  An implication of the vacuum degeneracy is that $\sigma = 0$ \cite{Gross:1995bp, Gross:1997mx, Komargodski:2020mxz, Dempsey:2021xpf}, and therefore the fundamental Wilson loop exhibits perimeter-law scaling.  As we will show explicitly for $N_c = 2, 3$, $\sigma$ vanishes only for $m=0$.  As we increase $m$, $\sigma$ increases and approaches the value $\frac{N_c^2 - 1}{4N_c} g^2$ at large $m$, as expected from the pure Yang-Mills theory.

For $m = g \sqrt{\frac{N_c}{2 \pi}}$, adjoint QCD$_2$ has $(1, 1)$ supersymmetry.  The supersymmetry was first observed in light-cone quantization \cite{Kutasov:1993gq, Boorstein:1993nd} (see also \cite{Popov:2022vud}) in the trivial flux tube sector.  It was later confirmed in the small circle expansion \cite{Dempsey:2024ofo}, and proven in \cite{Klebanov:2025mbu} by explicitly constructing a gauge-invariant and space-time covariant expression for the supercurrent.  The conservation of the supercurrent relies crucially on a certain quantum anomaly \cite{Klebanov:2025mbu}.   For $N_c = 2, 3$, the supersymmetry was also confirmed using Hamiltonian lattice gauge theory \cite{Dempsey:2023fvm, Dempsey:2024alw}.  The non-light-cone methods also give access to the non-trivial flux tube sectors, where the theory becomes gapless due to the fact that the supersymmetry is spontaneously broken and the low-energy physics is described by a massless Goldstino (which is a Majorana fermion).

In the vicinity of the supersymmetric point, the low-energy physics is governed by a massive Majorana fermion with mass $m_f$ proportional to $m - m_\text{SUSY}$.  It is well-known that the massless point of a Majorana fermion separates two distinct topological phases.\footnote{We thank Shu-Heng Shao for a discussion on this topic.}  Since a Majorana fermion is realized on the lattice by a Kitaev chain \cite{Kitaev:2000nmw}, the two phases are the continuum limits of the two topological phases of the Kitaev chain.  If $m_f > 0$ corresponds to the trivial phase of the Kitaev chain, as we will assume, then $m_f < 0$ corresponds the non-trivial phase.\footnote{One needs to impose boundary conditions at $\pm \infty$ in order to distinguish the two phases.  Equivalently, we can distinguish the two phases if the theory is placed on a finite interval or on a circle with periodic boundary conditions (PBC) or anti-periodic boundary conditions (APBC).  On an interval, the trivial phase has a non-degenerate ground state, while the non-trivial phase has a doubly-degenerate ground state.  On a circle, the trivial phase with APBC, the non-trivial phase with APBC, and the trivial phase with PBC each have unique ground states of the same fermion parity, while the non-trivial phase with PBC has a unique ground state of the opposite fermion parity.\label{footnote:Majorana}}  Thus, while in the trivial flux tube sector the ground state phase diagram corresponds to a trivial unique ground state for all $m>0$, in the non-trivial flux tube sectors the phase diagram consists of distinct topological phases separated by gapless points at $m = m_\text{SUSY}$, where we have a free Majorana CFT\@.   See Figure~\ref{fig:adjoint_phase_diagram} for the phase diagrams of the $N_c = 2$ and $3$ theories.

\tikzstyle{circle}=[fill=red!50!white, draw=black, shape=circle, minimum width=1.cm, minimum height=1.cm, inner sep=0pt, outer sep=0pt, thick]
\begin{figure}
    \centering
    \begin{subfigure}[t]{\textwidth}
        \centering
        \begin{tikzpicture}[xscale=4]
            \node at (0,2.3) {$\boxed{\grSU(2)}$};
            \node[left] at (-1.6,1) {\footnotesize $p = 0$};
            \draw[green!50!black,ultra thick] (-1.5,1) -- (-.5,1);
            \draw[-latex,ultra thick] (-.5,1) -- (1.5,1);
            \node[fill,circle,minimum size=5,inner sep=0] at (-.5,1) {};
            \node[below] at (-.5,.9) {$-m_\text{SUSY}$};
            \node[above] at (-.5,1.1) {\color{red!50} \scriptsize Gapless};
            \node[above,green!50!black,align=center] at (-1,1) {\footnotesize Non-trivial\\[-.3em]\footnotesize phase};
            \node[above,align=center] at (1,1) {\footnotesize Trivial\\[-.3em]\footnotesize phase};
            \node[left] at (-1.6,-1) {\footnotesize $p = 1$};
            \draw[green!50!black,ultra thick] (-1.5,-1) -- (.5,-1);
            \draw[-latex,ultra thick] (.5,-1) -- (1.5,-1);
            \node[fill,circle,minimum size=5,inner sep=0] at (.5,-1) {};
            \node[below] at (.5,-1.1) {$m_\text{SUSY}$};
            \node[above] at (.5,-.9) {\color{red!50} \scriptsize Gapless};
            \node[above,green!50!black,align=center] at (-.5,-1) {\footnotesize Non-trivial\\[-.3em]\footnotesize phase};
            \node[above,align=center] at (1,-1) {\footnotesize Trivial\\[-.3em]\footnotesize phase};

            \begin{scope}[yshift=-5.5cm]
            \node at (0,2.3) {$\boxed{\grSU(3)}$};
            \node[left] at (-1.6,1) {\footnotesize $p = 0$};
            \draw[-latex,ultra thick] (-1.5,1) -- (1.5,1);
            \draw[very thick] (0,1.2) node[above] {\scriptsize 1\textsuperscript{st} order} -- (0,.8) node[below] {$0$};
            \node[above,align=center] at (-1,1) {\footnotesize Trivial\\[-.3em] \footnotesize phase};
            \node[above,align=center] at (1,1) {\footnotesize Trivial\\[-.3em] \footnotesize phase};
            \node[left] at (-1.6,-1) {\footnotesize $p = 1,2$};
            \draw[green!50!black,ultra thick] (-.5,-1) -- (.5,-1);
            \draw[ultra thick] (-1.5,-1) -- (-.5,-1);
            \draw[-latex,ultra thick] (.5,-1) -- (1.5,-1);
            \node[fill,circle,minimum size=5,inner sep=0] at (-.5,-1) {};
            \node[fill,circle,minimum size=5,inner sep=0] at (.5,-1) {};
            \node[below] at (-.5,-1.1) {$-m_\text{SUSY}$};
            \node[below] at (.5,-1.1) {$m_\text{SUSY}$};
            \node[above] at (-.5,-.9) {\color{red!50} \scriptsize Gapless};
            \node[above] at (.5,-.9) {\color{red!50} \scriptsize Gapless};
            \node[above,green!50!black,align=center] at (0,-1) {\footnotesize Non-trivial\\[-.3em]\footnotesize phase};
            \node[above,align=center] at (1,-1) {\footnotesize Trivial\\[-.3em]\footnotesize phase};
            \node[above,align=center] at (-1,-1) {\footnotesize Trivial\\[-.3em]\footnotesize phase};
            \end{scope}
        \end{tikzpicture}        
    \end{subfigure}%
    \caption{Zero-temperature phase diagram of the flux tube sectors of the $\grSU(2)$ and $\grSU(3)$ theories.}
    \label{fig:adjoint_phase_diagram}
\end{figure}
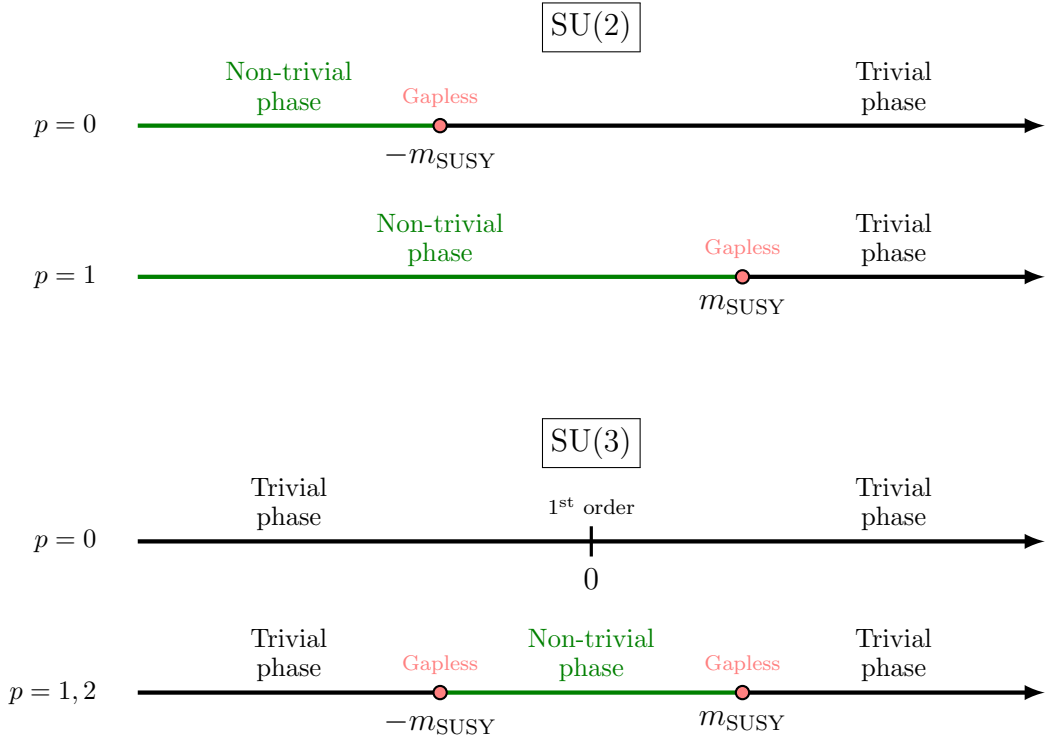

In these phase diagrams, we have also included the region $m<0$.  As explained above, the physics in the region $m<0$ can be obtained from that in the region $m>0$ by the $\Z_2$ axial transformation $\psi \to \gamma^5 \psi$, which maps $m \to -m$.  However, there is a mixed 't Hooft anomaly between this transformation and the $\Z_{N_c}$ one-form symmetry when $N_c$ is even \cite{Cherman:2019hbq}.  The effect of the anomaly is that the $\Z_2$ axial transformation also maps $p \to p + N_c /2$.  Then, for $N_c = 2$, the gapless point at $m = m_\text{SUSY}$ in the $p=1$ flux tube sector is mapped to $m = - m_\text{SUSY}$ in the $p=0$ sector;  for $N_c = 3$, the gapless points at $m = m_\text{SUSY}$ in the $p=1, 2$ flux tube sectors are mapped to $m = - m_\text{SUSY}$ in the same sectors.  When the theory is placed on a circle with periodic boundary conditions, the 't Hooft anomaly between the axial $\Z_2$ and the one-form $\Z_{N_c}$ symmetry is accompanied by a 't Hooft anomaly between the axial $\Z_2$ and fermion parity $(-1)^F$, present also only for even $N_c$ \cite{Cherman:2019hbq}.  The consequence of the latter anomaly (see also the discussion in Footnote~\ref{footnote:Majorana}) is that the axial transformation also interchanges the trivial and non-trivial phases of the topological theory.  These consequences are illustrated in the diagrams in Figure~\ref{fig:adjoint_phase_diagram}.

Another feature of the phase diagram in Figure~\ref{fig:adjoint_phase_diagram} is the first-order phase transition at $m=0$ in the $p=0$ flux tube sector of $\grSU(3)$. This phase transition occurs because there are two degenerate vacua in this flux tube sector at $m = 0$, and they have different values of $\langle \tr \bar\psi \psi\rangle$. From first-order perturbation theory, if we turn on a small positive mass, the ground state with the lowest value of $\langle \tr \bar\psi \psi\rangle$ becomes the unique vacuum; if we turn on a small negative mass, the ground state with the largest value of $\langle \tr \bar\psi \psi\rangle$ becomes the unique vacuum. In fact, the generic behavior for higher $N_c$ is to have a first-order phase transition of this kind in every flux tube sector, because at $m = 0$ each flux tube sector will contain at least two vacua with different values of $\langle \tr \bar\psi \psi\rangle$.  

We will now study various observables as functions of $m/g$ using the Hamiltonian lattice approach combined with the tensor network method introduced in the previous section.

\subsubsection{Numerical results}

The lattice Hamiltonian is given by \cite{Dempsey:2023fvm,Dempsey:2024alw}
\begin{equation}\label{eq:adjointH}
    H = \sum_n \left\lbrack \frac{g^2 a}{2}L^a_n L^a_n - \frac{i}{2}\left(a^{-1} + (-1)^n m\right)\chi_n^a \chi_{n+1}^a\right\rbrack\,,
\end{equation}
where $a=1,\ldots,\dim G$ is an adjoint index. On each site of the lattice, we have a Hilbert space transforming in the representation $\bm{R} = [11\cdots 1]$ of $\SU(N_c)$ along with $N_c - 1$ uncharged Majorana fermions $\lambda_{n,j}$ (with $j=1,\ldots,N_c - 1$). The fermion operators $\chi_n^a$ are represented as combinations of $\SU(N_c)$ invariant tensors and these Majorana fermions. For $\SU(2)$ and $\SU(3)$, the explicit expressions are \cite{Dempsey:2024alw}
\begin{equation}\label{eq:chi_defs}
    \chi^a_{\grSU(2),n} = \frac{1}{\sqrt{2}} \sigma^a_n \lambda_{n,1}, \qquad \chi^a_{\grSU(3),n} = \frac{1}{\sqrt{2}}F^a_n\lambda_{n,1} + \sqrt{\frac{3}{2}} D^a_n \lambda_{n,2} \,,
\end{equation}
respectively. Here, $\sigma^a_n$ are the usual Pauli matrices acting on the $\bm{R} = \bm{2}$ Hilbert space on site $n$, and $F^a_n$ and $D^a_n$ are given in terms of the $f$- and $d$-symbols\footnote{As in \cite{Dempsey:2024alw}, we define these symbols by
\begin{equation}
    [T^a, T^b] = if^{abc}T^c, \qquad \left\lbrace T^a, T^b\right\rbrace = \frac{1}{3}\delta^{ab}\mathds{1} + d^{abc}T^c\,,
\end{equation}
where the generators $T^a$ are normalized by $\tr(T^a T^b) = \frac{1}{2}\delta^{ab}$, and then take
\begin{equation}
    {(F^a)^b}_c = -i f^{abc}, \qquad {(D^a)^b}_c = d^{abc}\,.
\end{equation}
} of $\SU(3)$ and act upon the $\bm{R} = \bm{8}$ Hilbert space on site $n$.

To place the degrees of freedom on the lattice, we combine the Majorana fermions on sites $2n-1$ and $2n$ into a $2^{N_c - 1}$-dimensional Hilbert space (transforming trivially under $\SU(N_c)$) on an additional middle site.  Thus, if in the Hamiltonian \eqref{eq:adjointH} we have $N$ sites, our lattice implementation will have $3N/2$ sites. For the $\SU(2)$ theory, the on-site Hilbert spaces transform as $\bm{2},\bm{1}^{\oplus 2},\bm{2},\bm{2},\bm{1}^{\oplus 2},\bm{2},\ldots$ under $\grSU(2)$, and for the $\SU(3)$ theory they transform as $\bm{8},\bm{1}^{\oplus 4},\bm{8},\bm{8},\bm{1}^{\oplus 4},\bm{8},\ldots$ under $\grSU(3)$. Because of this prescription, the $\chi_{2n-1}^a\chi^a_{2n}$ and $\chi^a_{2n}\chi^a_{2n+1}$ terms have to be implemented differently. Details of the LEMPOs for the $\SU(2)$ and $\SU(3)$ theories can be found in Appendix~\ref{app:adjoint_lempos}. 

We can study a particular flux tube sector by restricting the irreps of $\SU(N_c)$ that can appear in the virtual spaces of the infinite MPS\@. After \eqref{eq:bdryH}, we describe how to restrict to a given flux tube sector on a finite chain. On an infinite chain, the flux tube sector with a probe quark in a representation with $N_c$-ality $p$ has irreps with $N_c$-ality $p$ on the link to the left (and to the right) of the unit cell of the infinite MPS\@. Note that for even $N_c$, some virtual spaces within the unit cell will not have $N_c$-ality equal to $p$; see \cite{Dempsey:2024alw}.

In \cite{Dempsey:2023fvm,Dempsey:2024alw}, the lattice model \eqref{eq:adjointH} was studied using exact diagonalization of the full Hilbert space (with a conservative truncation of the representations on links), and so it was only feasible to study up to $N = 12$ sites for $N_c = 2$ and up to $N = 6$ sites for $N_c = 3$. Now, using symmetric matrix product states and LEMPOs in combination with the VUMPS algorithm and the quasiparticle ansatz, we can study the theory on an infinite lattice and obtain far more precise results.

\begin{figure}
    \centering
    \begin{subfigure}[t]{0.85\linewidth}
        \includegraphics[width=\linewidth]{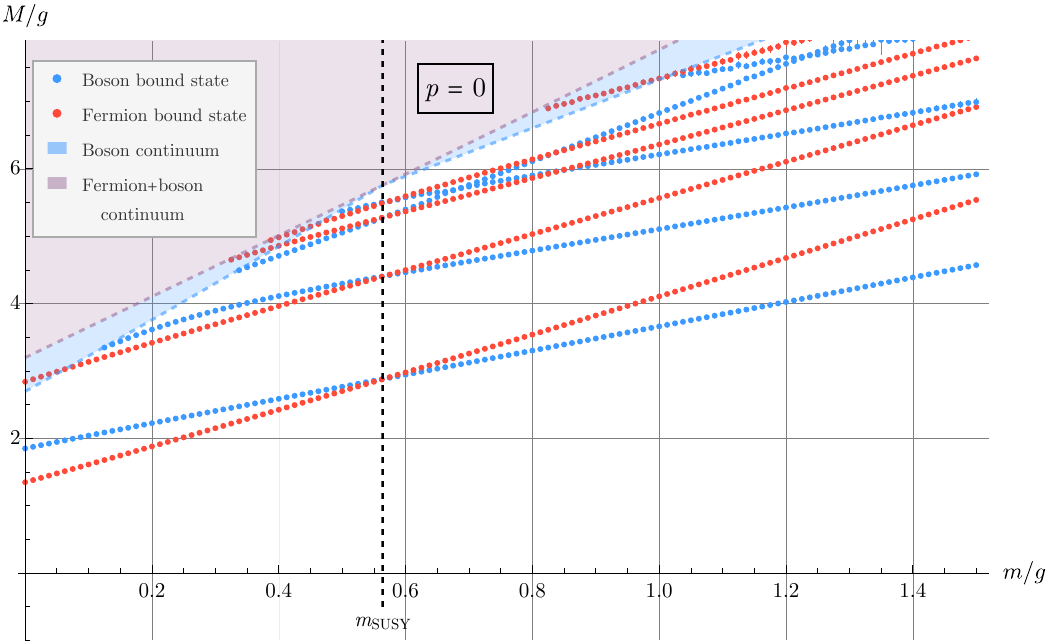}
        \caption{}
        \label{fig:su2_spectrum_p0}
    \end{subfigure}\\[1em]
    \begin{subfigure}[t]{0.85\linewidth}
        \includegraphics[width=\linewidth]{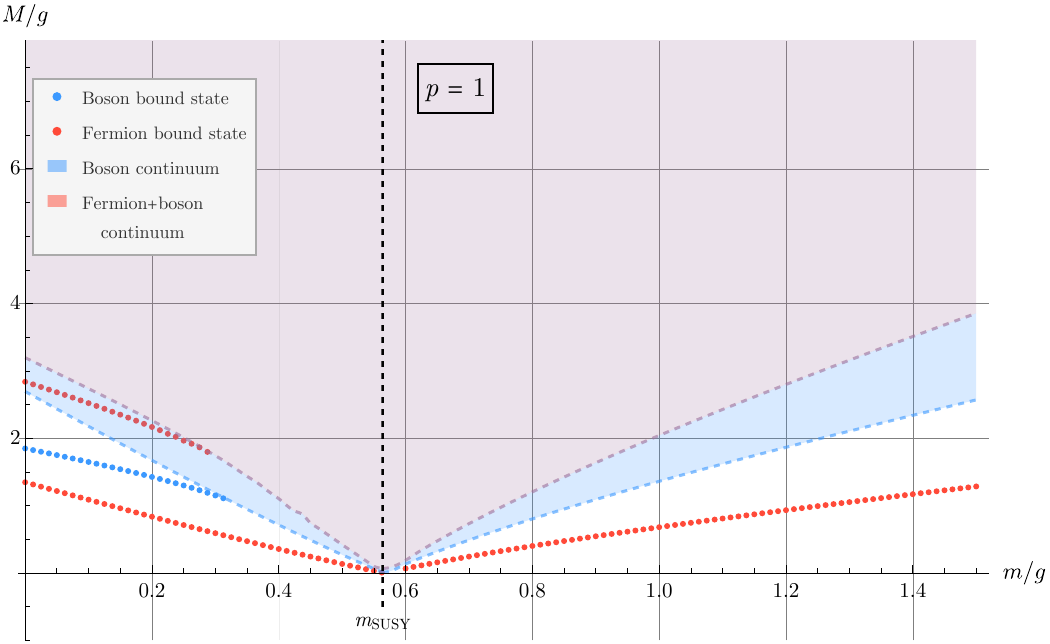}
        \caption{}
        \label{fig:su2_spectrum_p1}
    \end{subfigure}
    \caption{The low-lying spectrum of $\SU(2)$ adjoint QCD$_2$, including special behavior at $m = m_\text{SUSY} =  \frac{g}{\sqrt{\pi}}$.}
    \label{fig:adjoint_su2_spectrum}
\end{figure}

We will start with the $N_c = 2$ theory, which is the easiest case numerically. We will report values in units of the 't Hooft coupling $g^2 N_c$; as we will see shortly, many (but not all) quantities are nearly independent of $N_c$ when reported in these units. Our estimates for the masses of the bound states (two fermions and one boson, where fermion parity is defined relative to the vacuum using a $\mathbb{Z}_2$-charged quasiparticle ansatz \cite{Haegeman:2013}) and the chiral condensate at $m=0$ in the $p = 0$ universe are as follows (the second column is in the units traditionally used in DLCQ):
\begin{equation}\label{eq:su2_numbers}
\begin{aligned}
    \underline{N_c = 2}:& \\
    M_F^{(1)} &= 0.952774(7) \,g\sqrt{N_c}\,, & \left(M_F^{(1)}\right)^2 &= 5.70374(9)\,\frac{g^2 N_c}{2\pi}\,,\\
    M_B^{(1)} &= 1.30775(2) \,g\sqrt{N_c} \,, & \left(M_B^{(1)}\right)^2 &= 10.7455(3)\,\frac{g^2 N_c}{2\pi}\,,\\
    M_F^{(2)} &= 2.00745(6) \,g\sqrt{N_c} \,, & \left(M_F^{(2)}\right)^2 &= 25.320(2)\,\frac{g^2 N_c}{2\pi}\,,\\
    \langle \tr(\bar\psi\psi)\rangle_{p} &= (-1)^{p+1}\cdot 0.16135(8)\,\frac{g\sqrt{N_c}(N_c^2-1)}{2} \,.
\end{aligned}
\end{equation}
Note that the fermion bilinear condensate has a value close to that of the massless Schwinger model when rescaled in this way; this was first noticed in \cite{Bergner:2024ttq}. Errors are estimated using an ensemble of polynomial fits to extrapolate $a\to 0$, as in Section~\ref{subsec:schwinger_results}.

We can estimate the mass-dependent spectrum of the theory by performing similar extrapolations at other values of $m$. In Figure~\ref{fig:adjoint_su2_spectrum}, we plot the spectrum of $\SU(2)$ adjoint QCD$_2$ in both of its flux tube sectors for $m\ge 0$.\footnote{The 't Hooft anomalies among the symmetries of the $\SU(2)$ theory imply that the spectrum at fermion mass $m$ in the $p = 0$ flux tube sector is identical to the spectrum at fermion mass $-m$ in the $p = 1$ flux tube sector \cite{Cherman:2019hbq,Dempsey:2023fvm}.}  In the $p = 0$ flux tube sector, we clearly see boson-fermion degeneracy at the supersymmetric mass $m = m_\text{SUSY} = \frac{g}{\sqrt{\pi}}$ \cite{Dempsey:2022uie,Popov:2022vud}. In the $p = 1$ flux tube sector, the model is gapless at this mass; as mentioned above, the absence of a gap is due to a massless Goldstino of spontaneously broken supersymmetry \cite{Kutasov:1993gq}.

\begin{figure}
    \centering
    \includegraphics[width=0.7\linewidth]{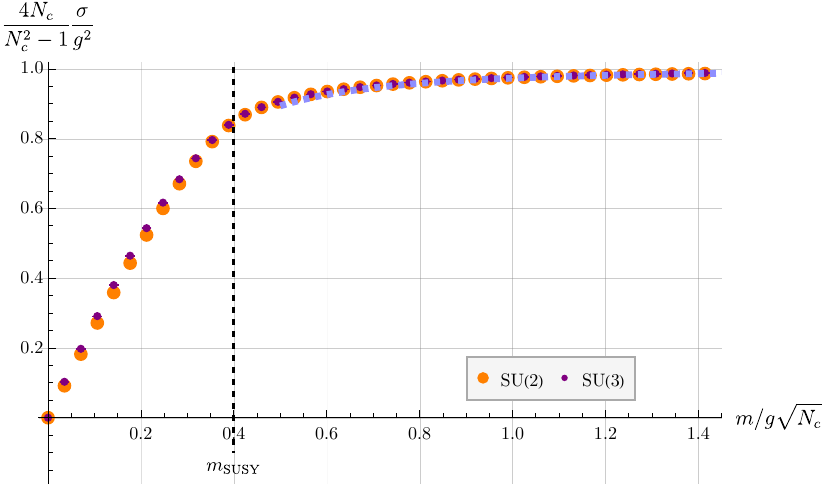}
    \caption{The fundamental string tension in the $\SU(2)$ and $\SU(3)$ theories. The blue dashed line is the weak-coupling expression \eqref{eq:string_tension_wc}.}
    \label{fig:adjoint_string_tension}
\end{figure}

It is also interesting to calculate the fundamental string tension \eqref{eq:string_tension}.  For each value of $m$ we can extrapolate $\sigma$ as computed on the lattice to the $a\to 0$ continuum limit, and obtain the data in Figure~\ref{fig:adjoint_string_tension}.  (See also \cite{Bergner:2024ttq} for a computation of the string tension using Euclidean lattice methods.)  At $m = 0$, the string tension for the $\SU(2)$ theory vanishes as a consequence of the 't Hooft anomalies \cite{Cherman:2019hbq}; at small $m$ we have 
\begin{equation}\label{eq:sigma_lowmass_su2}
    \sigma \sim m\left(\langle\tr(\bar\psi \psi)\rangle^{\grSU(2)}_{p=1} -  \langle\tr(\bar\psi \psi)\rangle^{\grSU(2)}_{p=0}\right) = 2m\left|\langle\tr(\bar\psi \psi)\rangle^{\grSU(2)}_{p=0}\right|,
\end{equation}
where $\langle\tr(\bar\psi \psi)\rangle^{\grSU(2)}_{p} \approx (-1)^{p+1} 0.34 g$ is the value of the chiral condensate at $m = 0$. At large $m$, the leading-order fundamental string tension for any $N_c$ is $\frac{N_c^2 - 1}{4N_c}g^2$ (i.e., $\frac{1}{2}g^2$ times the quadratic Casimir eigenvalue of the fundamental representation); the one-loop correction to the gauge coupling then gives
\begin{equation}\label{eq:string_tension_wc}
    \frac{4N_c}{N_c^2-1}\frac{\sigma}{g^2} = 1 - \frac{1}{12\pi}\frac{g^2 N_c}{m^2} + \ldots
\end{equation}
at large mass (see \cite{Dempsey:2023gib} for a very similar calculation). In Figure \ref{fig:adjoint_string_tension} we see that the lattice data agrees well with \eqref{eq:string_tension_wc} at large mass.

For the $\SU(3)$ theory, the dimensions of the virtual spaces needed to capture the ground state are significantly larger than for $\SU(2)$, and so our numerical accuracy is more limited. For the theory with $m = 0$, our estimates for the bound state masses in the $p=0$ flux tube sector and chiral condensate are as follows:\footnote{From Figure \ref{fig:su3_spectrum} below, it appears that at $m=0$ there is a third fermionic bound state of mass $\approx 3.3g$. However, based on the DLCQ results of \cite{Dempsey:2022uie}, this particle should be interpreted as part of a fermionic continuum that starts where the bosonic continuum does.}
\begin{equation}\label{eq:su3_numbers}
\begin{aligned}
    \underline{N_c = 3}:& \\
    M_F^{(1)} &= 0.954(2) \,g\sqrt{N_c}\,, & \left(M_F^{(1)}\right)^2 &= 5.72(3)\,\frac{g^2 N_c}{2\pi}\,,\\
    M_B^{(1)} &= 1.306(5) \,g\sqrt{N_c} \,, & \left(M_B^{(1)}\right)^2 &= 10.71(8)\,\frac{g^2 N_c}{2\pi}\,,\\
    M_F^{(2)} &= 1.651(3) \,g\sqrt{N_c} \,, & \left(M_F^{(2)}\right)^2 &= 17.13(5)\,\frac{g^2 N_c}{2\pi}\,,\\
    \langle \tr(\bar\psi\psi)\rangle_{p=0,\pm} &= \pm 0.160(1)\,\frac{g\sqrt{N_c}(N_c^2-1)}{2}\,,\\
    \langle \tr(\bar\psi\psi)\rangle_{p=1} &= \langle \tr(\bar\psi\psi)\rangle_{p=2} = 0\,.
\end{aligned}
\end{equation}
Note that the $\pm$ index of $\langle \tr(\bar\psi\psi)\rangle_{p=0,\pm}$ indicates which of the two degenerate vacua of the $p = 0$ flux tube sector that we calculate the expectation value in, as labeled by the sign of the expectation value of $\tr (\bar \psi \psi)$. In the $p=\pm 1$ flux tube sector at $m=0$, the bound state masses are degenerate with those reported in \eqref{eq:su3_numbers} and have multiplicities as indicated in Figure~\ref{fig:su3_spectrum}.

\begin{figure}
    \centering
    \begin{subfigure}[t]{0.48\textwidth}
        \centering
        \includegraphics[width=\linewidth]{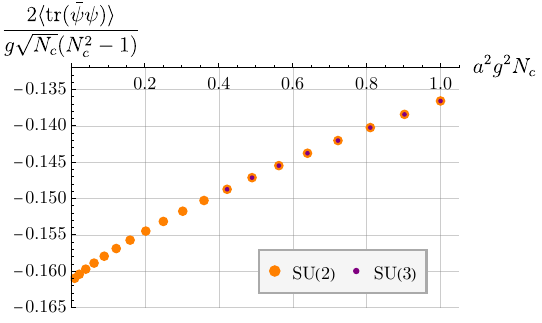}
        \caption{}
    \end{subfigure}%
    \hspace{.02\textwidth}%
    \begin{subfigure}[t]{0.48\textwidth}
        \centering
        \includegraphics[width=\linewidth]{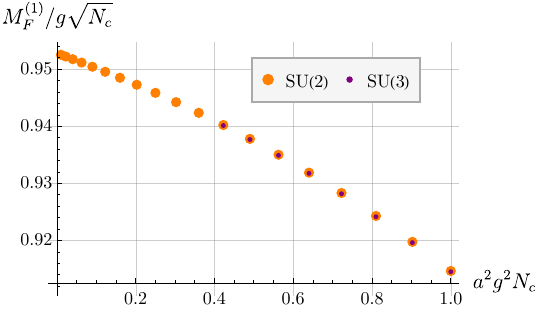}
        \caption{}
    \end{subfigure}%
    \caption{The chiral condensate and the lowest fermion mass for the $\SU(2)$ and $\SU(3)$ theories for $m=0$ in the $p = 0$ universe, showing that these quantities are nearly independent of $N_c$ when rescaled as shown and plotted as a function of $a^2 g^2 N_c$. For the $\SU(3)$ theory, there are two degenerate vacua in the $p = 0$ universe with opposite values of the chiral condensate; we work in the one for which it is negative.}
    \label{fig:su2_vs_su3}
\end{figure}

For some quantities, the values for the $\SU(2)$ and $\SU(3)$ theories are extremely close, even on the lattice, when the quantities themselves as well as the lattice spacing are expressed in units of the 't Hooft coupling. In Figure~\ref{fig:su2_vs_su3}, we show examples of this phenomenon for the lightest fermion mass and the chiral condensate in the massless theory. We would thus conjecture that these quantities in the $\SU(3)$ theory are very close to the values given in \eqref{eq:su2_numbers}; these results are consistent with observations made using other methods \cite{Bergner:2024ttq,Dempsey:2022uie}. The same is true of the lightest boson mass at $m = 0$. However, the second-lightest fermion differs substantially between $N_c = 2$ and $N_c \ge 3$. This was observed previously in \cite{Dempsey:2022uie} in the context of lightcone quantization. There, it was found that the second-lightest fermion for $N_c \ge 3$ is charge-conjugation odd, and since the $\grSU(2)$ theory does not have such a sector, there is no analogous particle in the $\grSU(2)$ theory.

The fundamental string tension in the $\SU(3)$ theory is plotted in Figure~\ref{fig:adjoint_string_tension}. For $m>m_\text{SUSY}$, it nearly coincides with the $\SU(2)$ value when rescaled as in the figure. At $m = 0$, the string tension again vanishes, which can be understood as a consequence of the non-invertible symmetries of the theory \cite{Komargodski:2020mxz}. At small masses, we have
\begin{equation}
    \sigma \sim m\left(\langle\tr(\bar\psi \psi)\rangle^{\grSU(3)}_{p=1} - \langle\tr(\bar\psi \psi)\rangle^{\grSU(3)}_{p=0,-}\right) = m\left|\langle\tr(\bar\psi \psi)\rangle^{\grSU(3)}_{p=0}\right|\,,
\end{equation}
where $\langle\tr(\bar\psi \psi)\rangle^{\grSU(3)}_{p=0,-}$ is the negative value given in \eqref{eq:su3_numbers}. Using this equation along with \eqref{eq:sigma_lowmass_su2}, and the fact that $\langle \tr(\bar\psi \psi)\rangle/(g\sqrt{N_c}(N_c^2-1))$ is nearly independent of $N_c$, we find that the slope of the $\SU(3)$ curve in Figure~\ref{fig:adjoint_string_tension} should be $\frac{9}{8} = 1.125$ times that of the $\SU(2)$ curve near $m = 0$. And indeed, at $m = 0.05g$ we find a ratio of $1.125(7)$. By contrast, for $m\gtrsim m_\text{SUSY}$ the string tension as rescaled in Figure~\ref{fig:adjoint_string_tension} is very similar between the $N_c = 2$ and $N_c = 3$ theories; for instance, at $m = m_\text{SUSY}$ we find
\begin{equation}
    \left.\frac{8}{3}\frac{\sigma}{g^2}\right|_{\grSU(2),\,m=g\sqrt{\frac{1}{2\pi}}} = 0.848(1)\,, \qquad \left.\frac{3}{2}\frac{\sigma}{g^2}\right|_{\grSU(3),\,m=g\sqrt{\frac{3}{2\pi}}} = 0.850(2) \,.
\end{equation}
The latter value is consistent with the estimation given in \cite{Dempsey:2024alw}, but now with far greater precision.

\begin{figure}[h!]
    \centering
    \begin{subfigure}[t]{0.85\textwidth}
        \centering
        \includegraphics[width=\linewidth]{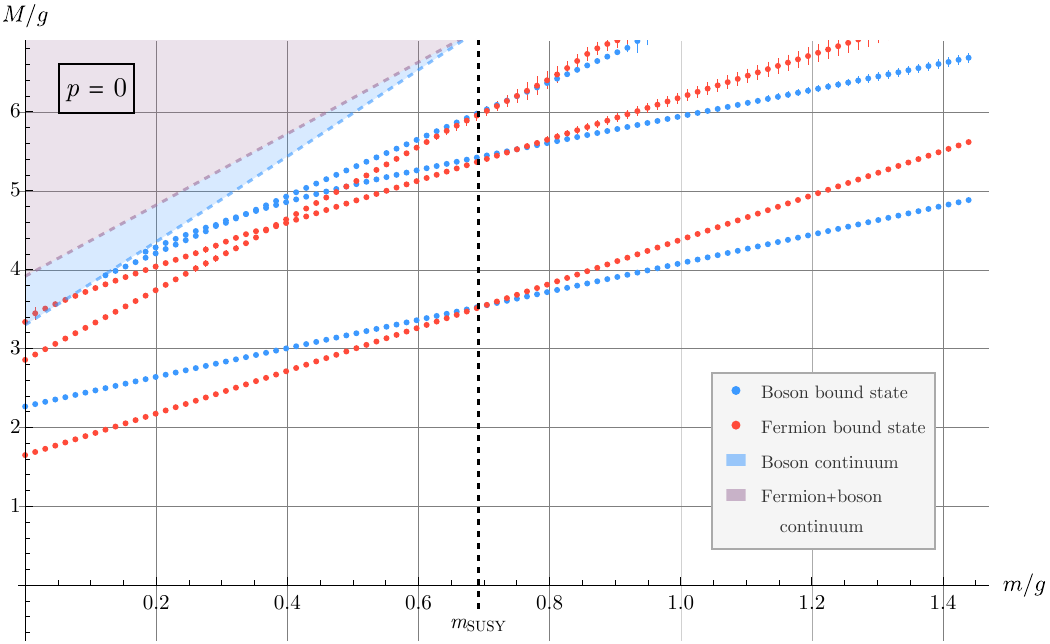}
        \caption{}
    \end{subfigure}\\[1em]
    \begin{subfigure}[t]{0.85\textwidth}
        \centering
        \includegraphics[width=\linewidth]{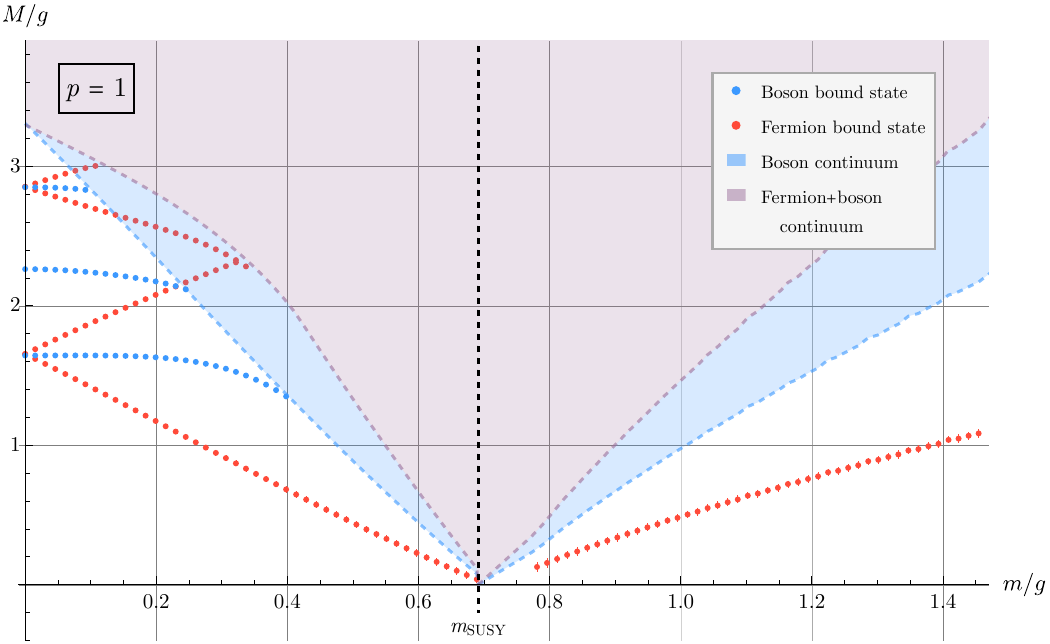}
        \caption{}
    \end{subfigure}%
    \caption{The low-lying spectrum of $\SU(3)$ adjoint QCD$_2$, including special behavior at $m =  m_\text{SUSY} =  g\sqrt{\frac{3}{2\pi}}$. Note that in the bottom panel, we have assumed the gaplessness at $m = m_\text{SUSY}$ in order to extend the continuum threshold curves through the near-critical region where the numerics become especially difficult.}
    \label{fig:su3_spectrum}
\end{figure}

We can also study the spectrum of the $\SU(3)$ theory as a function of mass. In Figure~\ref{fig:su3_spectrum}, we plot the masses of the lightest bound states. We see that in the $p = 0$ flux tube sector, there is boson-fermion degeneracy at $m = m_\text{SUSY} = \sqrt{\frac{3}{2\pi}}g$. In the $p = 1$ flux tube sector there is a gapless point at $m = m_\text{SUSY}$, again due to a massless Goldstino of spontaneously broken supersymmetry.

Note that extracting the correct spectrum is complicated by the non-trivial fermionic phases shown in Figure \ref{fig:adjoint_phase_diagram}. In particular, depending on the details of how the model \eqref{eq:adjointH} is bosonized (see Appendix \ref{app:adjoint_lempos}), the fermionic bound states may be represented in the bosonized lattice model as either a particle or a solitonic excitation. We plan to elaborate further on this subtlety in upcoming work.

\FloatBarrier

\section{Discussion and future directions}\label{sec:discussion}
In this work, we introduced LEMPOs, a generalization of MPOs that allow for operators to act on virtual bonds of MPSs. LEMPOs efficiently encode the Hamiltonian of lattice gauge theories in a local and translational-invariant manner. To demonstrate the efficacy of LEMPOs, we used uMPS with the VUMPS algorithm and the quasiparticle ansatz to perform a precision study of ground state properties and the bound state spectrum for the Schwinger model and adjoint QCD$_2$ with gauge groups $\SU(2)$ and $\SU(3)$.

As emphasized in the main text, various MPS algorithms can be augmented to include LEMPOs with relative ease. Thus, a natural direction for future work would be to apply some of the MPS methods that have been developed in a condensed matter context to $(1+1)$-dimensional gauge theories. One application is real-time evolution using algorithms such as the time-dependent variational principle \cite{Haegeman:2011zz,Haegeman:2016gfj}. Such a simulation would allow for the study of scattering events \cite{Jha:2024jan}, or for the calculation of dynamical two-point functions from which one can extract parton distribution functions in gauge theories \cite{Banuls:2025wiq}. It would also be interesting to adapt the methods of \cite{Vanderstraeten_smatrix}, where the $S$-matrix elements are directly extracted from an ansatz constructed out of two quasiparticle ans\"atze, to gauge theories.

Another promising future direction is the study of $(2+1)$-dimensional lattice gauge theories. This generalization involves both physical and computational difficulties relative to what we presented in Section~\ref{sec:lmpo}. Physically, in higher dimensions we have to contend with there being dynamical gauge degrees of freedom. Although our presentation in Section~\ref{sec:lmpo} relied upon fully gauge-fixing our theory, this is in fact not essential. Rather than eliminating the $\H_\text{gauge}$ part of the Hilbert space, as we did in Section~\ref{subsec:lgt}, we can keep the full Hilbert space. Indeed, in the $(1+1)$-dimensional setup of Section~\ref{subsec:lgt}, we can construct the corresponding state in the full Hilbert space by contracting the representation indices for the virtual spaces of the MPS tensors with the indices of the gauge kets, as follows:
\begin{align}\label{eq:lgt_mps}
    \ket{A_1,\ldots,A_N} &= \sum_{\substack{i_n,\beta_n\\m_{n,L},m_{n,R}}} (A_1)_{i_1}^{m_{0,R},(\beta_1,m_{1,L})} (A_2)_{i_2}^{(\beta_1,m_{1,R}),(\beta_2,m_{2,L})} \cdots (A_{N})_{i_{N}}^{(\beta_{N-1},m_{N-1,R}),m_{N,L}} \notag\\
    &\times  \ket{\bm{r}_0; m_{0, R}} \otimes \ket{\bm{R}_1;i_1}\otimes\cdots\otimes \ket{\bm{R}_N;i_N}\otimes\ket{\bm{r}_N; m_{N, L}}\\
    &\otimes\frac{\ket{\bm{r}(\beta_1);m_{1,L},m_{1,R}}}{\sqrt{\dim \bm{r}(\beta_1)}}\otimes\cdots\otimes\frac{\ket{\bm{r}(\beta_{N-1});m_{N-1,L},m_{N-1,R}}}{\sqrt{\dim \bm{r}(\beta_N)}} \,.\notag
\end{align}
\begin{figure}
		\centering
		\begin{subfigure}[t]{0.48\textwidth}
			\centering
			\begin{tikzpicture}[scale=1.1]
				\node[circle, draw, fill=black, minimum size=1.5mm] (a1) at (0,0) {};
				\node[circle, draw, fill=black, minimum size=1.5mm] (a2) at (2,0) {};
				\node[circle, draw, fill=black, minimum size=1.5mm] (a3) at (4,0) {};
				\draw[very thick] (-0.3,0) node[left,inner sep=0,] {$\dots$} -- (a1) -- (a2) node[midway, fill=white, inner sep=0,sloped] {\tiny$\ket{\bm r;m_{L},m_{R}}$} -- (a3) node[midway, fill=white, inner sep=0,sloped] {\tiny$\ket{\bm r;m_{L},m_{R}}$} -- (4.3,0) node[right,inner sep=0,] {$\dots$};
				\draw[very thick] (a1) -- ++(0,-0.3) node[below] {\tiny$\ket{\bm R;i}$};
				\draw[very thick] (a2) -- ++(0,-0.3) node[below] {\tiny$\ket{\bm R;i}$};
				\draw[very thick] (a3) -- ++(0,-0.3) node[below] {\tiny$\ket{\bm R;i}$};
			\end{tikzpicture}
			\caption{}
		\end{subfigure}%
		\hspace{.02\textwidth}%
		\begin{subfigure}[t]{0.48\textwidth}
			\centering
			\begin{tikzpicture}[scale=1.1]
				\node[circle, draw, fill=black, minimum size=1.5mm] (a11) at (0,0) {};
				\node[circle, draw, fill=black, minimum size=1.5mm] (a12) at (2,0) {};
				\node[circle, draw, fill=black, minimum size=1.5mm] (a21) at (-40:1.8) {};
				\node[circle, draw, fill=black, minimum size=1.5mm] (a22) at ($(2,0) + (-40:1.8)$) {};
				\draw[very thick] (a11) -- (a12) node[midway, fill=white, inner sep=0,sloped] {\tiny$\ket{\bm r;m_{L},m_{R}}$};
				\draw[very thick] (a21) -- (a22) node[midway, fill=white, inner sep=0,sloped] {\tiny$\ket{\bm r;m_{L},m_{R}}$};
				\draw[very thick] (a21) -- (a11) node[midway, fill=white, inner sep=0,sloped] {\tiny$\ket{\bm r;m_{L},m_{R}}$};
				\draw[very thick] (a22) -- (a12) node[midway, fill=white, inner sep=0,sloped] {\tiny$\ket{\bm r;m_{L},m_{R}}$};
				\draw[very thick] (a11) -- ++(0,-0.3) node[below] {\tiny$\ket{\bm R;i}$};
				\draw[very thick] (a12) -- ++(0,-0.3) node[below] {\tiny$\ket{\bm R;i}$};
				\draw[very thick] (a21) -- ++(0,-0.3) node[below] {\tiny$\ket{\bm R;i}$};
				\draw[very thick] (a22) -- ++(0,-0.3) node[below] {\tiny$\ket{\bm R;i}$};
				\draw[very thick] (a22) -- ++(0.3,0) node[right,inner sep=0,] {$\dots$};
				\draw[very thick] (a12) -- ++(0.3,0) node[right,inner sep=0,] {$\dots$};
				\draw[very thick] (a11) -- ++(-0.3,0) node[left,inner sep=0,] {$\dots$};
				\draw[very thick] (a21) -- ++(-0.3,0) node[left,inner sep=0,] {$\dots$};
				\draw[very thick] (a22) -- ++(-40:0.27) node[right,inner sep=0, rotate = -40] {$\dots$};;
				\draw[very thick] (a21) -- ++(-40:0.27) node[right,inner sep=0, rotate = -40] {$\dots$};
				\draw[very thick] (a11) -- ++(140:0.27) node[right,inner sep=0, rotate = 140] {$\dots$};
				\draw[very thick] (a12) -- ++(140:0.27) node[right,inner sep=0, rotate = 140] {$\dots$};
			\end{tikzpicture}
			\caption{}
            \label{fig:peps}
	\end{subfigure}%
		\caption{The contraction pattern in generalized MPS (a) and PEPS (b) for lattice gauge theory. The index contractions in the figure are schematic and emphasize that the gauge field states $\ket{\bm r;m_{L},m_{R}}$ are contracted into to neighboring MPS/PEPS tensor, see~\eqref{eq:lgt_mps} for the precise expression for (a).}
        \label{fig:mps_peps}
	\end{figure}
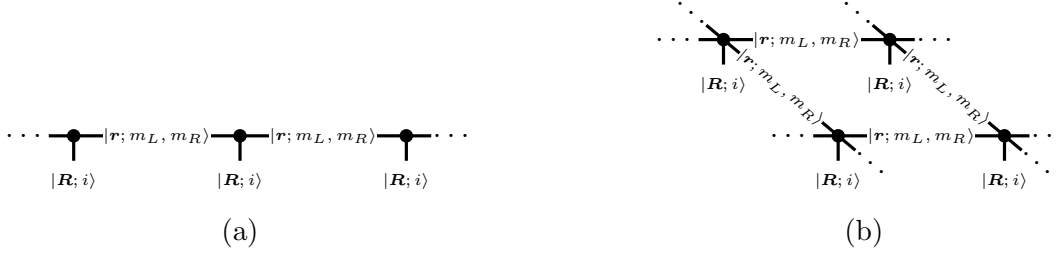
A state of the form \eqref{eq:lgt_mps} can be straightforwardly generalized to higher-dimensional tensor networks in the projected entangled pair state (PEPS) construction. In Figure~\ref{fig:mps_peps}, we show schematically how this generalization proceeds. An important caveat is that, when taking the inner product of two of the states in Figure~\ref{fig:peps}, one cannot simply contract the tensors as though they were ordinary PEPSs: it is important to account for the orthogonality of the gauge kets. By contrast, for the finite chains considered in this paper, the fact that we could fully gauge-fix the theory implies that there is no analogous subtlety.

In addition to the difficulties in calculating inner products when we are dealing with gauge theories, tensor network methods are already considerably more complicated in more than one spatial dimension. Although gradient-based schemes built on algorithms such as the corner transfer matrix renormalization group \cite{Nishino:1995yte} can be used to minimize the energy of a PEPS, they are not nearly as efficient or robust as analogous algorithms in one spatial dimension. For all of these reasons, we leave the implementation of our ideas in $(2+1)$ or higher dimensions to future work.

Another direction, which will be the topic of upcoming work, is to apply the methods described in this work to $(1+1)$-dimensional gauge theories that exhibit a rich vacuum structure. One example of this is adjoint QCD$_2$: for gauge group $\SU(N_c)$ with $N_c>3$, this theory has more vacua than can be explained by conventional 't Hooft anomalies, and this is understood as a consequence of non-invertible symmetries \cite{Komargodski:2020mxz}. In addition to properties of the vacua, recent developments also predict novel degeneracies between particles and solitons in certain gauge theories \cite{Cordova:2024nux}, which could be tested using the Hamiltonian lattice techniques described in this paper. We hope to report on lattice realizations of these predictions in a future publication.

\section*{Acknowledgments}

We thank Shu-Heng Shao for useful discussions, and Igor Klebanov for useful discussions, comments on a draft of this paper, and collaboration on related work.  All numerical calculations have been carried out using the Julia packages \texttt{MPSKit.jl} \cite{MPSKit2025} and \texttt{TensorKit.jl} \cite{Devos:2025yoj}.  This work is supported in part by the U.S.~Department of Energy under Award No.~DE-SC0007968\@.  RD is also supported in part by a Princeton University Charlotte Elizabeth Procter Fellowship.

\appendix

\section{Quantum mechanics of gauge fixing}\label{app:quantum_gauge}

In this appendix, we provide more details on the gauge-fixing procedure for the lattice Schwinger model.  We do so by adapting the arguments of \cite{Lenz:1994tb,Lenz:1994cv}, which studied the continuum $(3+1)$d QED and QCD theories.  (See also \cite{Lenz:1994du} for an application of the same method to $(1+1)$-dimensional adjoint QCD.) 

For the reader's convenience, we recall some key equations that already appeared in the main text. The Hamiltonian of the lattice Schwinger model is given by
\begin{equation}\label{eq:appH}
    H=\frac{ag^2}{2}\sum_{n=1}^{N-1} \left(L_n+\frac{\theta}{2\pi}\right)^2-\frac{i}{2a}\sum_{n=1}^{N-1}(c_n^\dagger U_n c_{n+1}-c_{n+1}U_n^\dagger c_n)+\frac{m}{2}\sum_{n=1}^{N}(-1)^n c^\dagger_n c_n\, ,
\end{equation}
and the Gauss law is
\begin{equation}\label{eq:appGauss}
    G_n \equiv L_n-L_{n-1} - Q_n=0\,, \qquad \text{with}\qquad Q_n \equiv c^\dagger_nc_n - \delta_{n,\text{odd}}\,.
\end{equation}
Here $L_0, L_{N}\in\mathbb Z$ are the left-incoming and right-outgoing electric fields, respectively, and we set them to definite values as boundary conditions. The full Hilbert space of theory is $\mathcal H = \mathcal H_\text{matter}\otimes \mathcal H_\text{gauge}$, where $\mathcal H_\text{matter}$ is the $2^N$-dimensional Hilbert space generated by acting with $c_n^\dagger$ on the Fock vacuum $\ket{0}$, and $\mathcal H_\text{gauge}$ is the space of wavefunctions on a circle on each link (spanned by a basis of momentum eigenstates $\ket{\ell_n}$):
\begin{equation}
    \mathcal H_\text{gauge} = \bigotimes_{n=1}^{N-1}\mathcal H_{B,n}\,,\qquad \mathcal H_{B,n}=\spn\left\{\ket{\ell_n} \;\Big\vert\; \ell_n\in\mathbb Z\right\}\,.
\end{equation}
Thus, we want to study the Hamiltonian in the physical subspace 
 \begin{equation}
     \mathcal H_\text{phys} = \left\{\ket{\psi}\in \mathcal H \;\Big\vert\; G_n\ket{\psi} = 0 \text{ for $n=1,\ldots,N$} \right\} \,.
 \end{equation}
The fermionic oscillators satisfy the canonical anti-commutation relations $\{c_n,c_m^\dagger\}=\delta_{nm}$, and the gauge operators satisfy the algebra $[L_n,U_m]=\delta_{nm}U_n$ and act on the basis states as
\begin{equation}
    U_n\ket{\ell_n} = \ket{\ell_n+1}\,, \qquad L_n\ket{\ell_n} = \ell_n\ket{\ell_n}\,.
\end{equation}
It will also be convenient to introduce the Hermitian operators $\phi_n$ such that $U_n= e^{i\phi_n}$, which implies the identification $\phi_n \sim \phi_n+2\pi$ and the commutation relation $[\phi_n,L_m]=i\delta_{nm}$.

To gauge fix \eqref{eq:appH}, we then act with the unitary operator
\begin{equation}
    \mathcal U = \exp{\left[i\sum_{n=1}^{N-1}Q_n\sum_{m=n}^{N-1}\phi_m\right]}= \exp{\left[i\sum_{n=1}^{N-1}\phi_n\sum_{m=1}^{n}Q_m\right]}\,,
\end{equation}
which yields the relations
\begin{equation}
\begin{aligned}
    \mathcal U^\dagger c_n\mathcal U&= e^{i\sum_{m=n}^{N-1}\phi_m} c_n = U_n\ldots U_{N-1}c_n\,,  &\text{for $1 \leq n\leq N$}\, ,\\
    \mathcal U^\dagger U_n\mathcal U&= U_n\,,&\text{for $1\leq n<N$}\,,\\
    \mathcal U^\dagger L_n\mathcal U&= L_n + \sum_{m=1}^n Q_m\,,  &\text{for $1\leq n<N$}\, .
\end{aligned}
\end{equation}
The Hamiltonian \eqref{eq:appH} and Gauss law \eqref{eq:appGauss} transform as
\begin{equation}
\begin{aligned}
    \mathcal U^\dagger H \mathcal U &=\frac{g^2a}{2}\sum_{n=1}^{N-1} \left(L_n+\frac{\theta}{2\pi}+\sum_{m=1}^nQ_m\right)^2\\
    &{}-\frac{i}{2a}\sum_{n=1}^{N-1}(c_n^\dagger c_{n+1}-c_{n+1}^\dagger c_n)+\frac{m}{2}\sum_{n=1}^N(-1)^n c^\dagger_n c_n\,, \\
    \mathcal U^\dagger G_n \mathcal U &= L_{n}-L_{n-1} \qquad \text{for $1\leq n<N$}\, ,\\
    \mathcal U^\dagger G_N \mathcal U &= L_N-L_{N-1} - \sum_{n=1}^NQ_n\,.
\end{aligned}
\end{equation}
After this transformation, the Gauss law is trivially solved by going to the subspace with $L_n=L_0$ for $n=1,2,\ldots N-1$. All that remains is the constraint on the total charge
\begin{equation}\label{eq:appTotal}
    L_N-L_0 = \sum_{n=1}^N Q_n\,,
\end{equation}
which follows from $G_N = 0$.

Thus, in summary, we find the fully gauge-fixed Hamiltonian
\begin{equation}
    H_\text{gauge-fixed} =\frac{ag^2}{2}\sum_{n=1}^{N-1} \left(\frac{\theta}{2\pi}+\sum_{m=1}^nQ_m\right)^2
-\frac{i}{2a}\sum_{n=1}^{N-1}(c_n^\dagger c_{n+1}-c^\dagger_{n+1} c_n)+\frac{m}{2}\sum_{n=1}^N(-1)^n c^\dagger_n c_n\,,
\end{equation}
subject to the constraint \eqref{eq:appTotal}.

\section{Factoring states into MPSs}\label{app:factoring}
In this appendix, we will describe how to factor an arbitrary quantum state on a one-dimensional chain into an MPS\@. For more details on the matrix factorization algorithms used, see the review \cite{Schollwoeck:2010uqf}.

Consider the following generic state $\ket{\psi}$, represented by a tensor with $N$ indices
\begin{equation}\label{eq:wavefn}
	\ket{\psi} =\begin{tikzpicture}[diagram]
		\draw[rounded corners] (-.35,-.35) rectangle (4.35,.35);
		\draw (0,-.35) -- ++(0,-.4);
		\draw (1.5,-.35) -- ++(0,-.4);
		\draw (4.,-.35) -- ++(0,-.4);
		\draw[fill=white, draw=none]  (2.25,-.4) rectangle (3.25,.4);
		\node at (2.75,0) {$\ldots$};
	\end{tikzpicture}
\end{equation}
that is understood to be contracted with the bases of the Hilbert spaces on each site. We seek to cast this state as an MPS\@. 

The main ingredient in the construction below is a matrix factorization algorithm, such as the singular value decomposition, which allows us to write a linear transformation $A$ (from some domain to some codomain) as a product of two others, $A=LR$. Given a tensor with many indices, we frame it as a matrix by choosing which indices correspond to the domain and which ones correspond to the codomain, and then apply this factorization. For example, a three-index tensor can be decomposed as
\begin{equation}
\begin{tikzpicture}[diagram]
		\node[style=square] at (0,0) (A) {$A$};
		\draw (A) -- ++(0.75,0) node[right] {domain};
		\draw (A) -- ++(0,-0.75) node[below] {domain};
		\draw (A) -- ++(-0.75,0) node[left] {codomain};
	\end{tikzpicture}=
	\begin{tikzpicture}[diagram]
		\node[style=square] at (0,0) (L) {$L$};
		\node[style=square] at (1.5,0) (R) {$R$};
		\draw (R) -- (L);
		\draw (R) -- ++(0.75,0);
		\draw (L) -- ++(-0.75,0);
		\draw (R) -- ++(0,-0.75);
	\end{tikzpicture}\,.
\end{equation}
To factor the state $\ket{\psi}$ in \eqref{eq:wavefn}, we treat the first physical index as corresponding to the codomain and the other $N-1$ physical indices as corresponding to the domain, and factorize. Then, in the resulting right block, we take the newly created virtual index and the second physical index to describe the codomain and the remaining $N-2$ physical indices as representing the domain. Repeating this procedure yields an MPS:
\begin{equation}\label{eq:iterate}
	\begin{aligned}
	\ket{\psi} &=\begin{tikzpicture}[diagram]
		\node[style=square] at (0,0) (A) {};
		\draw[rounded corners] (1.15,-.35) rectangle (4.35,.35);
		\draw (A) -- ++(1.15,0);
		\draw (0,-.35) -- ++(0,-.4);
		\draw (1.5,-.35) -- ++(0,-.4);
		\draw (4.,-.35) -- ++(0,-.4);
		\draw[fill=white, draw=none]  (2.25,-.4) rectangle (3.25,.4);
		\node at (2.75,0) {$\ldots$};
	\end{tikzpicture}\\
	&=\begin{tikzpicture}[diagram]
		\node[style=square] at (0,0) (A1) {};
		\node[style=square] at (1.5,0) (A2) {};
		\draw[rounded corners] (2.65,-.35) rectangle (4.35,.35);
		\draw (A1) -- (A2) -- ++(1.15,0);
		\draw (0,-.35) -- ++(0,-.4);
		\draw (1.5,-.35) -- ++(0,-.4);
		\draw (4.,-.35) -- ++(0,-.4);
		\draw[fill=white, draw=none]  (2.25,-.4) rectangle (3.25,.4);
		\node at (2.75,0) {$\ldots$};
	\end{tikzpicture}\\
	&\;\,\vdots\\
	&=\begin{tikzpicture}[diagram]
		\node[style=square] at (0,0) (A1) {};
		\node[style=square] at (1.5,0) (A2) {};
		\node[style=square] at (4.,0) (A3) {};
		\draw (A1) -- (A2) -- ++(.75,0);
		\draw (A3) -- ++(-.75,0);
		\draw (0,-.35) -- ++(0,-.4);
		\draw (1.5,-.35) -- ++(0,-.4);
		\draw (A3) -- ++(0,-.75);
		\node at (2.75,0) {$\ldots$};
	\end{tikzpicture}\,.
	\end{aligned}
\end{equation}
To make this into a symmetric MPS, we can consider the rank-$N$ tensor in \eqref{eq:wavefn} as an invariant tensor of some group $G$, with the physical spaces transforming in some representation $\mathcal H_n=\bm R$. To be more precise, in the first step of \eqref{eq:iterate}, we have the following domain and codomain:
\begin{equation}
	\mathcal V_\text{codomain} = \overline{\mathcal H_1}=\overline{\bm R}= \bigoplus_i \bm r_{\text{codomain},i}\, , \qquad \mathcal V_\text{domain} = \bigotimes_{n=2}^N\mathcal H_n = \bigoplus_i \bm r_{\text{domain},i}\,.
\end{equation}
The last equalities give the decompositions into irreps $\bm r_i$. Then, applying Schur's lemma to the linear map $M:\mathcal V_\text{domain}\rightarrow \mathcal V_\text{codomain}$, we find that as a matrix it must have the block structure 
\begin{equation}
	M = \left( \begin{array}{c|c|c}
		M_{\bm r_1} & &  \\
		\hline & M_{\bm r_2} & \\ \hline & & \ddots 
	\end{array}\right)\,,
\end{equation}
where $M_{\bm{r}_j}$ is a matrix mapping the copies of $\bm{r}_j$ in the domain to the copies of $\bm{r}_j$ in the codomain. Thus, the blocks are only non-trivial if $\bm{r}_j \in \V_\text{domain} \cap \V_\text{codomain}$. The matrix factorization procedure can then be modified to respect the group structure by factoring each block separately.

We can also extend this factorization method to states with non-trivial boundary conditions, which we represent with two additional uncontracted indices corresponding to the representation of a probe quark at the left and right boundary (see the discussion around \eqref{eq:bdryH}). Starting from this modified tensor,
\begin{equation}
	\ket{\psi} =\begin{tikzpicture}[diagram]
		\draw[rounded corners] (-.35,-.35) rectangle (4.35,.35);
		\draw (0,-.35) -- ++(0,-.4);
		\draw (1.5,-.35) -- ++(0,-.4);
		\draw (4.,-.35) -- ++(0,-.4);
		\draw[fill=white, draw=none]  (2.25,-.4) rectangle (3.25,.4);
		\node at (2.75,0) {$\ldots$};
		\draw (-.35,0) -- ++(-.4,0);
		\draw (4.35,0) -- ++(.4,0);
	\end{tikzpicture}\,,
\end{equation}
we can apply the same logic as in \eqref{eq:iterate} to factorize the state as
\begin{equation}\label{eq:bc}
	\ket{\psi} = \begin{tikzpicture}[diagram]
		\node[style=square] at (0,0) (A1) {};
		\node[style=square] at (1.5,0) (A2) {};
		\node[style=square] at (4.,0) (A3) {};
		\draw (A1) -- (A2) -- ++(.75,0);
		\draw (A3) -- ++(-.75,0);
		\draw (0,-.35) -- ++(0,-.4);
		\draw (1.5,-.35) -- ++(0,-.4);
		\draw (A3) -- ++(0,-.75);
		\node at (2.75,0) {$\ldots$};
		\draw (-.35,0) -- ++(-.4,0);
		\draw (4.35,0) -- ++(.4,0);
	\end{tikzpicture}\,.
\end{equation}
As described around \eqref{eq:bdryH}, this modification is important for studying non-trivial flux tube sectors.

\section{Some implementation details}\label{app:algo}
In this appendix, we will comment on a few technical aspects of LEMPOs and describe the use of LEMPOs in the context of one-site DMRG.

\subsection{MPS gauge transformations and LEMPOs}\label{app:mps_gauge}
A fundamental property of MPSs is the ability to transform their tensors via the introduction of invertible transformations of the virtual spaces, as follows:

\tikzstyle{circle}=[fill=white, draw=black, shape=circle, minimum width=1.cm, minimum height=1.cm, inner sep=0pt, outer sep=0pt, thick]

\begin{equation}\label{eq:gauge_trans}
\begin{split}
	\ket{ \ldots,A_n,A_{n+1},\ldots} &= 
	\begin{tikzpicture}[diagram]
		\node[style=square] at (0,0) (A2) {$A_{n}$};
		\node[style=square] at (1.5,0) (A3) {$A_{n+1}$};
		\node at (-1,0) {$\cdots$};
		\node at (2.5,0) {$\cdots$};
		\draw (A2) -- (A3);
		\draw (A2) -- ++(-0.75,0);
		\draw (A3) -- ++(0.75,0);
		\draw (A2) -- ++(0,-0.75);
		\draw (A3) -- ++(0,-0.75);
	\end{tikzpicture}=
	\begin{tikzpicture}[diagram]
	\draw[shade] (-.6,-.5) rectangle (1.6,1.);
	\draw[shade] (1.9,-.5) rectangle (4.1,1.);
	\node at (1,.65) {$A'_{n}$};
	\node at (3.5,.65) {$A'_{n+1}$};
	\node[style=square] at (0,0) (A2) {$A_{n}$};
	\node[style=circle] at (1,0) (X2) {$X_{n}$};
	\node[style=circle] at (2.5,0) (X3) {$X^{-1}_{n}$};
	\node[style=square] at (3.5,0) (A3) {$A_{n+1}$};
	\node at (-1,0) {$\cdots$};
	\node at (4.5,0) {$\cdots$};
	\draw (A2)--(X2)--(X3) -- (A3);
	\draw (A2) -- ++(-0.75,0);
	\draw (A3) -- ++(0.75,0);
	\draw (A2) -- ++(0,-0.75);
	\draw (A3) -- ++(0,-0.75);
	\end{tikzpicture} \\
    &= \ket{ \ldots,A_n',A_{n+1}',\ldots}\,.
\end{split}
\end{equation}
This transformation is often called a regauging, but it is not to be confused with physical gauge invariance.

Since \eqref{eq:gauge_trans} is a redundancy in the description of the physical state, no physical quantities should depend on the choice of MPS gauge. Nevertheless, in numerical implementations we use a specially-chosen transformation \eqref{eq:gauge_trans} to put the MPS in so-called mixed canonical form. This mixed canonical form is crucial for the numerical stability of various variational algorithms \cite{Vanderstraeten:2019voi}.  

The LEMPO defined in \eqref{eq:HtotLEMPO} inserts a tensor $W^L_n$ on the virtual bond, and so it is not immediately clear that the MPS gauge transformation \eqref{eq:gauge_trans} commutes with the action of a LEMPO\@. We argued below \eqref{eq:old_action} that for the matrices $X$ allowed for a symmetric MPS, there exist operators that commute with $X$, and that they are precisely the class of link operators that we need for a lattice gauge theory. Let us briefly repeat this argument in a more general setting.

For a symmetric MPS, $X$ must be a symmetric tensor with respect to the gauge group. In particular, by Schur's lemma, $X$ acts as a block-diagonal matrix (one block for each irrep appearing in the virtual space). Again by Schur's lemma, a matrix that commutes with all such matrices $X$ must act as a multiple of the identity on each of these blocks, but the coefficient of the identity can differ from block to block. Thus, as long as $W_n^L$ is built from matrices with this property, it will obey 
\begin{equation} 
	\begin{tikzpicture}[diagram]
		\node[style=circle] at (5,0) (X) {$X$};
		\node[style=circle] at (6,0) (W3) {$W^L_{n}$};
		\draw (X)-- (W3);
		\draw (X) -- ++(-0.75,0);
		\draw (W3) -- ++(0.75,0);
		\draw (W3) to[out=35, in=180] (6.75,0.45);
		\draw (4.25,.45) to[out=0, in=145] (W3);
	\end{tikzpicture} =
	\begin{tikzpicture}[diagram]
		\node[style=circle] at (6,0) (X) {$X$};
		\node[style=circle] at (5,0) (W3) {$W^L_{n}$};
		\draw (X)-- (W3);
		\draw (X) -- ++(0.75,0);
		\draw (W3) -- ++(-0.75,0);
		\draw (W3) to[out=35, in=180] (6.75,0.45);
		\draw (4.25,.45) to[out=0, in=145] (W3);
	\end{tikzpicture}
\end{equation}
and thus respect the regauging freedom of the MPS\@. And indeed, in this paper the link operators we consider are all built from the Casimir operator, which has this property.

\subsection{$W_n^L$ and dynamically-adjusted virtual spaces}\label{subapp:link}

In Section~\ref{subsec:LEMPO}, we described the matrices that form a LEMPO\@. The components of the matrices $W_n^\text{LEMPO}$ are operators acting on the physical Hilbert spaces $\H_n$, which are fixed. However, the components of the matrices $W_n^{L,\text{LEMPO}}$ are operators acting on the virtual spaces $\V_n$, which are typically adjusted dynamically by variational optimization algorithms.

Thus, we need to repeatedly update the operators in $W_n^{L,\text{LEMPO}}$ to be commensurate with the spaces $\V_n$. In the cases we consider in this paper, these updates are particularly simple because we are only interested in link operators that are multiples of the Casimir \eqref{eq:link_L2}. This class of operators acts by scalar multiplication on each irreducible representation in the decomposition of $\V_n$. Thus, rather than pre-computing and storing the tensors $W_n^{L,\text{LEMPO}}$, we simply compute the Casimir dynamically and act by scalar multiplication.

\subsection{Effective Hamiltonians for LEMPOs}
As an example of how LEMPOs can be implemented into MPS algorithms, we consider how to set up single-site DMRG (see \cite{Schollwoeck:2010uqf} for a review) using LEMPOs. A single step of this algorithm proceeds by minimizing the expectation value of the Hamiltonian with respect to a single MPS tensor while keeping the rest fixed. By sweeping back and forth along the chain and updating the MPS tensors one at time, the algorithm converges to the approximate ground state.

The expectation value of a Hamiltonian expressed as a LEMPO in a state represented by an unnormalized MPS is given by
\begin{equation}
    E(A_1,\ldots,A_N) = \frac{\begin{tikzpicture}[diagram]
		\node[style=square] at (0,0) (A1) {$A_{n-1}$};
		\node[style=circle] at (1,0) (WL1) {$W^L_{n-1}$};
		\node[style=square] at (2,0) (X) {$A_n$};
		\node[style=square] at (2,-2) (Xb) {$\bar A_n$};
		\node[style=circle] at (3,0) (WL2) {$W^L_{n}$};
		\node[style=square] at (4,0) (A2) {$A_{n+1}$};
		\node[style=circle] at (0,-1) (W1) {$W_{n-1}$};
		\node[style=circle] at (2,-1) (WX) {$W_{n}$};
		\node[style=circle] at (4,-1) (W2) {$W_{n+1}$};
		\node[style=square] at (0,-2) (Ab1) {$\bar A_{n-1}$};
		\node[style=square] at (4,-2) (Ab2) {$\bar A_{n+1}$};
		\draw (A2) -- ++(.75,0);
		\draw (W2) -- ++(.75,0);
		\draw (Ab2) -- ++(.75,0);
		\draw (A1) -- ++(-.75,0);
		\draw (W1) -- ++(-.75,0);
		\draw (Ab1) -- ++(-.75,0);
		\node at (-1.,0) {$\cdots$};
		\node at (-1.,-1) {$\cdots$};
		\node at (-1.,-2) {$\cdots$};
		\node at (5.,0) {$\cdots$};
		\node at (5.,-1) {$\cdots$};
		\node at (5.,-2) {$\cdots$};
		\draw (A1) -- (WL1) -- (X) -- (WL2) -- (A2);
		\draw (Ab1) -- (Xb) -- (Ab2);
		\draw (A1) -- (W1) -- (Ab1);
		\draw (A2) -- (W2) -- (Ab2);
		\draw (W1) to[out=0, in=-100] (WL1);
		\draw (WL1) to[out=-80, in=-180] (WX);
		\draw (WX) to[out=0, in=-100] (WL2);
		\draw (WL2) to[out=-80, in=-180] (W2);
		\draw (X) -- (WX) -- (Xb);
	\end{tikzpicture}}
	{\begin{tikzpicture}[diagram]
		\node[style=square] at (0,0) (A1) {$A_{n-1}$};
		\node[style=square] at (1.5,0) (X) {$A_n$};
		\node[style=square] at (1.5,-1.5) (Xb) {$\bar A_n$};
		\node[style=square] at (3,0) (A2) {$A_{n+1}$};
		\node[style=square] at (0,-1.5) (Ab1) {$\bar A_{n-1}$};
		\node[style=square] at (3,-1.5) (Ab2) {$\bar A_{n+1}$};
		\draw (A2) -- ++(.75,0);
		\draw (Ab2) -- ++(.75,0);
		\draw (A1) -- ++(-.75,0);
		\draw (Ab1) -- ++(-.75,0);
		\node at (-1.,0) {$\cdots$};
		\node at (-1.,-1.5) {$\cdots$};
		\node at (4.,0) {$\cdots$};
		\node at (4.,-1.5) {$\cdots$};
		\draw (A1) --  (X)  -- (A2);
		\draw (Ab1) -- (Xb) -- (Ab2);
		\draw (A1) -- (Ab1);
		\draw (A2) -- (Ab2);
		\draw (X) -- (Xb);
	\end{tikzpicture}}\,.
\end{equation}
To minimize the energy with respect to $A_n$, we treat $\bar A_n$ as an independent variable and require $\frac{\partial}{\partial \bar A_n}E(A_1,\ldots,A_n) = 0$, which yields the extremality criterion
\begin{equation}\label{eq:gen_eig}
\begin{tikzpicture}[diagram]
		\node[style=square] at (0,0) (A1) {$A_{n-1}$};
		\node[style=circle] at (1,0) (WL1) {$W^L_{n-1}$};
		\node[style=square] at (2,0) (X) {$A_n$};
		\node[style=circle] at (3,0) (WL2) {$W^L_{n}$};
		\node[style=square] at (4,0) (A2) {$A_{n+1}$};
		\node[style=circle] at (0,-1) (W1) {$W_{n-1}$};
		\node[style=circle] at (2,-1) (WX) {$W_{n}$};
		\node[style=circle] at (4,-1) (W2) {$W_{n+1}$};
		\node[style=square] at (0,-2) (Ab1) {$\bar A_{n-1}$};
		\node[style=square] at (4,-2) (Ab2) {$\bar A_{n+1}$};
		\draw (A2) -- ++(.75,0);
		\draw (W2) -- ++(.75,0);
		\draw (Ab2) -- ++(.75,0);
		\draw (A1) -- ++(-.75,0);
		\draw (W1) -- ++(-.75,0);
		\draw (Ab1) -- ++(-.75,0);
		\node at (-1.,0) {$\cdots$};
		\node at (-1.,-1) {$\cdots$};
		\node at (-1.,-2) {$\cdots$};
		\node at (5.,0) {$\cdots$};
		\node at (5.,-1) {$\cdots$};
		\node at (5.,-2) {$\cdots$};
		\draw (A1) -- (WL1) -- (X) -- (WL2) -- (A2);
		\draw (A1) -- (W1) -- (Ab1);
		\draw (A2) -- (W2) -- (Ab2);
		\draw (W1) to[out=0, in=-100] (WL1);
		\draw (WL1) to[out=-80, in=-180] (WX);
		\draw (WX) to[out=0, in=-100] (WL2);
		\draw (WL2) to[out=-80, in=-180] (W2);
		\draw (X) -- (WX) -- (2,-3.);
		\draw (Ab1) -- ++(1.5,0) to[out=0, in=0] ++(0,-.5);
		\draw (Ab2) -- ++(-1.5,0) to[out=180, in=180] ++(0,-.5);
	\end{tikzpicture}=E(A_1,\ldots,A_n)
    \begin{tikzpicture}[diagram]
		\node[style=square] at (0,0) (A1) {$A_{n-1}$};
		\node[style=square] at (1.5,0) (X) {$A_n$};
		\node[style=square] at (3,0) (A2) {$A_{n+1}$};
		\node[style=square] at (0,-1.5) (Ab1) {$\bar A_{n-1}$};
		\node[style=square] at (3,-1.5) (Ab2) {$\bar A_{n+1}$};
		\draw (A2) -- ++(.75,0);
		\draw (Ab2) -- ++(.75,0);
		\draw (A1) -- ++(-.75,0);
		\draw (Ab1) -- ++(-.75,0);
		\node at (-1.,0) {$\cdots$};
		\node at (-1.,-1.5) {$\cdots$};
		\node at (4.,0) {$\cdots$};
		\node at (4.,-1.5) {$\cdots$};
		\draw (A1) -- (X)-- (A2);
		\draw (A1) -- (Ab1);
		\draw (A2) -- (Ab2);
		\draw (X) -- (1.5,-2.5);
		\draw (Ab1) -- ++(1.,0) to[out=0, in=0] ++(0,-.5);
		\draw (Ab2) -- ++(-1.,0) to[out=180, in=180] ++(0,-.5);
	\end{tikzpicture}\,.
\end{equation}
We can treat this equation as a generalized eigenvalue problem by thinking of $A_n$ as a vector, and of the contractions on each side as different matrices acting on $A_n$; then, to find the minimizer, we compute the minimum-eigenvalue solution to this generalized eigenvalue problem. The only difference between \eqref{eq:gen_eig} and the usual setup of one-site DMRG with regular MPOs \cite{Schollwoeck:2010uqf} is the presence of the link operators $W^L_n$ on the left-hand side of \eqref{eq:gen_eig}. In view of Appendix~\ref{subapp:link}, it is straightforward to insert these link operators in existing routines. Moreover, it is possible to put the MPS into mixed canonical form despite the presence of the link operators (as argued in Appendix~\ref{app:mps_gauge}), which reduces \eqref{eq:gen_eig} to an ordinary eigenvalue problem \cite{Schollwoeck:2010uqf}.

Finally, we comment on how \eqref{eq:gen_eig} generalizes to two-site DMRG\@. In two-site DMRG, we treat two consecutive MPS tensors as one tensor with four indices and minimize the Hamiltonian with respect to this more general ansatz, and then put the minimizer back into MPS form using matrix factorization. This generalization is mostly straightforward, with the only subtlety being that the link operator $W_n^{L}$ is supposed to act on the virtual space inside the two-site block. This action can be implemented as
\begin{equation} 
	W_n^L:\quad \begin{tikzpicture}[diagram]
		\draw[rounded corners] (-.35,-.35) rectangle (1.35,.35);
		\draw (-.75,0) -- (-.35,0);
		\draw (1.35,0) -- (1.75,0);
		\draw (0,-.35) -- (0,-.75);
		\draw (1,-.35) -- (1,-.75);
		\node at (.5,0) {$X$};
	\end{tikzpicture}\quad \longmapsto \quad 
	\begin{tikzpicture}[diagram]
		\draw[rounded corners] (-.35,-.35) rectangle (1.35,.35);
		\node at (.5,0) {$X$};
		\node[style=circle] at (2.5,0) (L2) {$W_n^L$};
        \draw (L2) to[in=0,out=-120] ++(-.5,-.7);
        \draw (L2) to[in=180,out=-60] ++(.5,-.7);
		\draw (-.75,0) -- (-.35,0);
		\draw (0,-.35) -- (0,-.75);
		\draw (1.35,0) -- (2,0);
		\draw (1,-.35) to[out=-90, in=-90] (1.75,0) -- (L2) -- ++(1.75,0);
		\draw (3.5,0) -- ++(0,-0.75);
	\end{tikzpicture}\,.
\end{equation}
Here, the fusion of the two right-most indices of $X$ is an isomorphism given by a Clebsch-Gordan decomposition of their tensor product. The splitting to the right of $W_n^L$ is the inverse isomorphism. This implementation works because Schur's lemma guarantees that the irrep decomposition of the virtual space that would appear between the sites covered by $X$ is the same as the decomposition of the tensor product of the rightmost two indices.

\section{LEMPOs for $\SU(2)$ and $\SU(3)$ Adjoint QCD$_2$}\label{app:adjoint_lempos}

As described in Section~\ref{subsec:adjoint_results}, we arrange the degrees of freedom for the Hamiltonian lattice formulation of adjoint QCD$_2$ using a three-site unit cell; the $n$th unit cell corresponds to physical sites $2n-1$ and $2n$. Sites 1 and 3 of the unit cell have Hilbert spaces transforming in $\bm{R} = [11\cdots 1]$ while site 2 transforms in $2^{N_c-1}$ copies of the singlet representation. We will now describe LEMPOs on this unit cell that encode the Hamiltonian \eqref{eq:adjointH}.

For the $\SU(2)$ theory, we represent the Majorana fermion operators by
\begin{equation}
\begin{split}
    \lambda_{2n-1,1} &= Z_1Z_2\cdots Z_{n-1}X_n\,,\\
    \lambda_{2n,1} &= Z_1Z_2\cdots Z_{n-1}Y_n\,,
\end{split}
\end{equation}
where $\{X_n, Y_n, Z_n\}$ are Pauli matrices acting on the two-dimensional Hilbert space at the middle site of the $n$th unit cell. Using these expressions and applying \eqref{eq:chi_defs}, we find
\begin{equation}
    \chi_{2n-1}^a \chi_{2n}^a = \frac{i}{2}\sigma^a_{2n-1} Z_n \sigma^a_{2n}\qquad\text{and}\qquad \chi_{2n}^a \chi_{2n+1}^a = \frac{i}{2}X_n \sigma^a_{2n} \sigma^a_{2n+1} X_{n+1} \,.
\end{equation}
Thus, the matter matrices for the LEMPO are
\begin{equation}
\begin{split}
    W_{(1)}(n) &= \left(\begin{array}{c|c|c|c}
        \mathds{1}_{2n-1} &  & \frac{1}{4}\vec{\sigma}_{2n-1}  & \\
        \hline
         & \vec{\sigma}_{2n-1} &  &  \\
        \hline
         &  & &  \\
        \hline
         &  &  & \mathds{1}_{2n-1}
    \end{array} \right), \qquad W_{(2)}(n) = \left(\begin{array}{c|c|c|c}
        \mathds{1}_n & \frac{1}{4}X_n & & \\
        \hline
         & &  & X_n \\
        \hline
         &  & Z_n  &  \\
        \hline
         &  &  & \mathds{1}_n
    \end{array} \right),\\
    W_{(3)}(n) &= \left(\begin{array}{c|c|c|c}
        \mathds{1}_{2n} &  &  & \\
        \hline
         &  \vec{\sigma}_{2n} &  &  \\
        \hline
         &  &  & \vec{\sigma}_{2n} \\
        \hline
         &  &  & \mathds{1}_{2n}
    \end{array} \right)\,,
\end{split}
\end{equation}
with virtual spaces are $\W_1 = \W_2 = \W_3 = \bm{1} \oplus \bm{3} \oplus \bm{3} \oplus \bm{1}$. The link matrices are
\begin{equation}
\begin{split}
    W_{(1)}^L(n) &= \left(\begin{array}{c|c|c|c}
        I_{2n-1} &  &  & \frac{g^2 a}{2} \Q_{2n-1}^2 \\
        \hline
         & I_{2n-1} &  &  \\
        \hline
         &  & I_{2n-1} &  \\
        \hline
         &  &  & I_{2n-1}
    \end{array} \right), \qquad W_{(2)}^L(n) = \left(\begin{array}{c|c|c|c}
        I_n &  & & \\
        \hline
         & I_n & & \\
        \hline
         &  & I_n  &  \\
        \hline
         &  &  & I_n
    \end{array} \right),\\
    W_{(3)}^L(n) &= \left(\begin{array}{c|c|c|c}
        I_{2n} &  &  & \frac{g^2 a}{2} \Q_{2n}^2 \\
        \hline
         & I_{2n} &  &  \\
        \hline
         &  & I_{2n} &  \\
        \hline
         &  &  & I_{2n}
    \end{array} \right)\,.
\end{split}
\end{equation}
For the $\SU(3)$ theory, we have a completely analogous construction, but it will look more complicated because $\chi^a_n$ has two terms (see \eqref{eq:chi_defs}). We will represent the Majorana fermion operators by
\begin{equation}
\begin{split}
    \lambda_{2n-1,1} &= (ZZ)_1 (ZZ)_2 \cdots (ZZ)_{n-1} (XX)_n\,,\\
    \lambda_{2n-1,2} &= (ZZ)_1 (ZZ)_2 \cdots (ZZ)_{n-1} (XY)_n\,,\\
    \lambda_{2n,1} &= (ZZ)_1 (ZZ)_2 \cdots (ZZ)_{n-1} (YX)_n\,,\\
    \lambda_{2n,2} &= (ZZ)_1 (ZZ)_2 \cdots (ZZ)_{n-1} (YY)_n\,,
\end{split}
\end{equation}
where, for example, $(XY)_n$ denotes the Kronecker product of two Pauli matrices acting on the four-dimensional Hilbert space at the middle site of the $n$th unit cell. These expressions lead to
\begin{equation}
\begin{split}
    \chi^a_{2n-1}\chi^a_{2n} &= \frac{i}{2}F^a_{2n-1} (Z\mathds{1})_n F^a_{2n} - \frac{\sqrt{3}}{2} F^a_{2n-1} (ZZ)_n D^a_{2n} \\
    &\quad+ \frac{\sqrt{3}}{2} D^a_{2n-1} (ZZ)_n F^a_{2n} + \frac{3i}{4} D^a_{2n-1} (Z\mathds{1})_n D^a_{2n}\,,\\
    \chi^a_{2n} \chi^a_{2n+1} &= \frac{1}{2}(XY)_n F^a_{2n} F^a_{2n+1} (XX)_{n+1} + \frac{\sqrt{3}}{2}(XY)_n F^a_{2n} D^a_{2n+1} (XY)_{n+1} \\
    &\quad- \frac{\sqrt{3}}{2}(XX)_n D^a_{2n} F^a_{2n+1}(XX)_{n+1} - \frac{3}{4}(XX)_n D^a_{2n}D^a_{2n+1}(XY)_{n+1}\,.
\end{split}
\end{equation}
We can arrange these eight different kinds of terms in the Hamiltonian into $10\times 10$ LEMPO matrices. The link matrices are exactly as in the $\SU(2)$ case, except $10\times 10$, and the matter matrices are as follows:

\begingroup
\scriptsize
\begin{equation}
\begin{split}
    W_{(1)}(n) &= \left(\begin{array}{c|c|c|c|c|c|c|c|c|c}
    \mathds{1}_{2n-1} & & & & & \frac{i}{2}(F^a)_{2n-1} & -\frac{\sqrt{3}}{2} (F^a)_{2n-1} & \frac{\sqrt{3}}{2} (D^a)_{2n-1} & \frac{3i}{4}(D^a)_{2n-1} & \\
    \hline
    & (F^a)_{2n-1} & & & & & & & & \\
    \hline
    & & (D^a)_{2n-1} & & & & & & & \\
    \hline
    & & & (F^a)_{2n-1} & & & & & & \\
    \hline
    & & & & (D^a)_{2n-1} & & & & & \\
    \hline
    & & & & & & & & & \\
    \hline
    & & & & & & & & & \\
    \hline
    & & & & & & & & & \\
    \hline
    & & & & & & & & & \\
    \hline
    & & & & & & & & & \mathds{1}_{2n-1}
    \end{array}\right)\,,\\
    W_{(2)}(n) &= \left(\begin{array}{c|c|c|c|c|c|c|c|c|c}
    \mathds{1}_n & \frac{1}{2}(XY)_n & \frac{\sqrt{3}}{2}(XY)_n & -\frac{\sqrt{3}}{2}(XX)_n & -\frac{3}{4}(XX)_n & & & & & \\
    \hline
    & & & & & & & & & (XX)_n \\
    \hline
    & & & & & & & & & (XY)_n \\
    \hline
    & & & & & & & & & (XX)_n \\
    \hline
    & & & & & & & & & (XY)_n \\
    \hline
    & & & & & (Z \mathds{1})_n & & & & \\
    \hline
    & & & & & & (ZZ)_n & & & \\
    \hline
    & & & & & & & (ZZ)_n & & \\
    \hline
    & & & & & & & & (Z\mathds{1})_n & \\
    \hline
    & & & & & & & & & \mathds{1}_n
    \end{array}\right)\,,\\
    W_{(3)}(n) &= \left(\begin{array}{c|c|c|c|c|c|c|c|c|c}
    (\mathds{1})_{2n} & & & & & & & & & \\
    \hline
    & (F^a)_{2n} & & & & & & & & \\
    \hline
    & & (F^a)_{2n} & & & & & & & \\
    \hline
    & & & (D^a)_{2n} & & & & & & \\
    \hline
    & & & & (D^a)_{2n} & & & & & \\
    \hline
    & & & & & & & & & (F^a)_{2n} \\
    \hline
    & & & & & & & & & (D^a)_{2n} \\
    \hline
    & & & & & & & & & (F^a)_{2n} \\
    \hline
    & & & & & & & & & (D^a)_{2n} \\
    \hline
    & & & & & & & & & \mathds{1}_{2n}
    \end{array}\right)\,.
\end{split}
\end{equation}
\normalsize
\endgroup
The virtual spaces of these matrices are $\W_1 = \W_2 = \W_3 = \bm{1} \oplus \bm{8} \oplus \cdots \oplus \bm{8} \oplus \bm{1}$.

\bibliographystyle{JHEP}
\bibliography{refs.bib}

\end{document}